\documentclass[traditabstract]{aa}
\usepackage{natbib}
\usepackage{amssymb}
\usepackage{amsmath}
\usepackage{graphicx}
\usepackage{verbatim}
\usepackage{subfig}
\usepackage{epsfig}
\usepackage{tikz}
\usepackage{booktabs}
\usepackage{color}
\usepackage{multirow}
\newcommand{\Mg}{\ion{Mg}{\!}}
\newcommand{\Mgi}{\ion{Mg}{i}}
\newcommand{\Mgii}{\ion{Mg}{ii}}
\renewcommand{\H}{\ion{H}{\!}}

\newcommand{\teff}{$T_\mathrm{eff}$} 
\newcommand{\logg}{log {\it g}}

\newcommand{\fig}[1]{Fig.~\ref{#1}}
\newcommand{\taufh}{\tau_{_{500\mathrm{nm}}}}

\newcommand{\avsun}{$\langle 3\rm{D}_{\sun}\rangle_{1\rm{D}}$}
\newcommand{\avstar}{$\langle 3\rm{D}\rangle_{1\rm{D}}$}

\begin{document}

\title{\Mg\ line formation in late-type stellar atmospheres:\\I. The model atom
}
\author{
  Y. Osorio \inst{1,2} \and  
  P. S. Barklem \inst{1} \and 
  K. Lind \inst{1} \and 
  A. K. Belyaev \inst{1,3}\and
  A. Spielfiedel \inst{4,5} \and 
  M. Guitou \inst{6} \and
  N. Feautrier \inst{4,5}
}
\institute{
Theoretical Astrophysics, Department of Physics and Astronomy, Uppsala University, Box 516, SE-751 20 Uppsala, Sweden \and 
  Nordic Optical Telescope, Apartado 474, E-38700 Santa Cruz de La Palma, Spain \and 
  Department of Theoretical Physics, Herzen University, St. Petersburg 191186, Russia \and
  LERMA, Observatoire de Paris,  PSL Research University, CNRS, UMR 8112, F-75014, Paris, France  \and
 Sorbonne Universit\'es, UPMC Univ. Paris 6, UMR 8112, LERMA, F-75005, Paris, France \and
  Universit\'e Paris-Est, Laboratoire Mod\'elisation et Simulation Multi-Echelle, UMR 8208 CNRS, 5 Boulevard Descartes, Champs sur Marne, F-77454 Marne-la-Vall\'ee, France
}
\offprints{Yeisson Osorio}
\mail{yeisson.osorio@physics.uu.se}


   \date{}
  \abstract
   {Magnesium is an element of significant astrophysical importance, often traced in late-type stars using lines of neutral magnesium, which is expected to be subject to departures from local thermodynamic equilibrium (LTE). The importance of \Mg, together with the unique range of spectral features in late-type stars probing different parts of the atom, as well as its relative simplicity from an atomic physics point of view, makes it a prime target and test bed for detailed \emph{ab initio} non-LTE modelling in stellar atmospheres.  Previous non-LTE modelling of spectral line formation has, however, been subject to uncertainties due to lack of accurate data for inelastic collisions with electrons and hydrogen atoms.} 
  {In this paper we build and test a \Mg\ model atom for spectral line formation in late-type stars with new or recent inelastic collision data and no associated free parameters. We aim to reduce these uncertainties and thereby improve the accuracy of \Mg\ non-LTE modelling in late-type stars. }
   {For the low-lying states of \Mgi, electron collision data
were calculated using the $R$-matrix method. Hydrogen collision data, including charge transfer processes, were taken from recent calculations by some of us.  Calculations for collisional broadening by neutral hydrogen were also performed where data were missing. These calculations, together with data from the literature, were used to build a model atom.  This model was then employed in the context of standard non-LTE modelling in 1D (including average 3D) model atmospheres in a small set of stellar atmosphere models.  First, the modelling was tested by comparisons with observed spectra of benchmark stars with well-known parameters.  Second, the spectral line behaviour and uncertainties were explored by extensive experiments in which sets of collisional data were changed or removed.  }
   {The modelled spectra agree well with observed spectra from benchmark stars, showing much better agreement with line profile shapes than with LTE modelling. The line-to-line scatter in the derived abundances shows some improvements compared to LTE (where the cores of strong lines must often be ignored), particularly when coupled with averaged 3D models. The observed \Mg\ emission features at 7 and 12~$\mu$m in the spectra of the Sun and Arcturus, which are sensitive to the collision data, are reasonably well reproduced.  Charge transfer with \H\ is generally important as a thermalising mechanism in dwarfs, but less so in giants.  Excitation due to collisions with \H\ is found to be quite important in both giants and dwarfs.  The $R$-matrix calculations for electron collisions also lead to significant differences compared to when approximate formulas are employed.  The modelling predicts non-LTE abundance corrections $\Delta A(\Mg)_{_{\mathrm{NLTE}-\mathrm{LTE}}}$ in dwarfs, both solar metallicity and metal-poor, to be very small (of order $0.01$ dex), even smaller than found in previous studies.  In giants, corrections vary greatly between lines, but can be as large as 0.4 dex.   }
   {Our results emphasise the need for accurate data of  \Mg\ collisions with both electrons and \H\ atoms for precise non-LTE predictions of stellar spectra, but demonstrate that such data can be calculated and that \emph{ab initio} non-LTE modelling without resort to free parameters is possible.  In contrast to Li and Na, where only the introduction of charge transfer processes has led to differences with respect to earlier non-LTE modelling, the more complex case of Mg finds changes due to improvements in the data for collisional excitation by electrons and hydrogen atoms, as well as due to the charge transfer processes.  Grids of departure coefficients and abundance corrections for a range of stellar parameters are planned for a forthcoming paper.}


 \maketitle

\section{Introduction}

Neutral magnesium creates a broad range of spectral features in late-type stars, including some of the strongest lines in their spectra. Consequently, \Mg\ is detectable even in low-quality spectra and in metal-poor stars, making \Mg\ an excellent tracer of $\alpha$-element abundances.  The strong \Mgi\ b lines with pressure-broadened wings can be used as a surface gravity diagnostic \citep[e.g.][]{1988AaA...190..148E,1997AaA...323..909F}.   Additionally, the inter-combination line at {4571~\AA} is unique among lines of similar strength in the Sun in that it is expected to have an almost LTE line formation across the entire profile \citep[see][and discussion therein]{Carlsson12mu92}.  
Emission features due to \Mgi\ Rydberg transitions have been identified in the infrared (IR) region of the solar spectrum, particularly those near 12~$\mu m$  \citep{1981ApJ...247L..97M, 1983ApJ...275L..11C}.  The \Mgi\ emission features dominate other features arising from similar transitions in other atoms due to a combination of abundance and ionisation energy.  These emission lines have also been observed in other main-sequence and giant stars: Arcturus \citep{1996ASPC..109..723U}, Procyon \citep{2004ApJ...617..551R}, and Pollux \citep{2008AaA...486..985S}\footnote{The 12.3~$\mu$m feature is observed in absorption in Betelgeuse \citep{1996ASPC..109..723U}, but this is due to the blending with a strong water line \citep{2008Aaamp;A...486..985S}.}.

At $T \gtrsim 5000$~K, \Mgi\ is a minority species and as a result is expected to be sensitive to departures from LTE in stellar atmospheres, in particular due to photoionisation by non-local radiation. While in the Sun non-LTE effects on lines in the optical range are expected to be relatively weak \citep{1998AaA...333..219Z}, significant departures from LTE have been predicted in metal-poor dwarfs and giants due to the increase in UV radiation because of the decreased line-blanketing \citep{2000AaA...362.1077Z}.  Moreover, the 12~$\mu m$ features have been satisfactorily explained by non-LTE modelling with a photospheric origin \citep{Carlsson12mu92, 1998AaA...333..219Z, 2008AaA...486..985S}.  However, past studies have often been forced to use atomic collision data of questionable quality, frequently approximate formulae, and this is a significant source of uncertainty in the non-LTE modelling.  \citeauthor{Carlsson12mu92} showed that these uncertainties were the largest in their calculations of the solar 12~$\mu m$ lines (see Sect.~6.3 of their paper).  \citeauthor{1998AaA...333..219Z} further demonstrated this, as well as showing the particular sensitivity of the {8806~\AA} line to the rates for collisions with neutral hydrogen. 

The astrophysical importance of \Mg, together with the unique range of spectral features in late-type stars probing different parts of the atom, plus its relative simplicity from an atomic physics point of view, makes it a prime target and test bed for detailed \emph{ab initio} non-LTE modelling in stellar atmospheres.  In this paper, we describe calculations of non-LTE \Mg\ line formation based on a new model of the \Mg\ atom with significant improvements in the collision data for neutral \Mg.  Our focus here is on inelastic collision data, although some new data for collisional broadening are also presented.  We perform calculations for excitation of the lower-lying levels due to electron impacts using the $R$-matrix method.  Recent data for excitation and charge transfer due to hydrogen atom impacts involving low-lying levels calculated by some of us \citep{2011JPhB...44c5202G,2012PhRvA..85c2704B,2012AaA...541A..80B} are also employed.  Furthermore, we have made considerable efforts to use physically motivated methods for calculating radiative and collisional data involving high-lying and Rydberg states.

Additionally, in expectation of accurate stellar positions and motions from the GAIA satellite, several complementary ground-based spectroscopic surveys have been initiated. Their goal is to determine high-precision abundances for eventually hundreds of thousands of stars and thereby unravel the formation and evolution of the Milky Way. The ongoing Gaia-ESO Survey will determine [Mg/Fe] ratios in all Galactic components to better than $\sim0.1$\,dex, in order to distinguish between the thin and thick disk and between accreted and in-situ halo components, for instance \citep{2013sf2a.conf..147R,2012ASPC..458..147G}. The targeted wavelength regions of the VLT/GIRAFFE spectrograph contain two strong Mg lines, 5528\AA\ and 8806\AA , detectable in all FGK type stars down to low metallicities. For stars in the solar neighbourhood, the GALAH survey \citep{GALAH} aims at a higher precision ($\sim0.05$\,dex) for many elements with
the aim to identify dispersed star clusters \citep{2012ASPC..458..421Z}. The wavelength settings of the AAT/HERMES instrument contain two weak Mg lines, 5711\AA\ and 7691\AA , suitable for analysis of high-metallicity stars.

The paper is structured as follows.  In Sect.~\ref{sect:hist} we review  previous studies and particularly the collision data used. In Sect.~\ref{sect:atomic} we describe the model atom and present new data for inelastic electron collisions and collisional broadening due to hydrogen; in Sect.~\ref{sect:comp} the modelling is compared to observed spectra in benchmark stars. In Sect.~\ref{sect:IRlines} we compare our calculated profiles of the solar \Mgi\ IR emission lines with observations. In Sect.~\ref{sect:lines} we examine the effect of the different collisional processes in commonly used \Mg\ lines.  Finally in Sect.~\ref{sect:conc} we present our conclusions.

\section{Previous works}
\label{sect:hist}

As mentioned, a significant number of non-LTE \Mgi\ line formation studies have been conducted, and the main difference between these, especially the more recent and comprehensive studies since around 1990, and the work presented here lies with the atomic collision data for bound-bound processes.  Thus, as a basis for comparison, in this section we present an overview of some past studies, focussing on the atomic collisional data employed.  

The earliest investigation of \Mgi\ line formation by \cite{1969ApJ...156..695A}  examined the \Mg\ b lines in the Sun. They used a small model atom with constant values for the electron collisional cross sections based on the van Regemorter (vR) formula  \citep{1962ApJ...136..906V} -- an often-used semi-empirical interpolation formula based on the Bethe approximation. \cite{1974ApJ...194..733A} updated the collisional cross-sections to those calculated by \cite{1970JPhB....3..932V}. \cite{1988ApJ...330.1008M} used a 13-level model atom to study the {4571 \AA} line and the \Mg\ b line at \mbox{5173 \AA} in the quiet Sun employing updated electron collisional data collected from various experiments and calculations. 

Subsequent studies in the late 1980s and early 1990s focussed on the \Mgi\ IR emission features at 12 and \mbox{18 $\mu m$} observed in the Sun \citep[among others]{1987Aaamp;A...173..375L,1991ApJ...379L..79C,Carlsson12mu92}, where high-lying and Rydberg levels were included. \citeauthor{1991ApJ...379L..79C} used a model atom with 41 levels where $3s\; nl$ levels with $l=s,p,d$ and $n\leq 7$ are included, and $7f,7g,7h,7i$ are merged; $8\leq n \leq15$ are super levels.  \citeauthor{Carlsson12mu92} presented a model atom with 71 levels plus continuum, including up to $3s\; nl$ with $n\leq9,l\leq8$, while for $n=10$, $l=s,p$ levels are included separately and $l\geq2$ are included as single super levels.   For $l \ge 3$ singlet and triplet terms are collapsed to single levels.   These studies demonstrated the importance of \emph{comprehensive} models including levels up to the quasi-continuum, where levels are collisionally dominated and strongly coupled to the continuum.  This permits a population flow from the continuum to lower-lying levels.  The importance of this was demonstrated by the failure of Lemke \& Holweger's model to reproduce the 12 $\mu m$ emission, due to a lack of levels sufficiently close to the continuum to be collisionally dominated, in addition to incorrect collision rates (see Sect.~6.6 of \citeauthor{Carlsson12mu92}). \cite{2008AaA...486..985S} also showed the need to extend the model atoms to include $n\ge10$ for giant stars; the sparser atmosphere means that the levels become collisionally dominated at higher $n$.

\citeauthor{1991ApJ...379L..79C} and \citeauthor{Carlsson12mu92}
both took electron collisional excitation rates from \citeauthor{1988ApJ...330.1008M}.  For the optically allowed transitions not included in \citeauthor{1988ApJ...330.1008M}, \citeauthor{1991ApJ...379L..79C} used the vR formula, while \citeauthor{Carlsson12mu92} used the impact parameter (IP) method \citep{1962amp..conf..375S}.  For forbidden transitions, \citeauthor{1991ApJ...379L..79C} used the vR formula with an arbitrary oscillator strength, while \citeauthor{Carlsson12mu92} used an arbitrary scaling of nearby allowed transitions.  

Inelastic collisions with neutral hydrogen atoms were included in \cite{1991ApJ...379L..79C}. For \Mgi+\H\ collisional excitation the formula from \cite{Kaulakys} was used for transitions between Rydberg levels, but did not include data for the lower-lying \Mgi\ levels.  Charge-transfer processes involving Rydberg states and protons, ${\Mg}^{**}(nl) + \H^+ \rightleftarrows {\Mg}^+ + \H(n)$, are discussed, but suggested to be unimportant in the solar photosphere, despite the large rate coefficients, due to low number densities.  We note that we have been unable to find the source of these data quoted as being in preparation. \citeauthor{Carlsson12mu92} discussed the possible role of collisional excitation with neutral \H, arguing that they will lead to significant $l$-changing collisions among Rydberg states.  They included this effect by setting high electron collision rates between levels of equal $n$ and with small differences in quantum defects to ensure relative LTE between these levels. Various revisions of the model atom in \citeauthor{Carlsson12mu92} have been used for studying the 12 and 18~$\mu m$ features in stars other than the Sun  \citep[e.g.][]{2004ApJ...617..551R, 2008AaA...486..985S}, some including collisions with hydrogen atoms via the Drawin formula \citep{1969ZPhy..225..483D, 1984AaA...130..319S}.  The Drawin formula has been shown in recent years to be unable
to provide reliable estimates of hydrogen collision processes, and no estimates at all for optically forbidden transitions or charge transfer processes \citep[e.g.][]{2011AaA...530A..94B}. 

The study of \cite{1998AaA...333..219Z} used similar energy levels to those of \citeauthor{Carlsson12mu92}; some small differences and uncollapsed terms led to a 83-level atom. They adopted electron collisional data from the vR formula for all allowed  transitions, while for forbidden transitions they assumed a collision strength of unity.  Collisions with neutral hydrogen were included using the Drawin formula with an empirically determined scaling factor that changes exponentially with the excitation energy of the upper level. Later studies by the same group \citep{2004AaA...413.1045G,2008AaA...478..529M} abandoned this exponential scaling in favour of a constant factor. 

\cite{2001Aaamp;A...369.1009P} built an atomic model that includes the same levels as that of  \citeauthor{Carlsson12mu92} and added $n=11$ and 12 for \Mgi\ together with the \Mgii\ levels $2p^6\; nl $ with $n\leq10,l\leq4$ and the ground state of \ion{Mg}{iii}.  Electron collisional excitation data for \Mgi\ were calculated using the vR formula except for the transitions between the ground and the four lowest excited levels, which were taken from \cite{1991PhRvA..44.2874C}.  No hydrogen collisions were included by them since, as they mentioned, hydrogen collisions will be unimportant in A-stars, the subject of that study.

\cite{2011MNRAS.418..863M} used an updated model atom to study the non-LTE effects on lines to be observed by the GAIA satellite. The model atom included levels with $nl$ quantum numbers up to $n$=10 and $l$=9, including fine structure at low $l,$ leading to a 149 level model atom.   They also included $\sim$300 radiative inter-combination  transitions. Electron collisional data were stated to be taken from ``quantum mechanical'' calculations, although the sources are not specified; the IP formula was used when calculations were unavailable.  Hydrogen collisions were not included. 

Recently, \cite{2013AaA...550A..28M} performed tests on eight stars including the Sun using a model atom with the same levels as \cite{1998AaA...333..219Z} for \Mgi, 2 levels for \Mgii\ and the ground state of \ion{Mg}{iii}. The electron collisional data were the same as those of \citeauthor{1998AaA...333..219Z}, but they replaced the rates for electron collisions from vR with those of \citeauthor{1988ApJ...330.1008M} when the latter were available. \citeauthor{2013AaA...550A..28M} used the hydrogen collision data from \cite{2012AaA...541A..80B}. 

Finally, we note that all these studies used plane-parallel 1D atmospheres.  \cite{1994IAUS..154..309R} have investigated the effects of inhomogeneity in the 12~$\mu$m lines and demonstrated that the shapes of these lines are relatively insensitive to granulation-induced asymmetries.  That work is, however, to our knowledge the only attempt at 3D non-LTE Mg line formation to date.  We further note that \cite{1994IAUS..154..309R} provided an excellent summary of non-LTE mechanisms in \Mgi, and \cite{1994IAUS..154..297C} provided a summary of the corresponding atomic physics relating to the infrared lines.

\section{Non-LTE modelling}
\label{sect:atomic}

We performed standard 1D non-LTE modelling, where the coupled radiative transfer and statistical equilibrium equations based on our \Mg\ model atom are solved in 1D stellar model atmospheres using version 2.3 of the {\tt MULTI} code \citep{MULTIuppsala,MULTIrev}.  We employed the same background line opacity data  \citep{2005Aaamp;A...442..643C} used in the calculation of the MARCS theoretical stellar atmospheres \citep{2008AaA...486..951G}, but re-sampled to $10\, 300$ frequency points.  Turbulence was modelled by the usual micro- and macro-turbulence parameters\footnote{Macroturbulence is assumed to be Gaussian and treated according to the radial-tangential model.}. 

The {\tt MULTI} code treats the restricted non-LTE radiative transfer problem, in which the element of interest, here Mg, is treated as a trace element.  Thus, it is  assumed that there is no feedback on the atmospheric structure or background opacities, which are computed assuming LTE.  Only very few studies have treated the full non-LTE problem in cool stars \citep{2005ApJ...618..926S,2009ApJ...691.1634S}.  \cite{2012MNRAS.427...50L} considered the effects of increased electron densities and decreased UV opacity due to the increased ionisation of \ion{Fe}{i} on the non-LTE line formation of Fe, arguing that the the first effect was insignificant and the second weaker than typical non-LTE corrections.   However, such tests are limited, and as stated by the authors, only fully consistent calculations will resolve this definitively.  Such calculations are beyond the authors' present capabilities and beyond the scope of the present work.  We note that the abundance of Mg is almost always larger than Fe, especially in metal-poor stars (typically about a factor of 3), and Mg also has a somewhat lower ionisation energy, making it a more important electron donor.  On the other hand, non-LTE does not perturb the ionisation balance of Mg as much as for Fe.  It would be important to test the trace element assumption.

In this section we present the details of our atomic model, often referred to as the ``model atom''.  In Sect~\ref{sect:energy_levels} we discuss the energy levels and the size of the model atom.  In Sect.~\ref{ssec:rad} the radiative data are outlined, namely bound-bound (b-b) and bound-free (b-f) transition probabilities and line broadening cross sections, and in Sect.~\ref{sect:collision_data} the inelastic collision data are described.   In Sect.~\ref{sect:testatoms} we describe the modified model atoms used to test the influence of new data on the final results.

\subsection{Energy levels}
\label{sect:energy_levels}

Following earlier studies, we decided to build a model atom including \Mgi, \Mgii,\ and the ground state of \ion{Mg}{iii}. To build the structure of the model atom, energy level data were collected from the NIST database \citep{NIST}, supplemented with data for high $n$ (up to $n=80$) in the case of \Mgi\ from \cite{Kurucz}.  For high angular momentum levels absent from either the NIST or \citeauthor{Kurucz} collections, we made use of the Sommerfeld-Dirac expression to estimate the energies.  In principle, it is desirable to have the model atom as large and as detailed as possible; however, this is neither computationally practical nor required to achieve converged results for the spectral lines of interest.  Thus, to investigate the convergence of the model atom, a computer script was written to generate model atoms characterised by three parameters: 1) the largest principle quantum number for which levels are included, 2) the principle quantum number above which all levels are merged into ``super levels'' encompassing all states of same $n$, and 3) the term above which all fine-structure components $J$ are merged into a single level.  This permitted to test a range of atomic models in size and detail.  

\begin{figure*}[t]
\centering
\hspace{-0.06\textwidth}\begin{tikzpicture}
\node[anchor=south east, inner sep=0] (image) at (0,0) {
\subfloat{\includegraphics[width=0.6\textwidth]{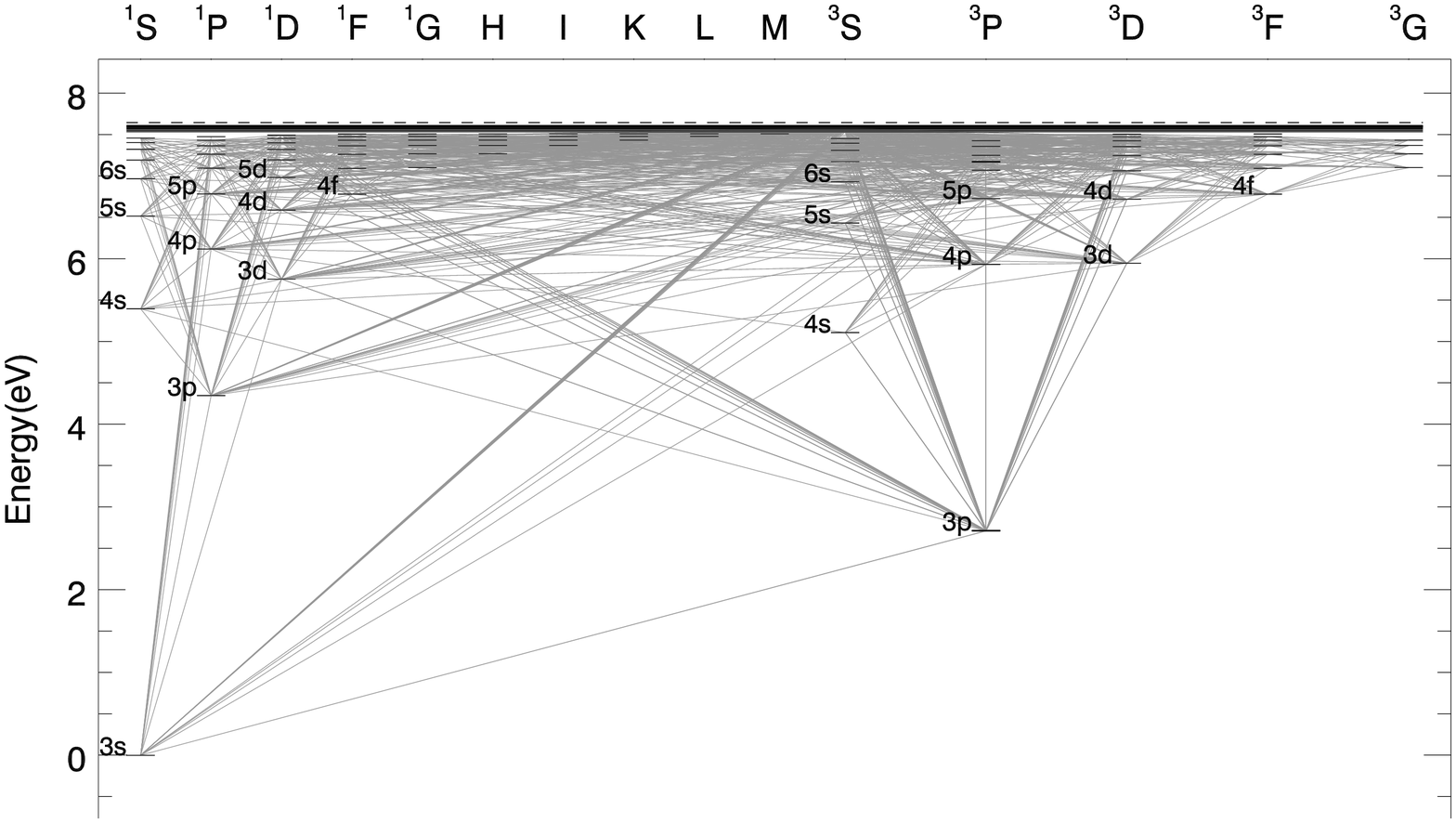}}
};
\node[anchor=south west] at (-0.06\textwidth,0.0) {
\subfloat{\includegraphics[width=0.5\textwidth]{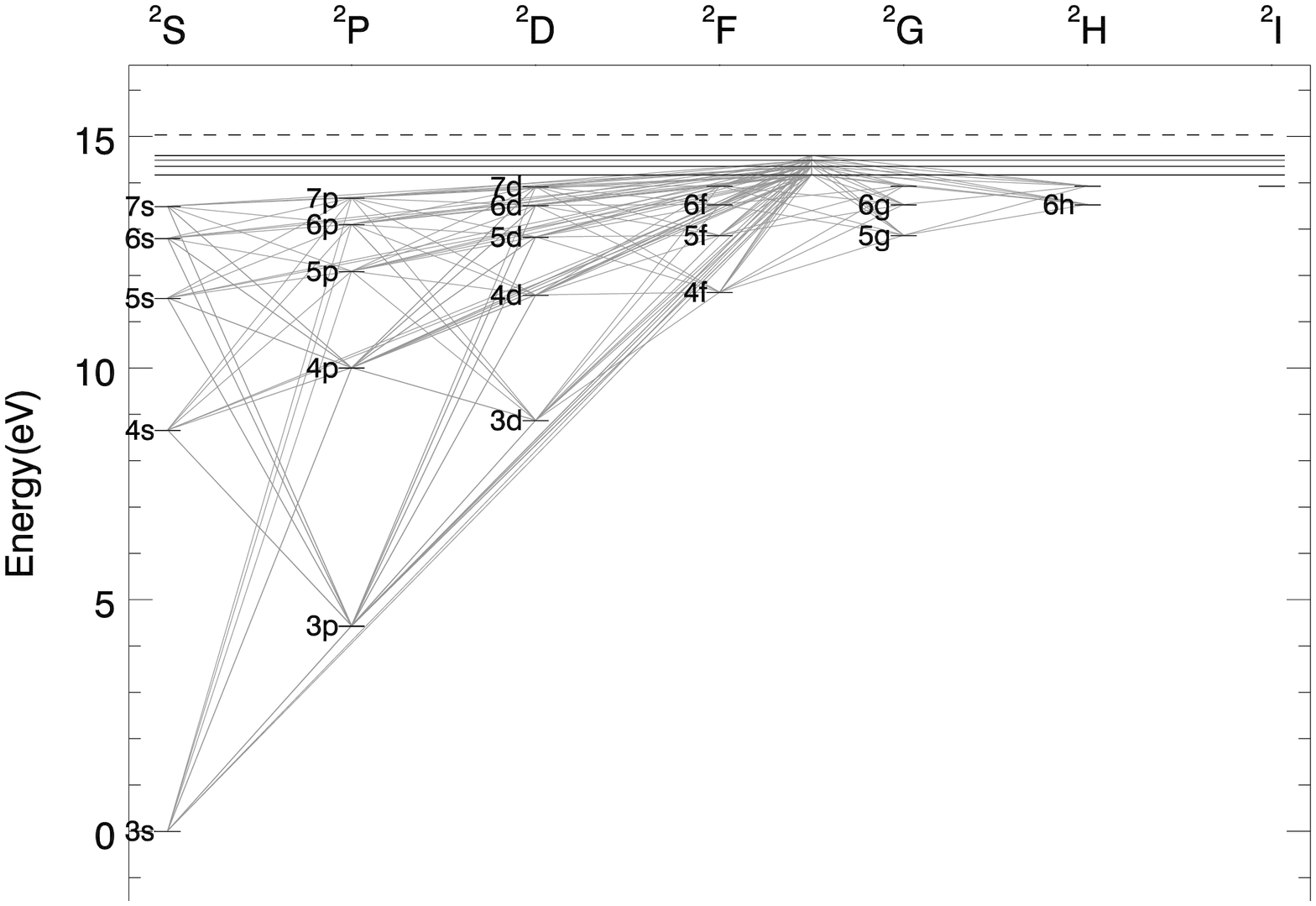}}
};
\node at (-0.1\textwidth,0.1\textwidth) {\Mgi};
\node at (0.35\textwidth,0.1\textwidth) {\Mgii};
\end{tikzpicture}
\caption{Grotrian diagrams of the \Mg\ levels and transitions representing the model atom adopted for this study. Energy is relative to the ground state in each case. The $nl$ configuration is shown for some levels. The dashed line in the diagrams is the ionised state. The vertical line in the $^3$P term connects the 3s\,3p $^3$P$^{\mathrm o}$ to the 3p$^2$ $^3$P level. Parity is not distinguished, and so these two terms are shown in the same column.}\label{grotrian}
\end{figure*}

Finally, after extensive testing, we found that the following model atom provided results that converged to better than 1\% for a range of late-type stellar parameters.  \Mgi\ includes states up to $n=20$ and uses super levels for $n>10$.  Fine-structure splitting was implemented for $^3$P states up to $3s\;7p \, ^3$P$_{\{2,3,4\}}$, as well as for the $3p^2 \, ^3$P$_{\{0,1,2\}}$ states. Splitting was not included for states $3s\; nl \; ^3L$ where $L\geq 2$.  \Mgii\ includes states up to $n=11$ and uses super levels for $n>7$. For states up to $4f \, ^2$F$^o_{\{5/2\,,\,7/2\}}$ fine-structure splitting was included.  This gives a final model atom covering \Mgi, \Mgii, with 108 and 34 states respectively, plus the ground state of \ion{Mg}{iii}. Grotrian diagrams for \ion{Mg}{i {\rm and} ii} are shown in \fig{grotrian}

\subsection{Radiative data}
\label{ssec:rad}

Radiative b-b transition probabilities were taken mainly from NIST and \cite{Kurucz}. When data for a transition were in both sources, the data from NIST were adopted.  For the \Mg\ b triplet, oscillator strengths were taken from \citet{2007Aaamp;A...461..767A}. When transitions involving Rydberg states of \Mgi\ were needed and data were unavailable, transition probabilities were calculated following \cite{Guseinov2012776}.  The data for the most important diagnostic lines are provided in Table~\ref{linestable}.  The final model atom has 1185 b-b transitions (962 \Mgi\ and 223 \Mgii).  In all cases, radiative damping was calculated consistently from the employed transition probabilities.

For most UV lines, S$_\nu >$B$_\nu$ in the line formation region contributing to UV overionisation that tends to depopulate lower levels. Thus, b-f processes are of considerable importance. Bound-free cross-sections for \Mgi\ and \Mgii\ were taken from TOPbase \citep{1992RMxAA..23..107C,1993Aaamp;A...275L...5C}.  The data for \Mgi\  were adopted from \cite{0953-4075-26-23-013}, and are available for $n\leq9$ and $l\le4$.  For \Mgii\ the original source is unpublished and data are available for $n\leq10$ and $l\le3$.  For larger $n$ and $l$, hydrogenic cross-sections were used. The threshold energies of the photoionisation cross-sections taken from TOPbase do not coincide exactly with the experimental ones used in our modelling, and so the TOPbase cross-sections were shifted appropriately.  Strong resonances in the b-f cross sections sometimes play an important role.
The photoionisation cross-sections for the ground and first excited state of \Mgi\ have no resonances strong enough to be important at typical temperatures in cool stellar atmospheres. The photoionisation from the $3s 3p \, ^1\mathrm{P}^o$ level shows a resonance at \mbox{$\lambda\sim$ 2900 \AA} due to the $3p^2 \, ^1\mathrm{S}$ auto-ionising level that is five times greater than the value of the cross-section at threshold, thus making it an important contributor. The photoionisation from the $3s 4s \, ^{1,3}\mathrm{S}$ levels show resonances due to the $3p 4s \, ^{1,3}\mathrm{P}^o$ auto-ionising levels that are more than 500 times greater than the value at threshold, making these resonances the main contributors to the total photoionisation rate at temperatures around 5000~K.  \Mgii\ photoionisation cross-sections do not have resonances in the energy range of interest.

\subsubsection{Line broadening}

In addition to turbulent and radiative broadening, both mentioned above, data on the effects of collisions on the modelled spectral lines were also required, in particular broadening of the lines due to collisions with charged particles and neutral hydrogen atoms.  For lines of diagnostic interest in this work (see table~\ref{linestable}), these data need to be as accurate as possible, while for other transitions this is probably not of great importance, and reasonable estimates suffice.

For some of the \Mgi\ lines we included collisional broadening due to electrons and protons using quadratic Stark broadening parameters from \cite{1996AaAS..117..127D}.  
We adopted the line width at 5000~K neglecting variation with temperature, which is typically weak. For the 18~$\mu$m lines we adopted widths from \cite{1993AaA...277..623V}.  For the inter-combination line $3p \,^3$P$^o - 3s \,^1$S at 4571~\AA\ we adopted the value used in \citet{2013AaA...550A..28M}\footnote{This value is stated to be taken from the VALD database, bit we were unable to find any such data for this line in the VALD.}.  For all other lines (i.e. those not in table~\ref{linestable}, or of \Mgii), Stark broadening was neglected.

\begin{table*}
\caption{Data for important transitions, including diagnostic lines. Columns 1 and 2 give the transition wavelength and labels for the states involved.  Columns 3 and 4 give the adopted oscillator strength, in the form $\log (gf)$, and the source of these data (see notes). Columns 5-8 give collisional broadening data (see text and notes). Columns 9-12 show, for each line of \Mgi, the collision process most responsible for the changes between model B and model F, and thus indicates the most important contributor of the new collision data introduced in this work (see Sect.~\ref{sect:lines} for more details).  Results are given for the four test atmospheric models described in Sect.~\ref{sect:lines}:  Dr = Dwarf, metal-rich; Dp = Dwarf, metal-poor; Gr = Giant, metal-rich; Gp = Giant, metal-poor. The collisional processes are labelled (see Sect.\ref{sect:lines}) H when the line has a sensitivity similar to charge transfer with \H\ and CH0; collisional excitation with \H, CH, and CE indicates the line is sensitive mostly to the electron collision rates. For \Mgii\ lines there are non-LTE effects in some lines, but the non-LTE abundance corrections are not sensitive to any particular collisional process. }\label{linestable}
\scriptsize
\center
\begin{tabular}{r r@{\:--\:} l r r r l r r r c c c c}
\hline\hline
 $\lambda$ &    \multicolumn{2}{c} {Transition}           & $\log{gf}$ & Source & \multicolumn{2}{c}{$\Gamma_6$} & $\log\Gamma_4/N_e$  & Source & Dr & Dp & Gr & Gp \\   
  &  \multicolumn{2}{c} { }  & &  &$\sigma$ & $\alpha$ &  (5000 K) & & & & & \\
  {[}\AA] & \multicolumn{2}{c} { }  & & & [a.u.] & & [rad s$^{-1}$~cm$^3$]  & & & & & \\
  \hline
  \\
& \multicolumn{2}{c} {\underline{\ion{Mg}{i}}}  & & \\
       3 829 & $3d\,^3$D & $3p\,^3\mathrm{P}^o_0$       & $-$0.23         & Fro/NIST      &  708  & 0.301 & $-$4.51       & D-SB & H & H & CH & CH    \\
       3 832 & $3d\,^3$D & $3p\,^3\mathrm{P}^o_1$       &    0.25    & Fro/NIST      &  708  & 0.301         & $-$4.51       & D-SB & H & H & CH & CH    \\
       3 838 & $3d\,^3$D & $3p\,^3\mathrm{P}^o_2$       &    0.47    & Fro/NIST      &  708  & 0.301         & $-$4.51       & D-SB & H & H & CH & CH    \\
       4 167 & $7d\,^1$D & $3p\,^1\mathrm{P}^o$         & $-$0.75         & C-T/NIST      &  222  & 0.249         & $-$3.49       & D-SB & H & H & CH & H       \\
       4 571 & $3p\,^3$P$^o_1$ & $3s\,^1\mathrm{S}$ & $-$5.62& Fro/NIST        &  222  &  0.249        & $-$6.51       & Mas   &    H     & H & CE & CE\\
       4 703 & $5d\,^1$D & $3p\,^1\mathrm{P}^o$         & $-$0.44         & C-T/NIST      & 2806 &  0.269$^*$&  $-$4.11& D-SB &    H     & H  & CH & H \\
       5 167 & $4s\,^3$S & $3p\,^3\mathrm{P}^o_0$       & $-$0.93         & Ald                   &  728  &  0.238        &  $-$5.40         & D-SB &    H     & CH& CH & CH \\ 
       5 173 & $4s\,^3$S & $3p\,^3\mathrm{P}^o_1$       & $-$0.45         & Ald                   &  728  &  0.238        &  $-$5.40         & D-SB &  CH   & CH & CH  & CH \\
       5 184 & $4s\,^3$S & $3p\,^3\mathrm{P}^o_2$       & $-$0.24         & Ald                   &  728  &  0.238        &  $-$5.40         & D-SB & CH & CH & CH  & CH \\
       5 528 & $4d\,^1$D & $3p\,^1\mathrm{P}^o$         & $-$0.50         & C-T/NIST      & 1460  &  0.312        &  $-$4.56      & D-SB & H & H & CH & CH \\
       8 710 & $7d\,^3$D & $4p\,^3\mathrm{P}_0^o$       & $-$1.57         & But/NIST      &               &               &  $-$2.72         & D-SB & CH0 & CH0 & CH & H \\ 
       8 713 & $7d\,^3$D & $4p\,^3\mathrm{P}_1^o$       & $-$1.09         & But/NIST      &               &               &  $-$2.72         & D-SB & CH0 & CH0 & CH & H \\ 
       8 718 & $7d\,^3$D & $4p\,^3\mathrm{P}_2^o$       & $-$0.87         & But/NIST      &               &               &  $-$2.72         & D-SB & CH0 & CH0 & CH & H \\ 
       8 736 & $7f\,^3$F$^o$ &  $3d\,^3\mathrm{D}$      & $-$0.53         & But/NIST      &               &               &  $-$2.95         & D-SB & CH0 & CH0 & CH & H\\ 
       8 806 & $3d\,^1$D &  $3p\,^1\mathrm{P}^o$        & $-$0.13  & Fro/NIST     & 529   &  0.277        &  $-$5.39      & D-SB & CH  & CH & CH & CH \\
       73 700  & $6h$ H$^o$ & $5g\,^3\mathrm{G}$        &    1.34    & Civ           & 4950 & 1.549$^*$      & $-$3.06  & vR-HB & CH & CE & CH & CH \\
      122 200 & $7h$ H$^o$ &  $6g\,\mathrm{G}$  &    1.62       & Civ     & 5191 & 1.738$^*$      &  $-$2.39      & D-SB & CH & CH & CH & CH \\
      123 200 & $7i$  I &  $6h\,\mathrm{H}^o$           &    1.95    & Hydro         & 4657 & 1.752$^*$      &  $-$2.55      & D-SB  & H & CH0 & CH & CH0 \\
      188 300 & $8h$  H$^o$ &  $7g\,\mathrm{G}$         &    0.49    & Hydro         & 4497 & 1.764$^*$      & $-$2.92       & vR-HB & CH0 & CH0 & CH& CH \\
      189 500 & $8i$  I &  $7h\,\mathrm{H}^o$           &   1.90         & Hydro         & 4304  & 1.778$^*$ & $-$2.12   & vR-HB & H & H & CH & CH \\
\\
& \multicolumn{2}{c} {\underline{\ion{Mg}{ii}}}  & & \\
       4 385 & $5d\,^2$D & $4p\,^2\mathrm{P}^o_{1/2}$                   & $-$0.78         & Sie/NIST      &   &   &            &  &  &  &  &           \\
       4 391 & $5d\,^2$D & $4p\,^2\mathrm{P}^o_{3/2}$                   & $-$0.48         & Sie/NIST      &   &   &               &   &  &  &  &  \\
       4 481 & $4f\,^2$F$^o$ & $3d\,^2$D                                                        &    0.76         & Fro/NIST           &   &   &                  &   &  &  &  &  \\
       5 402 & $7g\,^2$G & $4f\,^2\mathrm{F}^o$                                         &    0.06         &  K-P                  &       &   &                   &   &  &  &  &   \\
       7 877 & $4d\,^2$D$_{3/2} $ & $4p\,^2\mathrm{P}^o_{1/2}$         &    0.39       & Sie/NIST      &   &   &               &   &  &  &  &   \\
       7 896 & $4d\,^2$D$_{5/2} $ & $4p\,^2\mathrm{P}^o_{3/2}$         &    0.64       & Sie/NIST      &   &   &               &   &  &  &  &    \\
       \\
       \hline
\end{tabular}
\tablefoot{Oscillator strengths ($f$-values) were collected mostly from the NIST database \citep{NIST}. The original sources of the NIST data are \cite{MCHF-NIST-2003}~[Fro], \cite{Chang1990207}~[C-T] \cite{0953-4075-26-23-013}~[But], \cite{SIEGEL1998303}~[Sie].  Other sources are \cite{2007Aaamp;A...461..767A}~[Ald], \cite{refId0}~[Civ] and \cite{1975SAOSR.362.....K}~[K-P].  For the \Mgi\ IR lines with no data found in the literature we calculated the $f$-values using the hydrogenic formula from \cite{Guseinov2012776}~[Hydro]. To calculate the van der Waals line widths $\Gamma_6$ we used data in the ABO theory format where $\sigma$ is the broadening cross-section in atomic units and $\alpha$ is the velocity parameter. These were calculated with the ABO theory except for those marked with an asterisks, which were calculated in this work (see Table~\ref{tab:br_lines} for the IR lines). Stark broadening line widths, $\Gamma_4$, were taken from \cite{1996AaAS..117..127D}~[D-SB], \cite{1993AaA...277..623V}~[vR-HB] and \cite[including only electrons]{2013AaA...550A..28M}~[Mas].  }
\end{table*}

For transitions of \Mgi\ involving low-lying states, collisional broadening due to collisions with neutral \ion{H}{} is described via cross-sections and velocity parameters interpolated in tables from the ABO theory \citep{1995MNRAS.276..859A,1997MNRAS.290..102B,1998MNRAS.296.1057B,1998PASA...15..336B}.  In the specific cases of the 4167 and 8806~\AA\ lines, these lie slightly outside the range of the tables and so specific calculations have been made; the calculation for 8806~{\AA} was presented in \cite{1997MNRAS.290..102B}.  As such, calculations are at the edge of the validity of the theory, and so these data must be expected to be more uncertain than for other lines.  For transitions involving Rydberg states, the ABO theory is inappropriate for various reasons, most importantly because inelastic processes play a very important role.  \citet{1995JPhB...28.3147H}, hereafter HBvR, have performed calculations for the 12~$\mu$m lines in the impulse approximation.  In this work we investigate these and several other lines in the infrared that involve Rydberg states (at 7 and 18~$\mu$m), and thus we performed broadening calculations for these lines using a similar formalism described below.  We show below that our results agree reasonably well with HBvR for the 12~$\mu$m lines.

In the impact approximation, the broadening cross-section for an isolated line corresponding to the transition $i\rightarrow f$ can be written \citep[e.g.][]{1962amp..conf..493B,1995JPhB...28.3147H}
\begin{eqnarray}\label{eqn:br1}
 \lefteqn{\sigma^\mathrm{br}_{i\rightarrow f} = }  \\
&& \frac{1}{2} \left( \sum_{i'} \sigma_{i\rightarrow i'} +  \sum_{f'} \sigma_{f\rightarrow f'} + \int |f_i(\Omega) - f_f(\Omega)|^2 d\Omega \right), \nonumber 
\end{eqnarray}
where $\sigma_{i\rightarrow i'}$ and $\sigma_{f\rightarrow f'}$ are inelastic cross-sections, and $f_i$ and $f_f$ are elastic scattering amplitudes.  The broadening of the initial and final states due to inelastic scattering adds incoherently, and thus is simply the sum of the two contributions. The elastic scattering involving the two levels, however, subtracts coherently and must be treated together at the scattering amplitude level.  The inelastic cross-sections may for convenience be split into $l$-changing and $n$-changing collisions, such that

\begin{equation}
\sigma^\mathrm{inel}_{nl} = \sum_{n'l'} \sigma_{nl\rightarrow n'l'} = \sum_{l'} \sigma_{nl\rightarrow nl'} + \sum_{n' \ne n} \sum_{l'} \sigma_{nl\rightarrow n'l'}.
\end{equation} 

The cross-sections for levels in transitions of interest were calculated and are presented in Table~\ref{tab:br_levels}.  The $l$-changing inelastic cross-sections were calculated using the method for inelastic Rydberg-hydrogen collisions due to \cite{1991JPhB...24L.127K} described in Sect.~\ref{sect:hydrogen}.  The $n$-changing inelastic cross-sections were calculated in the scattering length approximation using the analytic expressions of \citet[Eq. 25]{1987JPhB...20.6041L} and \citet[eqns 4-5]{Kaulakys}, which give practically identical results, as pointed out by Lebedev and Marchenko.  We could calculate the $n$-changing cross-sections in the same way as done for the $l$-changing; however, the computational time is significant, and as seen in Table~\ref{tab:br_levels}, the $n$-changing contribution is very weak, as expected \citep[e.g.][]{1995JPhB...28.3147H}.  HBvR showed that the ratio of the elastic to the inelastic component is expected to be of order $\Delta n/n^3$, and thus we did not
include the elastic contribution.

The total broadening cross-sections are then calculated according to Eq.~\ref{eqn:br1} and are presented in Table~\ref{tab:br_lines}.  The cross-sections for a collision velocity of $10^4$~m/s are presented, together with a velocity parameter $\alpha$ that was calculated by fitting a power-law behaviour of the cross-section with velocity following \citet{1995MNRAS.276..859A}.  Line widths are also presented and compared with the results of HBvR for the 12~$\mu$m lines.  As seen from Table~\ref{tab:br_levels}, our cross-sections are generally larger than those of HBvR, although not quite as large as their calculations using the method of \cite{1991JPhB...24L.127K}.  Note that we have included the $n$-changing component, which was neglected by HBvR. Our cross-sections for the 12.2 and 12.3~$\mu$m lines differ from theirs by 1 and 21 per cent, respectively.  The line widths differ by considerably more because HBvR used an approximate method for calculating the line widths from a cross-section at a single velocity, rather than an integration over a Maxwellian velocity distribution.  In any case, we conclude that the uncertainty in the broadening cross-sections for these lines involving Rydberg states is at least 20 per cent, probably larger due to approximations inherent to both methods.  

\begin{table*}
\center
\caption{Inelastic $l$-changing ($\sum_{l'} \sigma$) and $n$-changing ($\sum_{n'} \sigma$) cross-sections for levels of interest, in atomic units.   Columns 2 and 3 give results from this work at $v = 10^4 \; \mathrm{m/s}$.   For direct comparison with $l$-changing crosssections from Table 4 of HBvR, results are also given for two states at $v = 10\,457 \; \mathrm{m/s}$ corresponding to the average collision velocity at $T=5000~K$.  Column 4 gives our results and Col. 5 the results from HBvR.}
\label{tab:br_levels}
\begin{tabular}{ccccc}
\hline \hline
  &  \multicolumn{2}{c}{$v = 10^4 \mathrm{m/s}$} &  \multicolumn{2}{c}{$v = 10\,457 \mathrm{m/s}$} \\
  \cmidrule(r){2-3} \cmidrule(l){4-5}
  Level & $\sum_{l'} \sigma$ & $\sum_{n'} \sigma $  & $\sum_{l'} \sigma$ & $\sum_{l'} \sigma$ \\
     & &  & & (HBvR) \\ 
        & [au] & [au] & [au] & [au]  \\
\hline
   5g   &  4752    &   78   & & \\
   6g   &  5456    &  130   & & \\
   6h   &  4938    &  132   & 4553 & 4036 \\
   7g   &  4744    &  196   & & \\
   7h   &  4600    &  196   & & \\
   7i   &  4048    &  196   & 3714 & 3088 \\
   8h   &  3786    &  270   & & \\  
   8i   &  3542    &  270   & & \\
   8k   &  3146    &  270   & & \\
\hline
\end{tabular}
\end{table*}

\begin{table*}
\center
\caption{Broadening cross-sections, velocity parameters, and line widths for the lines of interest.  Columns 3 and 4 give broadening parameters, the cross-section $\sigma^\mathrm{br}$ at $v = 10^4 \; \mathrm{m/s,}$ and the velocity parameter $\alpha$ (the ABO theory format).  Columns 5 and 6 compare the cross-sections from this work with those of HBvR at $v = 10457 \; \mathrm{m/s}$.  The broadening cross-section can be determined directly from their line width ($(\gamma_\H/N_\H) / 2 \times 1/\langle v_\H\rangle$ in their notation).  In Cols. 7 and 8 the line widths (half-width at half-maximum) at $T=5000$~K are compared.}
\label{tab:br_lines}
\begin{tabular}{cccccccc}
\hline \hline
Wavelength & Transition & $\sigma^\mathrm{br}$ & $\alpha$ & \multicolumn{2}{c}{$\sigma^\mathrm{br}(10\,457 \mathrm{m/s})$} & \multicolumn{2}{c}{$w/N_\H$} \\
 \cmidrule(r){5-6}
 \cmidrule(l){7-8}
 &  &  &  &  & (HBvR) &  & (HBvR) \\
 $[\mu \mathrm{m}]$ &  & [au] &  & [au]& [au]& \multicolumn{2}{c}{[cm$^3$ rad s$^{-1}$]}  \\
\hline
            7 &  $5g-6h$     &     4950     &   1.549 &  &  & 14.9 & \\
         12.2 &  $6g-7h$     &     5191     &   1.738 & 4804 & 4840 & 16.3 & 14.2 \\ 
         12.3 &  $6h-7i$     &     4657     &   1.752 & 4303 & 3560 &14.7 & 10.5 \\ 
         18.8 &  $7g-8h$     &     4497     &   1.764  &  &  & 14.3 & \\
        18.95 &  $7h-8i$     &     4304     &   1.778  &  &  & 13.7 & \\
        18.96 &  $7i-8k$     &     3831     &   1.789  &  &  & 12.2 & \\
\hline
\end{tabular}        
\end{table*}

For transitions covered neither by the ABO theory nor by the calculations for infrared lines described above, the Uns\"old formula was used with an enhancement factor of 2.5. This includes all transitions of \Mgii.

\subsection{Inelastic collision data}
\label{sect:collision_data}

In this subsection we describe the data adopted for inelastic collision processes: collisional excitation and ionisation due to electrons, and collisional excitation and charge transfer processes due to hydrogen atoms.  This includes the description of new calculations for collisional excitation of \Mgi\ using the $R$-matrix method.  

\begin{figure*}[ht!]
\centering
\begin{tikzpicture}
\node[anchor=south west, inner sep=0] (image) at (0.05\textwidth,0.15\textwidth) {
  \subfloat{\includegraphics[width=1.1\textwidth]{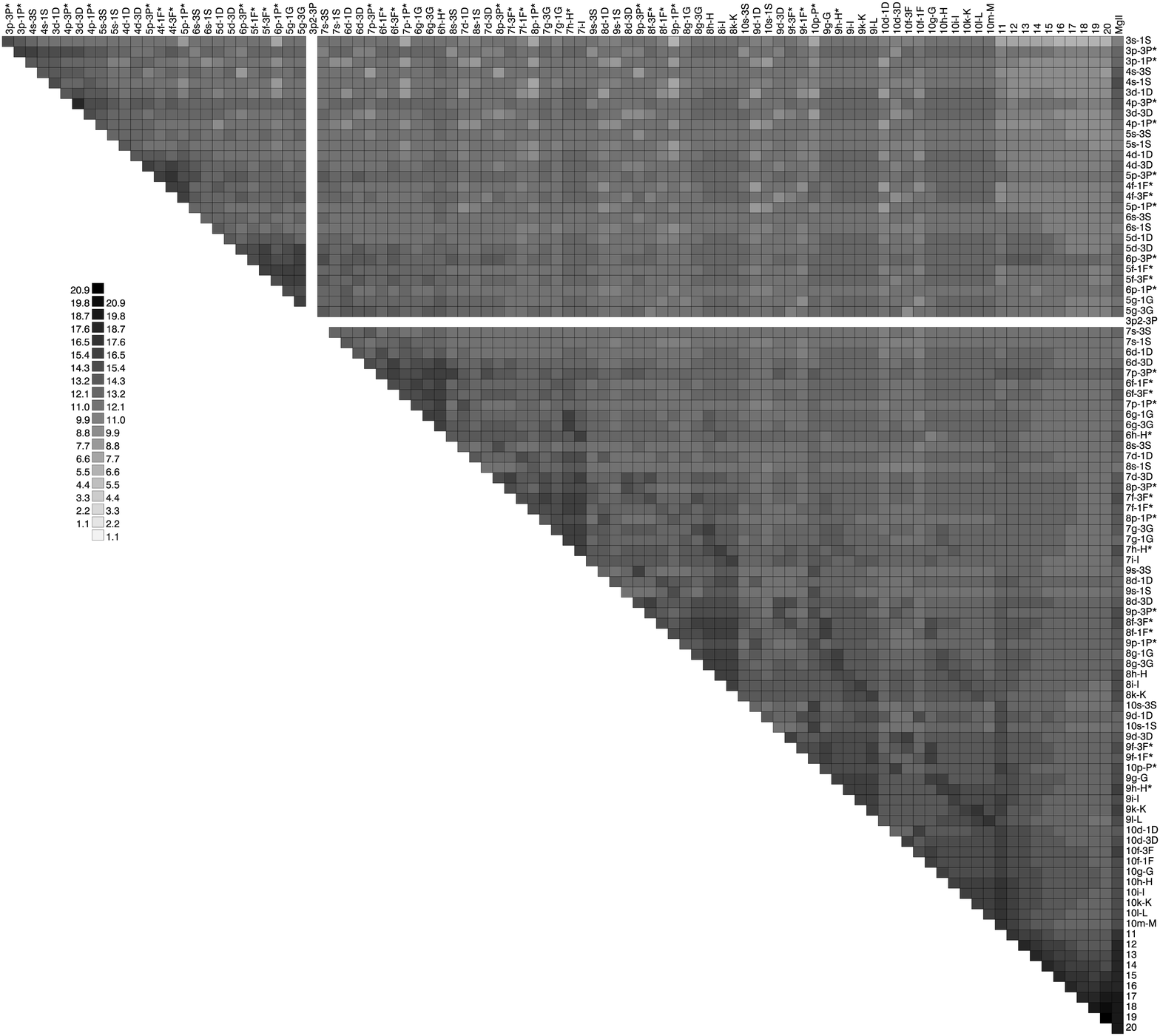} }};
\node at (0.53\textwidth,1.01\textwidth) {\scalebox{1.5}{\ion{Mg}{i}}};

\node[anchor=south west, inner sep=0] (image) at (0.2\textwidth,0) {
  \subfloat{\includegraphics[width=0.5\textwidth]{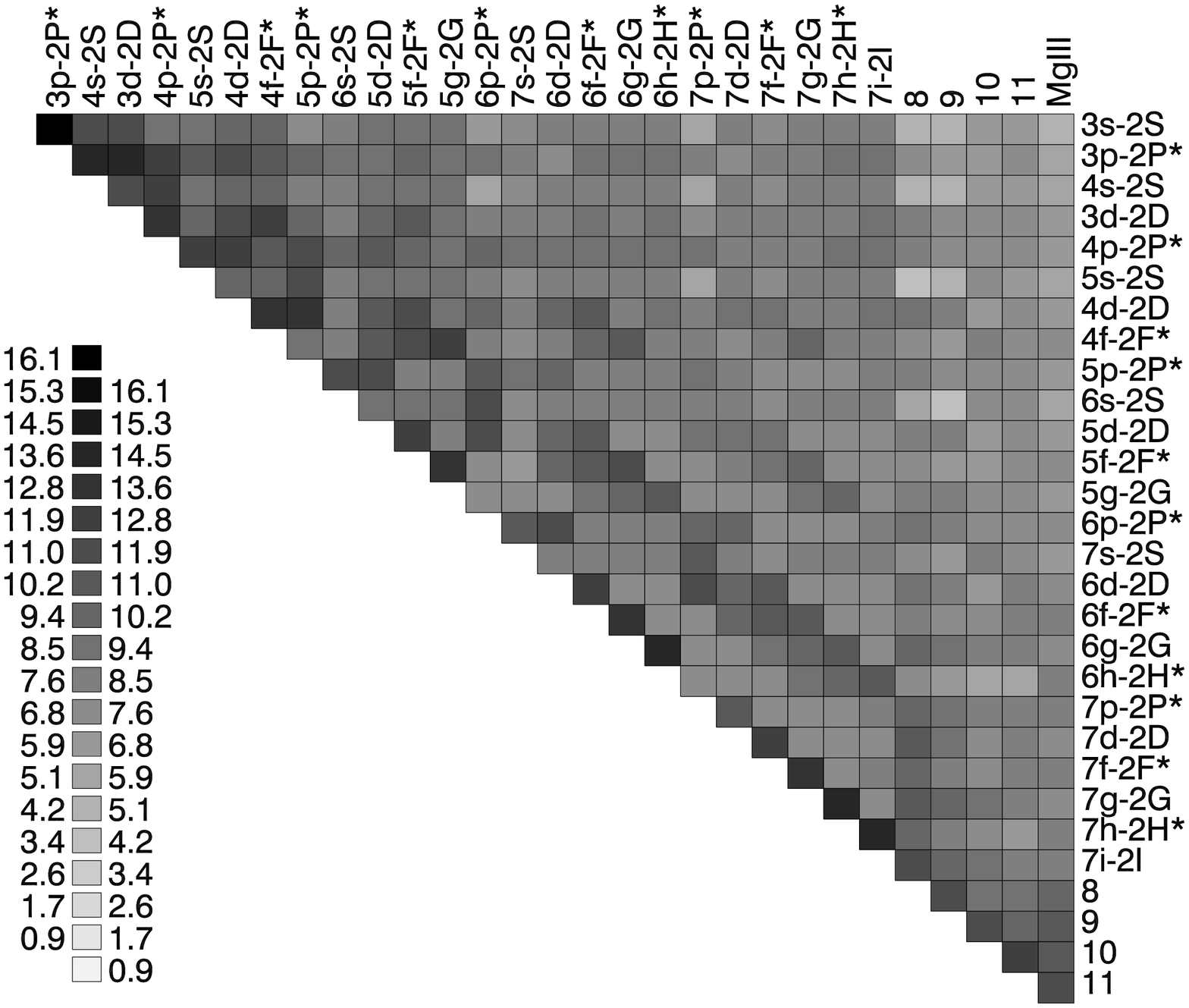} }};
\node at (0.5\textwidth,0.38\textwidth) {\scalebox{1.0}{\ion{Mg}{ii}}};

\end{tikzpicture}
\caption{Collision matrices for \Mgi\ and \Mgii\! at $\log\taufh = -0.23$ in the solar \avsun\ atmospheric model, which corresponds to a temperature of 6000~K. The legend in each case defines a range of $\log{C_{ij}}$, where $C_{ij}$ is the total collision rate in $s^{-1}$, such that darker squares represent higher rates. Note the $3p^2$ level of \Mgi\ is not collisionally coupled to other levels; see Sect.~\ref{sect:collision_data}.}
\label{bbmatrix}
\end{figure*}

Figure~\ref{bbmatrix} visualises the total downward collision rates calculated in the \avsun\ atmospheric model (see Sect.~\ref{sect:comp}) at a depth corresponding to roughly $T=6000$~K ($\log \taufh$=$-$0.23) using our final model atom.  The collision matrix for \Mgi\ has a structure very similar to that seen in \cite{Carlsson12mu92}.  In particular, we see the block structure of the very strongest transitions  along the diagonal caused by the strong $l$-changing collisions where $\Delta n=0$, where collisions with $\Delta l=1$ are strongest of all.  We also see structure off the diagonal due to the strength of collisions with $\Delta l=1$ and $\Delta n=1$.  A new feature in this work is the structure seen in the collisional coupling of \Mgi\ states to the \Mgii\ ground state.   Inclusion of charge transfer processes leads to rates between the levels around $4s\,\,^1\mathrm{S}$ and the \Mgii\ ground state having the same order of magnitude as the collisional rates from these levels to neighbouring levels.

Before we describe the data, a word on the $3p^2\,^3\mathrm{P}$ level is warranted.  This is the only level included in our modelling where the dominant configuration is not $3s\; nl$.  This level is too highly excited to be included in the detailed quantum mechanical calculations, and it is questionable whether the approximate methods are applicable to transitions involving two electrons\footnote{The 3s3p $^{1,3}$P$^{o}-$ 3p$^2$ $^3$P transition involves only one electron and is an exception. However, we still neglect the collisional coupling between these levels, which can be expected to be weak
as a result of the large energy difference.}.  Our expectation is that collisional processes involving this state will be relatively weak, and thus we ignored collisional coupling to this state (as can be seen in Fig.~\ref{bbmatrix}). This causes this level to behave quite differently from other levels of \Mgi\ in the non-LTE modelling and to depart from equilibrium very deep in the atmosphere ($\log{\taufh} \sim 1$ in the solar case).  However, this is not important for any lines observed in stellar spectra.

\subsubsection{Electrons}
\label{sect:electrons}

Collisional excitation and de-excitation between low-lying states of \Mgi\ are accounted for with $R$-matrix calculations, which are presented in detail below.  These calculations provide data for the transitions between the lowest ten states of \Mgi, up to $5p \, ^3$P$^o$.  For transitions involving higher lying states, including Rydberg levels, we must use more approximate methods.  For allowed transitions, two such methods are in common use in stellar astrophysics:  the vR formula \citep{1962ApJ...136..906V}, which is based on the Born and Bethe approximations with an empirical effective Gaunt factor, and the semi-classical impact parameter (IP) method from \citet{1962PPS....79.1105S}.  The IP method has been extended to the case of positive ions, where electrons have hyperbolic trajectories in the semi-classical picture, by \cite{1976MNRAS.174..345B}, but this seems to be little used.

\begin{figure}[t]
\centering
\begin{tikzpicture}
\node[anchor=south west, inner sep=0] (image) at (0,0) {
\subfloat{\hspace{-3em}\includegraphics[width=0.5\textwidth]{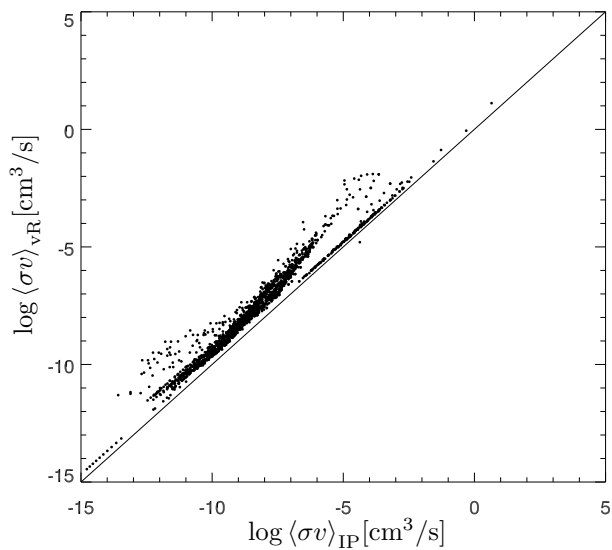} }};
\node at (0.23\textwidth,0.025\textwidth) {$\log{\langle\sigma v\rangle}_\mathrm{IP}[\rm{cm}^3/\rm{s}]$};
\node[rotate=90] at (0.0\textwidth,0.24\textwidth) {$\log{\langle\sigma v\rangle}_\mathrm{vR}[\rm{cm}^3/\rm{s}]$};
\end{tikzpicture}
\caption{Comparison of the \Mgi\ electron collision rate coefficients at T=5000~K using the impact parameter (IP) method and the van Regemorter formula (vR).}
\label{ipvr}
\end{figure}

It is well known that the Born approximation overestimates cross-sections at low energies and the IP method gives better results, as was
reported for instance by \citet{1962PPS....79.1105S}; see also \citet{1970ARAaA...8..329B} for a discussion.  Figure~\ref{ipvr} compares results from the two methods for allowed transitions in \Mgi. The vR results are larger than those from IP.  We also performed tests using either IP or vR in the atomic model, and we found that while we can reproduce the IR \Mgi\ emission lines observed in the Sun when using the IP method, we could not do so when using the vR formula (see \fig{IRlines}).  Thus, there is both physical and astrophysical support for preferring the IP method data over vR when \mbox{$R$-matrix} calculations are not available.  

Collisional ionisation of low-lying states by electrons was implemented according to the hydrogenic approximation presented in \citet[Sect~3.6.1]{allens}, which is based on semi-empirical expressions taken from \citet{1970ARAaA...8..329B}, which originate from \citet{Percival:1966tr}. For Rydberg levels we used the formula from \citet{eionRyd}, where the dependency of the collisional cross-sections on the total electronic orbital angular momentum of the Rydberg atom is taken into account.

\paragraph{$R$-matrix calculations.}
We now give the details of the electron collision calculations for the lowest ten states of \Mgi\ using the $R$-matrix (RM) method.  All calculations were performed in $L$-$S$ coupling.  Electronic structure calculations were performed using the code {\tt CIVPOL} (\citealt{CIVPOL}; see also \citealt{2004JPhB...37.2979P}), which employs the configuration interaction method.  {\tt CIVPOL} is an adaptation of {\tt CIV3} \citep{CIV3}, allowing the construction of polarisation pseudo-states.  
Orbitals for core electrons, $1s,\,2s,\,2p$ and $3s$, were taken from the ground-state Hartree-Fock calculations of \cite{Clementi}.  Valence orbitals, $4s,\,5s,\,3p,\,4p,\,3d,\,4d$ and $4f$, were optimised with {\tt CIVPOL}, providing energy levels that agree
well with experimental values taken from NIST; see Table~\ref{elevs}.  

We calculated the excitation cross-sections with the RM method \citep{1971JPhB....4..153B, 1976AdAMP..11..143B,1974CoPhC...8..149B,1978CoPhC..14..367B}.  This method separates the calculation into two cases: when the colliding electron is in the \emph{internal} region, that is, close to the target atom where interactions are strong, and when the colliding electron is in the \emph{external} region, far enough from the target atom such that interactions are weak and correlation and exchange effects can be neglected.  The $R$-matrix ensures continuity between wave functions at the boundary between the two regions. To perform the RM calculations for the {\it internal region} we used the code \mbox{{\tt RMATRIX I}}  \citep{InnerRmatrix}, and for the {\it external region} a version of the code {\tt STGF} \citep{OuterRmatrix} modified to treat collisions with neutral atoms \citep{Badnell}. 

\begin{table}
  \caption{Energies for \Mgi\ states from electronic structure calculations and comparison with experimental values.}\label{elevs}
  \centering
  \begin{tabular}{l l c c}\hline
    \multicolumn{2}{c}{State}    & \multicolumn{2}{c}{Energy (eV)} \\
    &         &    Experiment & {\tt CIVPOL}     \\\hline\hline\\[-0.2cm] 
    $2p^6$$3s^2$            & $^1$S   &      ---  &    --- \\
    \phantom{$2p^6$}$3s3p $ & $^3$P$^o$ &    2.71  &    2.67 \\
    \phantom{$2p^6$}$3s3p $ & $^1$P$^o$ &    4.34  &    4.38 \\
    \phantom{$2p^6$}$3s4s $ & $^3$S   &    5.11  &    5.10 \\
    \phantom{$2p^6$}$3s4s $ & $^1$S   &    5.39  &    5.36 \\
    \phantom{$2p^6$}$3s3d $ & $^1$D   &    5.75  &    5.73 \\
    \phantom{$2p^6$}$3s4p $ & $^3$P$^o$ &    5.93  &    5.88 \\
    \phantom{$2p^6$}$3s3d $ & $^3$D   &    5.94  &    6.01 \\
    \phantom{$2p^6$}$3s4p $ & $^1$P$^o$ &    6.12  &    6.05 \\
    \phantom{$2p^6$}$3s5s $ & $^3$S &    6.43 &    6.43 \\\hline\hline
  \end{tabular}
  \tablefoot{The energies shown are relative to the ground state of \Mgi. The experimental values were taken from NIST \citep{NIST}.}
\end{table}

After exploring the effects of the input data on the final collisional cross-sections, we adopted a target atom consisting of 25 spectroscopic states (i.e., energy levels), with ten additional pseudo-states states to represent the Rydberg levels and the continuum, giving 35 states in total.  Target atoms with more levels were also tested with no significant changes in the resulting cross-sections. 

The maximum occupation numbers in each $nl$ orbital used to build the configurations for both $N$ and $N+1$ atomic structures were\\[-0.2cm] 

\noindent {\small \begin{tabular}{l c c c c c c c c}
      Orbital     & $3s$ & $4s$ & $5s$ & $3p$ & $4p$ & $3d$ & $4d$ & $4f$  \\
    $N$ electrons   &  2   &  2   &  1   &  2   &  1   &  2   &  1   &  1    \\
    $N+1$ electrons &  2   &  2   &  2   &  3   &  2   &  3   &  2   &  2.    
  \end{tabular} } \\

\noindent The boundary between the internal and external regions was set at 30.4 atomic units.  The final calculation included partial waves $L\le20$.  We made test calculations using partial waves up to $L\le30$ to check convergence.  $R$-matrix calculations using $L\leq10$ provided rate coefficients within $20\%$ of those calculated using $L\le20$ except for the $3d\, ^3$D - $4p\, ^3$P$^o$ transition, where the rate coefficient increased by a factor of 2. There were no significant differences between the resulting rate coefficients when using $L\le30$ and $L\le20$ in the temperature range typical of cool stellar atmospheres.

The final collisional cross-sections obtained were compared with experiments and with other calculations where available. \cite{1742-6596-194-4-042029} calculated electron collisional excitation cross-sections for \Mgi\ from the ground state $3s\,^1S$ to the excited states $3p\,^{1,3}$P$^o$, $3d\,^1$D, $4s\,^1$S and $4p\,^1$P$^o$ and compared them with other calculations and with experiments. The cross-sections calculated in this work agree with those within $\sim$20\%. Rate coefficients were obtained by integrating the final cross-sections weighted by a Maxwellian velocity distribution and are presented in Table~\ref{tab:rates}. 

\begin{figure}[t]
\centering
\begin{tikzpicture}
\node[anchor=south west, inner sep=0] (image) at (0,0) {
\subfloat{\hspace{-3em}{\includegraphics[width=0.5\textwidth]{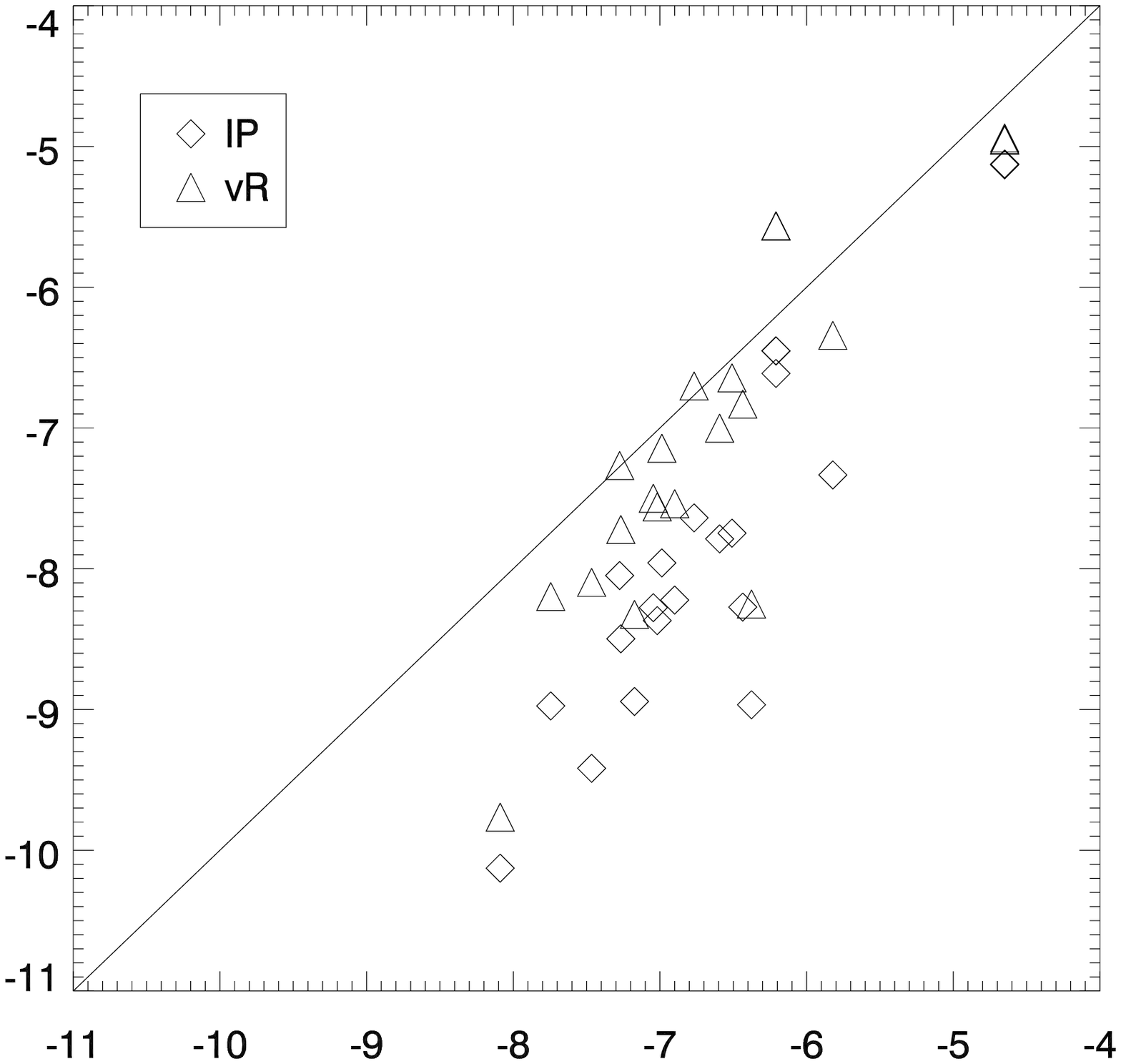}}}};
\node at (0.24\textwidth,0.02\textwidth) {$\log{\langle\sigma v\rangle}_\mathrm{This\,\,work}[\rm{cm}^3/\rm{s}]$};
\node[rotate=90] at (0.0\textwidth,0.23\textwidth) {$\log{\langle\sigma v\rangle}_\mathrm{IP\,,\,vR}[\rm{cm}^3/\rm{s}]$};
\end{tikzpicture}
\caption{Comparison between the electron collisional rates at $T=5000$~K calculated by us and those obtained with the impact parameter method (IP, diamonds) and the van Regermorter formula (vR, triangles).}
\label{RMipvr}
\end{figure}

\begin{figure}[t]
\centering
\begin{tikzpicture}
\node[anchor=south west, inner sep=0] (image) at (0,0) {
\subfloat{\hspace{-3em}{\includegraphics[width=0.5\textwidth]{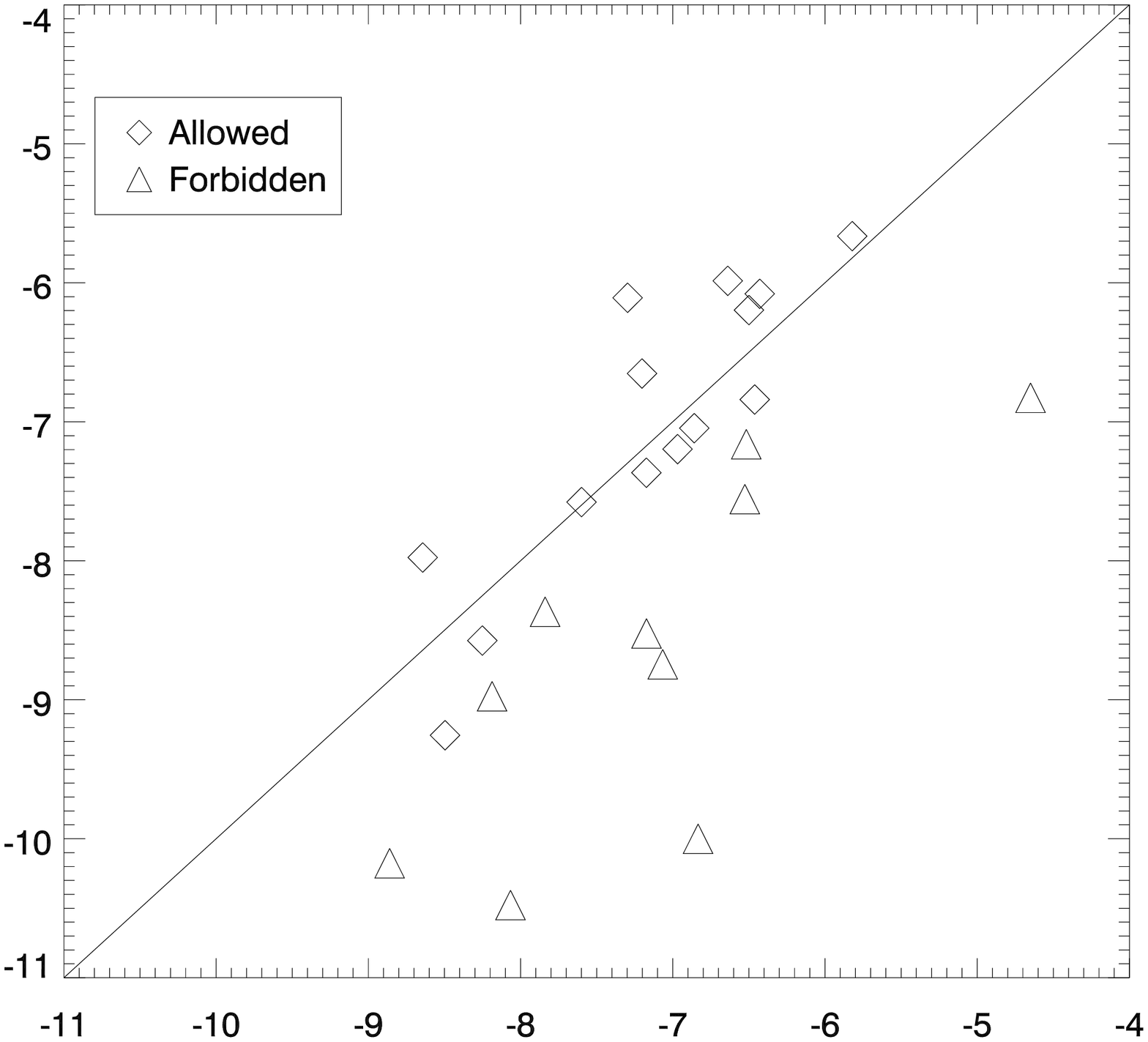}}}};
\node at (0.24\textwidth,0.02\textwidth) {$\log{\langle\sigma v\rangle}_\mathrm{This\,\,work}[\rm{cm}^3/\rm{s}]$};
\node[rotate=90] at (0.0\textwidth,0.24\textwidth) {$\log{\langle\sigma v\rangle}_\mathrm{Mauas\,\,et\,\,al.}[\rm{cm}^3/\rm{s}]$};
\end{tikzpicture}
\caption{Comparison between the electron collisional rates at $T=5000$~K calculated by us and those used by \cite{1988ApJ...330.1008M}. Diamonds correspond to the allowed transitions used by them (Table 1 in \citeauthor{1988ApJ...330.1008M}) and triangles are the forbidden transitions (Table 2 in \citeauthor{1988ApJ...330.1008M}).}
\label{mauascmp}
\end{figure}

Figure~\ref{RMipvr} compares our results with those obtained
with the IP and vR formulas. Both approximate formulas typically
yield lower values than the RM results.  Most past studies of \Mg\ in non-LTE have, however, employed the electron collisional data for transitions between low-lying levels that were presented in \cite{1988ApJ...330.1008M}. \citeauthor{1988ApJ...330.1008M} studied \Mgi\ non-LTE effects in the Sun using a 13-level model atom, and their compilation of electron collision data, adopting data from different sources (see that paper for details), has been employed extensively since.  Thus, it is of interest to compare our results with their compilation, which is done in
\fig{mauascmp}.  Allowed transitions (circles) have results similar to ours, but for forbidden transitions (diamonds) our results are significantly larger. For forbidden transitions \citeauthor{1988ApJ...330.1008M} mostly used rates equal to 0.1 times the collisional rate of an allowed transition with a similar energy difference calculated using the vR formula. 

For optically forbidden transitions, there are no simple  approximate formula similar to vR or IP, and we require some method to estimate the rates of forbidden transitions between highly excited levels.  Common approaches to circumvent this problem are to adopt an arbitrary collision strength $\Omega_{ij}$ of the order of unity, or to assume collision strengths similar to nearby optically allowed transitions.  We adopted a more nuanced approach.  First, from a physical point of view, it makes sense to divide optically forbidden transitions into two groups: those involving electron exchange (i.e. corresponding to intersystem transitions with change of spin state) and those not involving electron exchange.  At high energy, transitions not involving electron exchange may be described in the Born approximation, while transitions involving electron exchange are described in the Ochkur approximation.  This leads to significantly different behaviour of the cross-sections with energy, at least at high energy \citep[e.g.][]{Bransden:2003un,1992Aaamp;A...254..436B}.  Second, we then assume that transitions of the same type (exchange or non-exchange) and same transition energy will have similar rate coefficients.  Using Eq.~20 of  \citet{1992Aaamp;A...254..436B}, this implies a constant value of $\Upsilon_{ij}/g_i$ at a given temperature, where $\Upsilon_{ij}$ is the thermally averaged collision strength \citep[e.g.][eqn. 21]{1992Aaamp;A...254..436B}, and $g_i$ is the statistical weight of the initial state $i$.  

Indeed, we found using the results of the RM calculations that the behaviour of $\Upsilon_{ij}/g_i$ for a given transition type was similar, but significantly different between the two groups. Among the calculated RM collisions we have 4 non-exchange forbidden transitions and 40 spin-exchange transitions. 
We calculated the mean value for $\Upsilon_{ij}/g_i$ for the two groups from the RM calculations, and the results are presented in Table~\ref{tab:upsilon}.
The scatter of the spin-exchange transitions about the mean value is greatest at low temperatures, becoming smaller at higher temperatures. At 5\,000~K the values span over three orders of magnitude ($0.01\lesssim\Upsilon_{ij}/g_i\lesssim10$) with half of the values within a factor of 2 of the mean. The mean behaviour for exchange transitions with temperature is rather flat (except at the lowest temperature), while for non-exchange transitions the values steadily increase over the temperature range of interest.  To estimate rate coefficients for optically forbidden transitions not included in the RM calculations, we interpolate in this table, and then use Eq.~20 of  \citet{1992Aaamp;A...254..436B}, which relates the thermally averaged collision strength to the rate coefficient.  

\begin{table}
\center
\caption{Average values of $\Upsilon_{ij}/g_i$ (dimensionless) calculated from RM calculations as a function of temperature $T$.}
\label{tab:upsilon}
\begin{tabular}{rcc}
\hline
$T$ [K]       &  non-exchange & exchange \\
\hline

1\,000  & 0.086 & 1.020 \\
2\,000  & 0.174 & 0.654 \\
3\,000  & 0.257 & 0.635 \\
4\,000  & 0.336 & 0.642 \\
5\,000  & 0.413 & 0.650 \\
6\,000  & 0.492 & 0.655 \\
7\,000  & 0.568 & 0.658 \\
8\,000  & 0.646 & 0.660 \\
10\,000 & 0.797 & 0.658 \\ 
\hline
\end{tabular} 
\end{table}

\begin{table*}[ht!]
  \caption{Rate coefficients for de-excitation of \Mgi\ by electron collisions.  Rate coefficients, $\langle\sigma v\rangle$, are given in cm$^3$s$^{-1}$.}\label{tab:rates}
  \centering
  \begin{tabular}{ l  c c c c c c c c c}\hline\hline
  $_{\mathrm{initial}}$ \texttt{\char`\\ } $^{\mathrm{final}}$          & $3s\,^1$S & $3p\,^3$P$^o$ & $3p\,^1$P$^o$ & $4s\,^3$S & $4s\,^1$S$^o$ & $3d\,^1$D & $4p\,^3$P$^o$ & $3d\,^3$D & $4p\,^1$P$^o$   \\\hline
  \\
  &  &   &  & \multicolumn{2}{c}{1\,000 K}    &   &   &   &   \\  
$3p\,^3$P$^o$   & 5.01E$-$09  &   &   &   &   &   &   &   &   \\ 
$3p\,^1$P$^o$   & 3.95E$-$08  & 2.18E$-$06  &   &   &   &   &   &   &   \\ 
$4s\,^3$S           & 4.25E$-$09  & 7.69E$-$07  & 5.13E$-$08  &   &   &   &   &   &   \\ 
$4s\,^1$S           & 1.82E$-$09  & 2.59E$-$07  & 9.94E$-$08  & 7.94E$-$08  &   &   &   &   &   \\ 
$3d\,^1$D           & 5.52E$-$10  & 8.21E$-$08  & 2.58E$-$08  & 6.35E$-$08  & 1.25E$-$08  &   &   &   &   \\ 
$4p\,^3$P$^o$    & 1.72E$-$09  & 5.69E$-$07  & 4.73E$-$08  & 5.42E$-$07  & 1.13E$-$07  & 1.73E$-$06  &   &   &   \\ 
$3d\,^3$D           & 3.05E$-$10  & 7.52E$-$08  & 6.83E$-$09  & 7.95E$-$08  & 1.56E$-$08  & 1.64E$-$07  & 8.61E$-$05  &   &   \\ 
$4p\,^1$P$^o$    & 1.58E$-$09  & 1.24E$-$07  & 4.14E$-$08  & 5.18E$-$08  & 1.64E$-$07  & 3.29E$-$06  & 1.23E$-$07  & 4.94E$-$07  &   \\ 
$5s\,^3$S               & 1.56E$-$06  & 3.11E$-$04  & 2.56E$-$05  & 2.51E$-$04  & 1.17E$-$04  & 1.19E$-$03  & 9.01E$-$05  & 1.11E$-$03  & 1.55E$-$04  \\ 
     &  &   &  &  &   &   &   &   &   \\
     &  &   &  & \multicolumn{2}{c}{2\,000 K}    &   &   &   &   \\  
$3p\,^3$P$^o$           & 1.79E$-$08  &   &   &   &   &   &   &   &   \\ 
$3p\,^1$P$^o$     & 5.65E$-$08  & 6.79E$-$07  &   &   &   &   &   &   &   \\ 
$4s\,^3$S          & 1.11E$-$08  & 4.57E$-$07  & 7.91E$-$08  &   &   &   &   &   &   \\ 
$4s\,^1$S             & 7.37E$-$09  & 1.74E$-$07  & 2.22E$-$07  & 1.31E$-$07  &   &   &   &   &   \\ 
$3d\,^1$D             & 2.77E$-$09  & 9.74E$-$08  & 8.36E$-$08  & 9.25E$-$08  & 2.47E$-$08  &   &   &   &   \\ 
$4p\,^3$P$^o$         & 3.10E$-$09  & 2.68E$-$07  & 4.95E$-$08  & 3.98E$-$07  & 5.87E$-$08  & 6.24E$-$07  &   &   &   \\ 
$3d\,^3$D          & 8.93E$-$10  & 6.68E$-$08  & 1.53E$-$08  & 9.98E$-$08  & 1.45E$-$08  & 1.22E$-$07  & 4.25E$-$05  &   &   \\ 
$4p\,^1$P$^o$         & 4.24E$-$09  & 8.25E$-$08  & 6.73E$-$08  & 4.27E$-$08  & 1.93E$-$07  & 1.70E$-$06  & 1.40E$-$07  & 2.61E$-$07  &   \\ 
$5s\,^3$S             & 1.10E$-$07  & 5.30E$-$06  & 1.21E$-$06  & 5.57E$-$06  & 2.58E$-$06  & 1.98E$-$05  & 4.17E$-$06  & 2.26E$-$05  & 6.83E$-$06  \\ 
    &  &   &  &  &   &   &   &   &   \\  
  &  &   &  & \multicolumn{2}{c}{5\,000 K}    &   &   &   &   \\  
$3p\,^3$P$^o$         & 3.18E$-$08  &   &   &   &   &   &   &   &   \\ 
$3p\,^1$P$^o$     & 6.71E$-$08  & 3.04E$-$07  &   &   &   &   &   &   &   \\ 
$4s\,^3$S         & 1.54E$-$08  & 2.55E$-$07  & 8.09E$-$08  &   &   &   &   &   &   \\ 
$4s\,^1$S             & 1.45E$-$08  & 1.18E$-$07  & 3.67E$-$07  & 1.42E$-$07  &   &   &   &   &   \\ 
$3d\,^1$D             & 6.50E$-$09  & 8.58E$-$08  & 1.26E$-$07  & 7.86E$-$08  & 3.85E$-$08  &   &   &   &   \\ 
$4p\,^3$P$^o$         & 3.85E$-$09  & 1.62E$-$07  & 4.37E$-$08  & 3.08E$-$07  & 3.08E$-$08  & 2.87E$-$07  &   &   &   \\ 
$3d\,^3$D         & 1.38E$-$09  & 5.31E$-$08  & 1.98E$-$08  & 1.42E$-$07  & 9.67E$-$09  & 8.60E$-$08  & 2.24E$-$05  &   &   \\ 
$4p\,^1$P$^o$         & 8.14E$-$09  & 6.72E$-$08  & 9.58E$-$08  & 3.48E$-$08  & 3.10E$-$07  & 1.51E$-$06  & 1.33E$-$07  & 1.82E$-$07  &   \\ 
$5s\,^3$S             & 2.02E$-$08  & 4.21E$-$07  & 1.99E$-$07  & 4.20E$-$07  & 2.57E$-$07  & 1.84E$-$06  & 6.18E$-$07  & 1.81E$-$06  & 1.36E$-$06  \\  
   &   &   &   &   &   &   &   &   &   \\ 
     &  &   &  & \multicolumn{2}{c}{8\,000 K}    &   &   &   &   \\  
$3p\,^3$P$^o$          & 3.19E$-$08  &   &   &   &   &   &   &   &   \\ 
$3p\,^1$P$^o$      & 6.93E$-$08  & 2.25E$-$07  &   &   &   &   &   &   &   \\ 
$4s\,^3$S          & 1.36E$-$08  & 1.95E$-$07  & 6.84E$-$08  &   &   &   &   &   &   \\ 
$4s\,^1$S              & 1.51E$-$08  & 9.31E$-$08  & 4.21E$-$07  & 1.24E$-$07  &   &   &   &   &   \\ 
$3d\,^1$D              & 7.81E$-$09  & 7.51E$-$08  & 1.25E$-$07  & 5.98E$-$08  & 4.14E$-$08  &   &   &   &   \\ 
$4p\,^3$P$^o$          & 3.65E$-$09  & 1.41E$-$07  & 3.92E$-$08  & 2.69E$-$07  & 2.16E$-$08  & 2.04E$-$07  &   &   &   \\ 
$3d\,^3$D           & 1.38E$-$09  & 4.53E$-$08  & 1.84E$-$08  & 1.85E$-$07  & 7.41E$-$09  & 6.64E$-$08  & 1.66E$-$05  &   &   \\ 
$4p\,^1$P$^o$          & 1.02E$-$08  & 6.05E$-$08  & 1.12E$-$07  & 2.98E$-$08  & 4.13E$-$07  & 1.55E$-$06  & 1.13E$-$07  & 1.56E$-$07  &   \\ 
$5s\,^3$S              & 1.33E$-$08  & 2.25E$-$07  & 1.52E$-$07  & 1.83E$-$07  & 1.54E$-$07  & 1.03E$-$06  & 3.47E$-$07  & 8.21E$-$07  & 1.03E$-$06  \\ 
     &   &   &   &   &   &   &   &   &   \\ 
     &  &   &  & \multicolumn{2}{c}{10\,000 K}    &   &   &   &   \\  
$3p\,^3$P$^o$          & 3.04E$-$08  &   &   &   &   &   &   &   &   \\ 
$3p\,^1$P$^o$      & 7.05E$-$08  & 1.95E$-$07  &   &   &   &   &   &   &   \\ 
$4s\,^3$S          & 1.22E$-$08  & 1.73E$-$07  & 6.09E$-$08  &   &   &   &   &   &   \\ 
$4s\,^1$S              & 1.49E$-$08  & 8.22E$-$08  & 4.44E$-$07  & 1.13E$-$07  &   &   &   &   &   \\ 
$3d\,^1$D             & 8.22E$-$09  & 6.96E$-$08  & 1.23E$-$07  & 5.10E$-$08  & 4.19E$-$08  &   &   &   &   \\ 
$4p\,^3$P$^o$         & 3.46E$-$09  & 1.35E$-$07  & 3.68E$-$08  & 2.51E$-$07  & 1.81E$-$08  & 1.74E$-$07  &   &   &   \\ 
$3d\,^3$D          & 1.33E$-$09  & 4.21E$-$08  & 1.72E$-$08  & 2.10E$-$07  & 6.50E$-$09  & 5.75E$-$08  & 1.44E$-$05  &   &   \\ 
$4p\,^1$P$^o$         & 1.13E$-$08  & 5.70E$-$08  & 1.21E$-$07  & 2.73E$-$08  & 4.70E$-$07  & 1.56E$-$06  & 1.02E$-$07  & 1.44E$-$07  &   \\ 
$5s\,^3$S             & 1.16E$-$08  & 1.80E$-$07  & 1.49E$-$07  & 1.31E$-$07  & 1.34E$-$07  & 8.60E$-$07  & 2.75E$-$07  & 5.98E$-$07  & 9.61E$-$07  \\ 
\hline 
  \end{tabular}
\end{table*}

\subsubsection{Hydrogen atoms}
\label{sect:hydrogen}

The rate coefficients, $\langle \sigma \upsilon \rangle$, for excitation and de-excitation processes,
\begin{equation}
\mathrm{Mg }(3s \, nl\;^{2S+1}L)+\mathrm{H}(1s) \rightleftharpoons \mathrm{Mg }(3s \, n^\prime l^\prime\; ^{2S^\prime +1}L^\prime)+\mathrm{H}(1s) ,
\end{equation}
and for the charge transfer processes, ion-pair production and mutual-neutralisation, involving the ionic state, 
\begin{equation}
\mathrm{Mg }(3s \, nl\;^{2S+1}L) + \mathrm{H}(1s) \rightleftharpoons \mathrm{Mg}^+(3s\;^2\mathrm{S})+\mathrm{H}^- ,
\end{equation}
are taken from \cite{2012AaA...541A..80B}, which are based on calculations of \citet{2012PhRvA..85c2704B}; see also \citet{Guitou:2010hr,2011JPhB...44c5202G}.  These \emph{ab initio} quantum chemical calculations on which the scattering calculations are based were obtained using the
code {\tt MOLPRO}   \citep{molpro} and cover the seven lowest lying states of \Mgi\ and the ground state of \Mgii.  

We wish to emphasize that with regard to hydrogen collisions in general, the accurate quantum \citep{2003AaA...409L...1B,2010AaA...519A..20B,2012AaA...541A..80B} as well as the more approximate model calculations \citep{2013AaA...560A..60B,2014Aaamp;A...572A.103B}, based on a correct physical background, provide non-zero rate coefficients for both optically allowed and optically forbidden transitions, as well as for charge transfer processes, which generally have the highest rate coefficients among all hydrogen collision processes. Thus, the Drawin formula fails to provide reliable estimates for important inelastic processes in collisions with hydrogen.  In the case of Mg, the quantum eight-channel calculations \cite{2012PhRvA..85c2704B} provide 28 non-zero cross-sections for the endothermic processes (excitation and ion-pair production), while the Drawin formula gives only five non-zero (overestimated) cross-sections for the same 28 transitions \citep[see][]{2012AaA...541A..80B}. This is a general feature of the Drawin formula, as discussed in detail by \cite{2011AaA...530A..94B}.

For collisional excitation involving higher lying and Rydberg states of \Mgi,\ we employed the free electron model of \cite{Kaulakys:1986tl,1991JPhB...24L.127K}, which considers a binary encounter of the Rydberg electron with a neutral perturber.  This approach allows writing the cross-section for the inelastic transition $nl \rightarrow n'l'$  in terms of the elastic differential cross-section for the electron-perturber scattering and the momentum-space wave functions of the initial state \cite[eqs. 6--8]{1991JPhB...24L.127K}.  For the wave functions $g_{nl}(p)$, where $p$ is the electron momentum, we calculated non-hydrogenic wave functions from quantum-defect theory following \cite{1997JPhB...30.2403H}\footnote{The required Hankel transforms were derived analytically for $l$ ranging from 0 to 13 using Mathematica.  We note that the denominators in Eqns. 32 and 33 of  \cite{1997JPhB...30.2403H} contain a misprint where an exponent is incorrectly written as a multiplicative factor - the denominator of Eqn. 32 should be $[\nu^{-2}+q^2]^{(\nu-t+1)/2}$.  
}.  For neutral hydrogen perturbers, the relevant elastic differential cross-section $|f_e(p,\theta)|^2$ is that for e$^-$+\H\ collisions, where $f_e(p,\theta)$ is the scattering amplitude.  This total cross-section can be written in terms of the singlet and triplet amplitudes, $f^+$ and $f^-$, which is found by averaging over initial spins and summing over final spins to obtain the well-known expression \citep[e.g.][]{1962amp..conf..375S}
\begin{equation}
|f_e(p,\theta)|^2 = \frac{1}{4} |f^+(p,\theta)|^2 + \frac{3}{4} |f^-(p,\theta)|^2 .
\end{equation} 
The elastic singlet and triplet scattering amplitudes for e$^-$+\H\ collisions are known from variational calculations of the phase shifts for $s$- and $p$-wave scattering \citep{1961PhRv..124.1468S, 1968PhRv..171...91A}.  We make use of the form introduced by HBvR, namely
\begin{equation}
|f^{\pm}(p,\theta)|^2 = A^{\pm} + B^{\pm} \cos(\theta)
,\end{equation} 
where $A^{\pm}$ is the $s$-wave contribution and $B^{\pm}$ the $s$-$p$ coupling.  The coefficients $A^{\pm}$ and $B^{\pm}$ are listed as a function of $p$ in Table 3 of HBvR, derived from the published phase shifts mentioned above.  From this we calculated the total cross-section for the transition $nl \rightarrow n'l'$, which we denote $\sigma_{nl \rightarrow n'l'}$. 

In our model atom, singlet and triplet terms are often separated, and thus we would like to estimate how this total cross-section is distributed among final spin states.  
We considered a representation where the spin of the Rydberg electron is coupled to the spin of the other electrons in the target atom $S_c$ (assuming this to be a good quantum number) to give the total electronic spin of the Rydberg atom, $S$.  Recoupling of angular momenta gives the required spin-changing scattering amplitude in terms of the singlet and triplet scattering amplitudes (i.e. in the representation in which the Rydberg electron spin and the hydrogen atom spin are coupled), and the differential cross-section for change of spin $S \rightarrow S'$, $S\ne S'$, is found to be
\begin{eqnarray}
|f_e(p,\theta,S \rightarrow S')|^2 & = & \frac{(2S'+1)}{8(2S_c+1)} |f^+ - f^-|^2 \\
 & = &\frac{(2S'+1)}{2(2S_c+1)} \sigma_{exch} 
\end{eqnarray}
where $\sigma_{exch}$ is the differential cross-section for electron exchange in elastic e$^-$+\H\ collisions \citep[e.g.][]{1958PIRE...46..240F,1962PhRv..126..163B}.  The derivation essentially follows \cite{1965RSPSA.286..519D}, except that in our case the electron spin is coupled to the core electronic spin and not to the nuclear spin.  We note that $S' = S_c \pm 1/2$ and the factor $(2S'+1)/2(2S_c+1)$ is always in the range 0 to 1 and represents the probability that an electron exchange leads to a change in the total electronic spin of the Rydberg atom.  In the limit $p \rightarrow 0$, $-f^\pm \rightarrow a^\pm$, where $a^\pm$ are the scattering lengths, which have values of $a^+ = 5.965$ and $a^- = 1.769$ atomic units \citep{1961PhRv..124.1468S}.  In this limit, for the case of $S_c=1/2$, via comparison of the expressions for the cross sections we find that
\begin{equation}
\sigma_{nl,S=0 \rightarrow n'l',S'=0} = 0.706 \: \sigma_{nl \rightarrow n'l'},
\end{equation} 
\begin{equation}
\sigma_{nl,S=0 \rightarrow n'l',S'=1} = 0.294 \: \sigma_{nl \rightarrow n'l'},
\end{equation} 
\begin{equation}
\sigma_{nl,S=1 \rightarrow n'l',S'=0} = 0.098 \: \sigma_{nl \rightarrow n'l'},
\end{equation} 
\begin{equation}
\sigma_{nl,S=1 \rightarrow n'l',S'=1} = 0.902 \: \sigma_{nl \rightarrow n'l'}.
\end{equation}

\subsection{Test model atoms}
\label{sect:testatoms}

To understand the uncertainties in the modelling, we constructed various model atoms in which sets of collisional data were changed or removed compared to the final model. For ease of discussion, we adopted the following abbreviations\footnote{These notations are adopted from {\tt MULTI}.} for the various collision processes: \\

\begin{tabular}{l l}
CE & Collisional (de)excitation by electrons \\
CI & Collisional ionisation by electrons \\
CH &  Collisional (de)excitation by hydrogen \\
CH0 &  Charge transfer with hydrogen. \\
\end{tabular}\\

For reference, our final model adopted for CE data RM data wherever possible, IP data for all remaining allowed transitions, and we used the recipe described in Sect.~\ref{sect:electrons} and defined by Table~\ref{tab:upsilon} for forbidden transitions (hereafter referred to as the ``$\Upsilon$'' recipe).  The final model, which we give the label F, also includes data for CI, CH and CH0 as described in Sect.~\ref{sect:collision_data}.    

The first additional test model represents a basic model without any of the new data introduced here.  This model contains only CE and CI, and no hydrogen collision processes.  For CE it employs IP data for all allowed transitions and uses the $\Upsilon$ recipe for all forbidden transitions\footnote{The $\Upsilon$ recipe is a new addition, based on new RM data. However, its importance is limited and we chose to use it in the basic model to limit the number of degrees of freedom in our tests.  Its effects are tested via the F(IP,$\Omega=1$) model described later in this section.}. Thus, we refer to it, where appropriate, as the basic model, model B. 

The models F and B therefore represent two extremes, and it is interesting to investigate models that lie between these cases.  In particular, we constructed models B$+$RM (introduces RM), B$+$RM$+$CH (further introduces CH), B$+$RM$+$CH0 (introduces RM and CH0), B$+$H (introduces CH and CH0), which may be used to explore the relative importance of RM, CH, CH0, or hydrogen collisions in total, H, in causing the differences between model B and model F.  To isolate the effects of spin changing collisions, we constructed versions of the final model where spin changing collisions (i.e. between singlet and triplet states in \Mgi) are removed for collisions with electrons (F$-$SCE), hydrogen (F$-$SCH) or both (F$-$S).  Transitions between levels of different spin can only be removed when the spin of each level is defined, but not in merged super levels.

Finally, we constructed slightly modified versions of model F, in which the IP data were replaced with vR data and/or the $\Upsilon$ recipe was replaced by a typical approximation of setting the collision strength to unity, $\Omega=1$.  These models are labelled F(vR,$\Upsilon$), F(vR,$\Omega=1$), F(IP,$\Omega=1$), noting that F(IP,$\Upsilon$) would be equivalent to the final model F in this notation, and these models can be used to investigate the impact of various approximations for CE among high-lying levels not covered by RM data.

Table \ref{tab:testatoms} gives an overwiew of these models. By examining all of these models, we can determine for a given model atmosphere which of the newly introduced data or processes has produced the most important changes in each line, and this is reported in the last four columns of Table~\ref{linestable} and is part of our discussions of below, where we discuss the results for the various groups of lines. 

\begin{table}[h]
\caption{Description of the collisional data for the different versions of model atoms described in Sect.\,\ref{sect:testatoms}}\label{tab:testatoms}
{\tiny
\hspace{-0.01\textwidth}
\begin{tabular}{| l | c | c | c | c |}\hline
   \multirow{2}{*}{Name}        &        \multicolumn{2}{c |}{CE}                         &   \multirow{2}{*}{CH}                 &    \multirow{2}{*}{CH0}    \\\cline{2-3}
                                &   Allowed             &  Forbidden                         &                               &                            \\\hline\hline
        B                                       &      IP                       &     $\Upsilon$          &                                       &                                 \\\hline
        B+RM                    &    RM+IP              &  RM+$\Upsilon$         &                                       &                            \\\hline
  B+RM+CH               &    RM+IP              &  RM+$\Upsilon$         &  \checkmark           &                               \\\hline 
  B+RM+CH0              &    RM+IP              &  RM+$\Upsilon$         &                                       &   \checkmark         \\\hline 
    B+H                         &      IP                       &     $\Upsilon$          &   \checkmark          &  \checkmark         \\\hline 
     F $\equiv$ F(IP,$\Upsilon$)        &   RM+IP               & RM+$\Upsilon$   &       \checkmark      &    \checkmark         \\\hline
                                                 \multicolumn{5}{| c | }{}                         \\\hline 
  \multirow{2}{*}{F$-$SCE}      & \multirow{2}{*}{RM+IP}        & RM+$\Upsilon-$spin      &       \multirow{2}{*}{\checkmark}          &  \multirow{2}{*}{\checkmark}          \\
                                                &               &     exchange               &                                        &                                  \\\hline 
  \multirow{2}{*}{F$-$SCH}      & \multirow{2}{*}{RM+IP}        & \multirow{2}{*}{RM+$\Upsilon$} & No spin        & \multirow{2}{*}{\checkmark}             \\
                                                &                                       &                                                 &     exchange      &                  \\\hline
\multirow{2}{*}{F$-$S}  & \multirow{2}{*}{RM+IP}        & RM+$\Upsilon-$spin & No spin        &     \multirow{2}{*}{\checkmark}          \\
                                                &                            &      exchange                 &     exchange      &        \\\hline 
                        \multicolumn{5}{| c |}{}                \\\hline 
   F(IP,$\Omega$=1)     &   RM+IP       & RM+($\Omega=1$)       &         \checkmark      &    \checkmark         \\\hline
   F(vR,$\Upsilon$)     &   RM+vR       & RM+$\Upsilon$         &         \checkmark      &    \checkmark         \\\hline
   F(vR,$\Omega$=1)&   RM+vR    & RM+($\Omega=1$)       &       \checkmark         &    \checkmark         \\\hline
\end{tabular}
}
\tablefoot{RM+x means RM when available, otherwise method x.}
\end{table}

\section{Comparisons with observations}
\label{sect:compobs}

We now test our modelling by comparison with observed spectra.  First, we compare our results with optical spectra of standard benchmark late-type stars, including the Sun.  Then we compare
them with the IR emission lines in the solar spectrum.

\subsection{Optical absorption lines}
\label{sect:comp}
We tested the performance of our non-LTE modelling using high-quality spectra of six stars with reasonably well-known fundamental parameters. This sample of benchmark stars was designed to cover a wide parameter range and includes the Sun, Procyon, Arcturus, and three very metal-poor stars ($\rm{[Fe/H]}=-2.5\ldots-2.0$); a turn-off star (HD\,84937), a sub-giant (HD\,142083), and a red giant (HD\,122563). For the Sun and Arcturus, spectra from the Kitt Peak observatory with $R\approx150\,000$ and $S/N\approx1000$ are available \citep{Hinkle00}. For the other stars, we retrieved VLT/UVES spectra at $R\approx47\,000$ from the POP archive of bright field stars \citep{Bagnulo03}.

\begin{figure*}[t]
\centering
\begin{tikzpicture}
\node[scale=0.65, rotate=90, anchor=south west, inner sep=0] (image) at (0,0) {
\subfloat{\includegraphics[width=\textwidth]{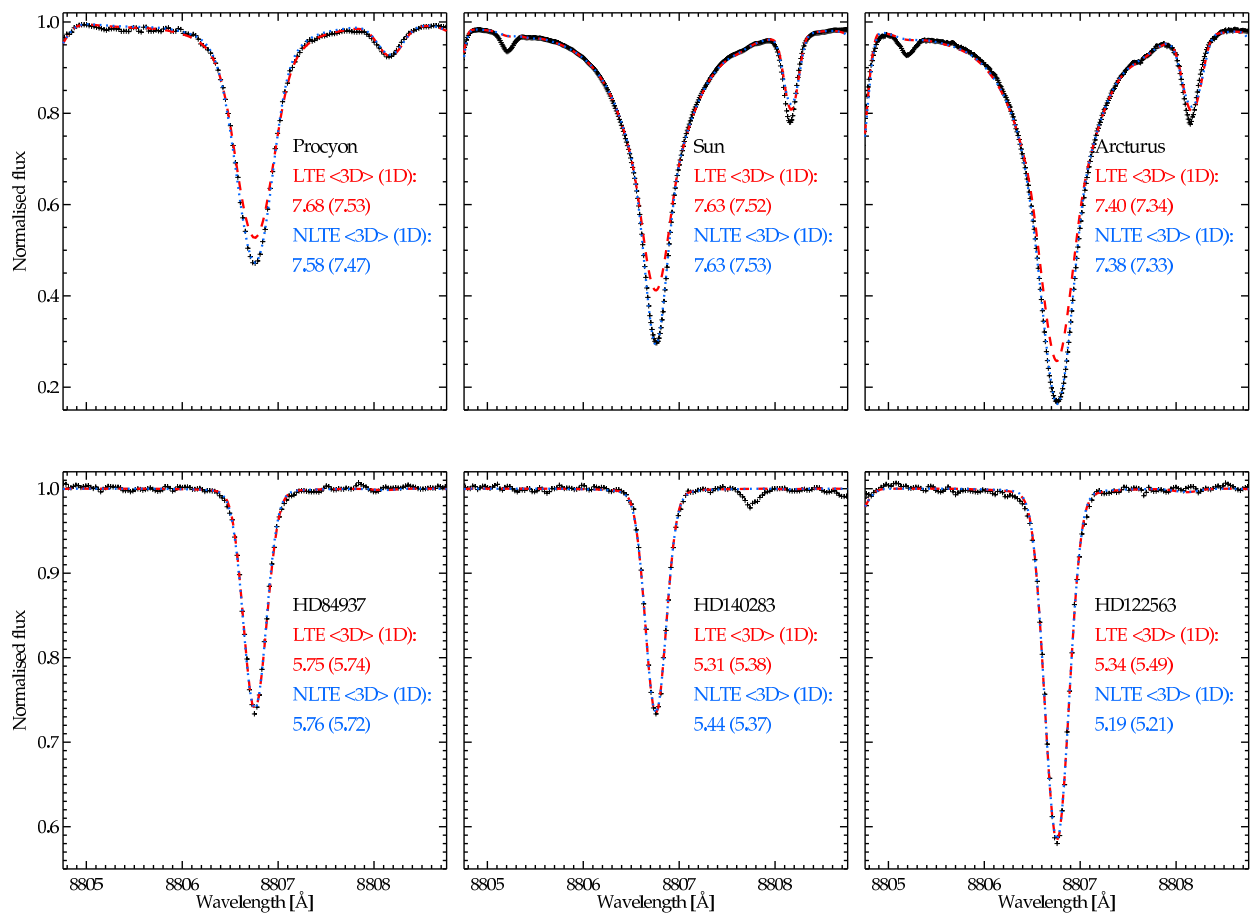}}
};
\end{tikzpicture}
\caption{Observed spectra of the non-LTE sensitive \Mgi\ 8806~\AA\  line \emph{(pluses)}. The best-fitting \avstar LTE \emph{(red dashed line)} and non-LTE profile \emph{(blue dotted line)} are overplotted and the corresponding \Mg\ abundances stated in each panel.}\label{8806}
\end{figure*}

Accurate effective temperatures for Procyon, Arcturus, and HD\,122563 were derived from direct measurements of their angular diameters (see adopted parameters in Table \ref{benchmark}). \cite{2014arXiv1410.4780C} presented fundamental parameters of HD\,140283, although \teff\ is uncertain, and for HD\,84937 no such measurements have been published, so we relied instead on the infra-red flux method \citep{Casagrande10}. For all stars except for the Sun, \textsc{Hipparcos} parallax measurements were used to calculate the surface gravity \citep{vanLeeuwen07}. Reference metallicities were taken from \citet{Jofre14} for Arcturus and \citet{Bergemann12} for all other stars. Microturbulence values were also taken from the literature studies and held fixed in the line formation calculations.

\begin{table*}
  \caption{Stellar parameters and average \Mg\ abundances, including line-to-line 1$\sigma$ dispersions, for the benchmark stars.}\label{benchmark}
  \centering
  \begin{tabular}{l r rr r r r rrrrrrr}\hline\hline\\[-0.2cm]
    Name  &  \teff  & \logg & [Fe/H] & $v_{\rm mic}$ & \multicolumn{4}{c}{1D} & & \multicolumn{4}{c}{\avstar}\\
          \cline{6-9}\cline{11-14}\\[-0.2cm] 
          &   [K]   &          &        & [km/s] & \multicolumn{2}{c}{LTE} & \multicolumn{2}{c}{non-LTE}& & \multicolumn{2}{c}{LTE} & \multicolumn{2}{c}{non-LTE}\\\hline\\[-0.2cm] 
    Sun     $^{(a)}$  & 5777   & 4.44   &  0.00  &  1.09 &7.57&0.08&7.57&0.08&&7.67&0.07&7.66&0.07\\
    Procyon $^{(b)}$  & 6545   & 3.99   & -0.03  &  2.00 &7.47&0.04&7.45&0.04&&7.60&0.05&7.56&0.06\\
    Arcturus$^{(b)}$  & 4247   & 1.59   & -0.52  &  1.63 &7.42&0.07&7.33&0.07&&7.52&0.10&7.38&0.06\\
    HD\,84937 $^{(a)}$  & 6408   & 4.13   & -2.03  &  1.40 &5.73&0.04&5.75&0.03&&5.78&0.05&5.80&0.03\\
    HD\,140283$^{(a)}$  & 5777   & 3.67   & -2.40  &  1.23 &5.36&0.10&5.39&0.10&&5.35&0.16&5.45&0.09\\
    HD\,122563$^{(b)}$  & 4608   & 1.61   & -2.64  &  1.50 &5.29&0.13&5.26&0.13&&5.28&0.08&5.27&0.08\\
    \hline\\[-0.2cm] 
  \multicolumn{12}{l}{Model parameters taken from $^{(a)}$ \citet{Bergemann12}, $^{(b)}$ \citet{Jofre14}}
  \end{tabular}
\end{table*}

We obtained 1D and \avstar\ model atmospheres for each star by interpolating in the MARCS \citep{2008AaA...486..951G} and STAGGER \citep{2013AaA...560A...8M} grids. Minor extrapolation in the \avstar\ grid was sometimes necessary to produce the desired model parameters. The 1D spectrum synthesis code {\tt SME} \citep[][and subsequent updates]{Valenti96} was used to compute synthetic spectra including blending lines. To synthesise the lines in non-LTE, grids of pre-computed departure coefficients\footnote{Defined as $b_i=n_i/n_i^*$ where $n_i$ is the non-LTE population and $n_i^*$ the LTE population of level $i$.} $b_i$ from {\tt MULTI} (using model F) were read in and interpolated before they
were applied to the LTE level populations computed by {\tt SME}. For all interpolation and line-formation, the continuum optical depth at 5000~\AA\ was used as reference scale, which is consistent with the reference scale used when computing the spatial and temporal average of the full 3D models.  We compared the spectral line profiles calculated by MULTI and SME used in this fashion, and they give practically identical results.

Between eight and eleven \Mg\ lines, selected to be relatively blend-free and situated between $4000-9000$\AA\  were synthesised per star. Whenever possible, the best-fitting abundance was found by simultaneously varying the \Mg\ abundance and the intrinsic line broadening, such as  caused by macro-turbulence and rotation. It was often found that neglecting non-LTE effects in the cores of strong lines prevented a match between the LTE synthetic spectrum and the observations without artificially reducing all broadening, even the instrumental, to unphysical values. In such cases, we instead fixed the broadening parameters to the values found in non-LTE and disregarded the line core when optimising the fit. By doing so, the agreement between LTE and non-LTE abundances is typically better than if the same equivalent widths had been enforced, for instance. For lines weaker than 150\,m\AA, the LTE and non-LTE line profiles are similar, and a good fit to the full observed line profile can always be found. Sample \avstar\  fits are seen in \fig{8806} for Mg\,I 8806~\AA .

The choice of model atmosphere in 1D or \avstar\ and line-formation method in LTE or non-LTE affects the derived \Mg\ abundances to various degrees and in various directions. Average abundances and line-by-line $1\sigma$ dispersions are presented in Table\,\ref{benchmark}\footnote{The abundance has the usual definition $ A(\Mg)=\log{(N_{\Mg}/N_{\H})}+12 $, where $N_{\Mg}$ is the number density of magnesium atoms and ions.}. We stress that our main goal is to illustrate the impact of the modelling method on a variety of different lines, not only those that are most suitable for abundance analysis. The average abundances we present should therefore not be taken too literally. For the Sun, non-LTE effects are very weak, of the order of 0.01\,dex, while \avstar\ abundances are higher by $0.05-0.15$\,dex than the corresponding 1D abundances because of the hotter temperatures ($\sim1\%$) in the line-formation region. Only two lines overlap with the recent study by \citet{Scott14}, but the average solar abundances nevertheless match well (A(Mg)$_\mathrm{Scott\, et.\, al.}=7.59\pm$0.04). 

In Arcturus and Procyon the effects work in the same direction as in the Sun, but the sensitivity to non-LTE is higher. The over-population of \Mgi\  penetrates to deeper layers and abundance corrections reach up to $-0.1$\,dex in Procyon and up to $-0.3$\,dex in Arcturus. In the metal-poor stars, most lines are subjected to a similar weakening, that is, abundance strengthening, in \avstar\ LTE as in 1D LTE because of the somewhat hotter temperatures at $\taufh\approx1$. However, lines that form relatively far out, such as the intercombination line 4571~\AA\ and the near IR 8806~\AA , are affected by the strong cooling of the surface layers that is characteristic of metal-poor radiation-hydrodynamical models \citep{2001Aaamp;A...372..601A}. Consequently, \avstar\ LTE abundances are significantly lower for these lines, up to $-$0.25\,dex in HD122563. However, the effects of the surface cooling on line strengths is counteracted by non-LTE effects in the opposite direction. The behaviour is qualitatively similar to that found for resonance lines of \ion{Ca}{i} in \citet{Lind13} and \ion{Fe}{i} in \citet{Bergemann12}. As also described in these studies, the steeper temperature gradients of metal-poor \avstar\ models than in 1D models strengthens the overionisation of neutral species, such as \Mgi, resulting in positive non-LTE corrections for all lines. Reassuringly, the line-to-line abundance scatter in \avstar non-LTE is overall an improvement with respect to 1D LTE.
 
\begin{figure}[t]
\centering
\includegraphics[height=0.45\textwidth,angle=90]{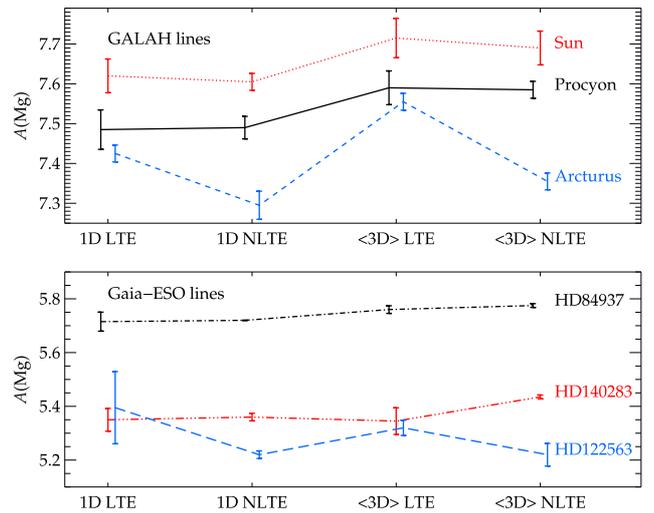}
\caption{Top panel: \Mg\ abundances of stars in the solar neighbourhood affected by different types of modelling. The spectral lines used are those covered by the GALAH survey: 5711\AA\ and 7691\AA . The bottom panel shows the same for metal-poor halo stars, based on lines observed in the Gaia-ESO survey: 5528\AA\ and 8806\AA .}\label{surveys}
\end{figure}

In \fig{surveys} we illustrate how abundances of the benchmark stars compare when limited to the diagnostics available to the GALAH and Gaia-ESO survey for metal-rich and metal-poor stars. Evidently, systematic uncertainties are comparable to or even exceed the targeted precision (0.1 and 0.05 dex for Gaia-ESO and GALAH, respectively). The subgiant and dwarf stars in our sample do not appear to suffer from strong differential effects, and sufficiently precise relative abundances can probably be obtained also with traditional modelling of Mg, although they may not necessarily be accurate in an absolute sense. However, giants show strong differential non-LTE effects ($0.1-0.2$\,dex), which must be corrected for to allow comparisons to less evolved stars.    

\subsection{Infrared emission lines}
\label{sect:IRlines}
As discussed in Sect.~\ref{sect:hist}, the formation of IR \Mgi\ emission lines in the solar spectrum, discovered in the early 80s, took a decade and considerable effort to understand.  Two main hypotheses arose: they form in the chromosphere under LTE conditions, or they form in the photosphere with a non-LTE source function increasing outwards.  This debate was finally and unambiguously
resolved by \cite{Carlsson12mu92} in favour of a non-LTE line formation scenario of photospheric origin.  Their modelling made several reasonable, yet \emph{ad hoc}, assumptions regarding collision rates.  As pointed out by \citet{1994IAUS..154..309R}, the 12~$\mu$m lines have non-LTE emissions peaks \emph{because} of the strong collisional coupling between the pertinent levels.  Based on the expectation that hydrogen collisions would provide such high rates \citep{Omont:1977vw}, \citeauthor{Carlsson12mu92} introduced high electron collisional rates for $l$-changing collisions among Rydberg states to ensure detailed balancing between levels of equal $n$.  Attempts at modelling since that time, such as \citet{1998AaA...333..219Z} and \citet{2008AaA...486..985S}, have described hydrogen collisions using the highly debated Drawin formula scaled by a free parameter chosen to best fit observations.  Importantly, however, these studies showed that the IR lines in the Sun and in red giants such as Arcturus are rather sensitive to the description of hydrogen collisions, and as these lines probe the highly excited parts of the atom, they are important tests of our \emph{ab initio} modelling.

\begin{figure*}[ht!]
\begin{tabular}{lll}
\vspace{-0.04\textwidth}\begin{tikzpicture}
\node[anchor=south west, inner sep=0] (image) at (0,0) {
\subfloat{\includegraphics[width=0.32\textwidth]{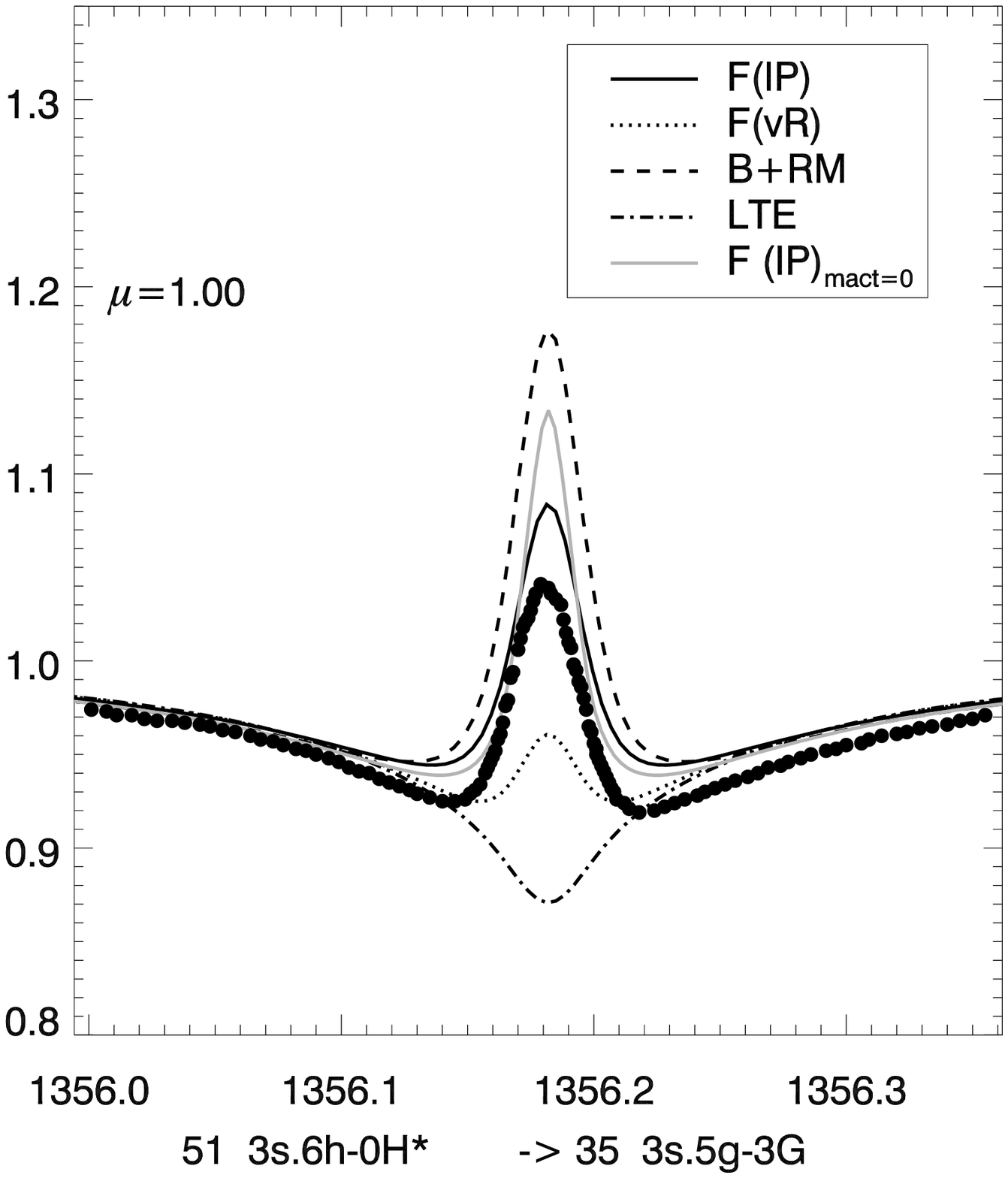}}
};
\draw [fill=white,white] (0.03\textwidth,0.00\textwidth) rectangle (0.25\textwidth,0.02\textwidth);
\node[left,rotate=90] at (0.0\textwidth,0.25\textwidth) {\scalebox{0.7}{Normalized Intensity}};   
\end{tikzpicture}
 & & \\
\vspace{-0.04\textwidth}\hspace{-0.01\textwidth}\begin{tikzpicture}
\node[anchor=south west, inner sep=0] (image) at (0,0) { 
\subfloat{\includegraphics[width=0.32\textwidth]{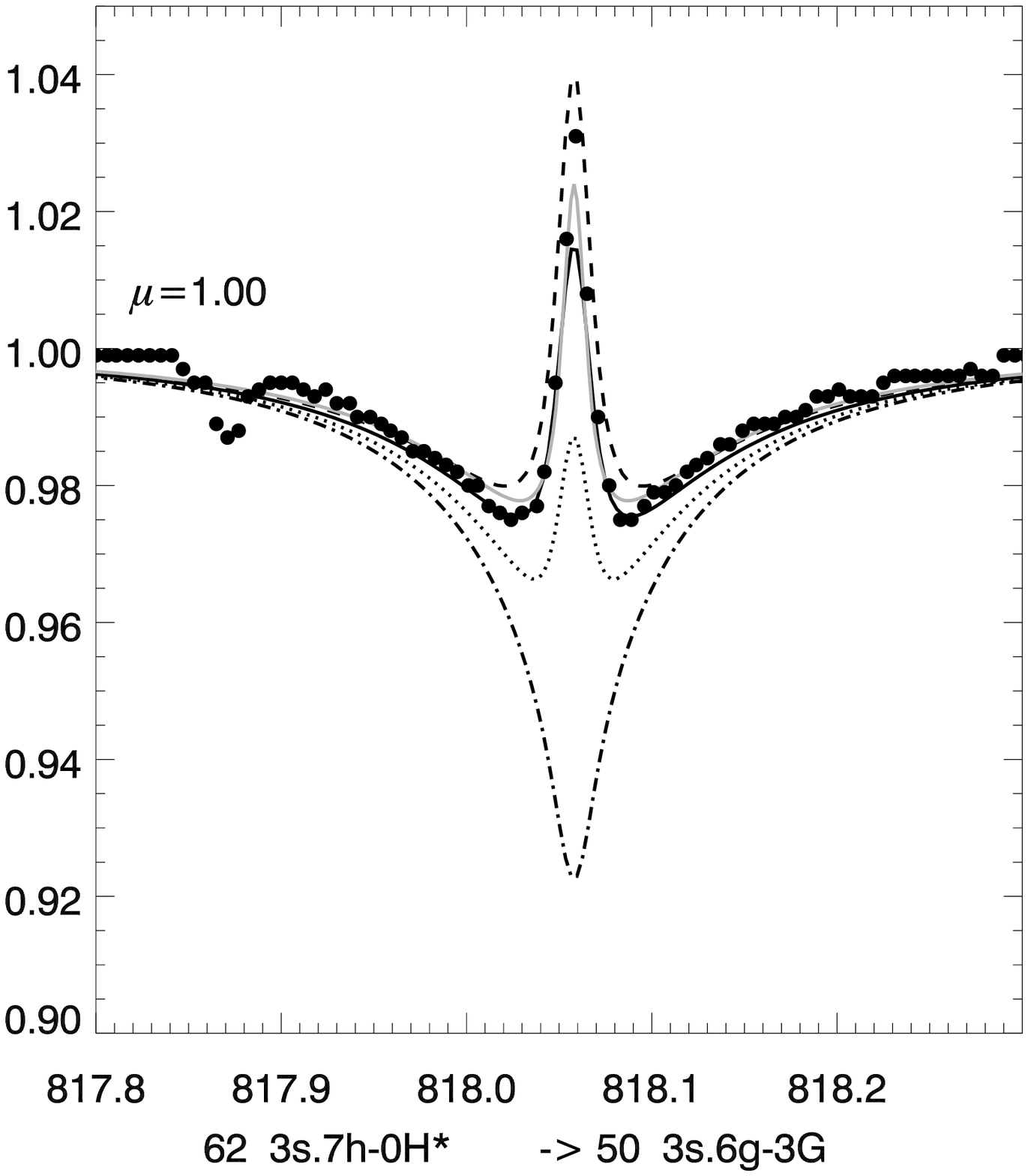}}
};
\draw [fill=white,white] (0.03\textwidth,0.00\textwidth) rectangle (0.25\textwidth,0.02\textwidth);
\node[left,rotate=90] at (-0.01\textwidth,0.25\textwidth) {\scalebox{0.7}{Normalized Intensity}};   
\end{tikzpicture}
&
\hspace{-0.02\textwidth}\begin{tikzpicture}
\node[anchor=south west, inner sep=0] (image) at (0,0) {
\subfloat{\includegraphics[width=0.32\textwidth]{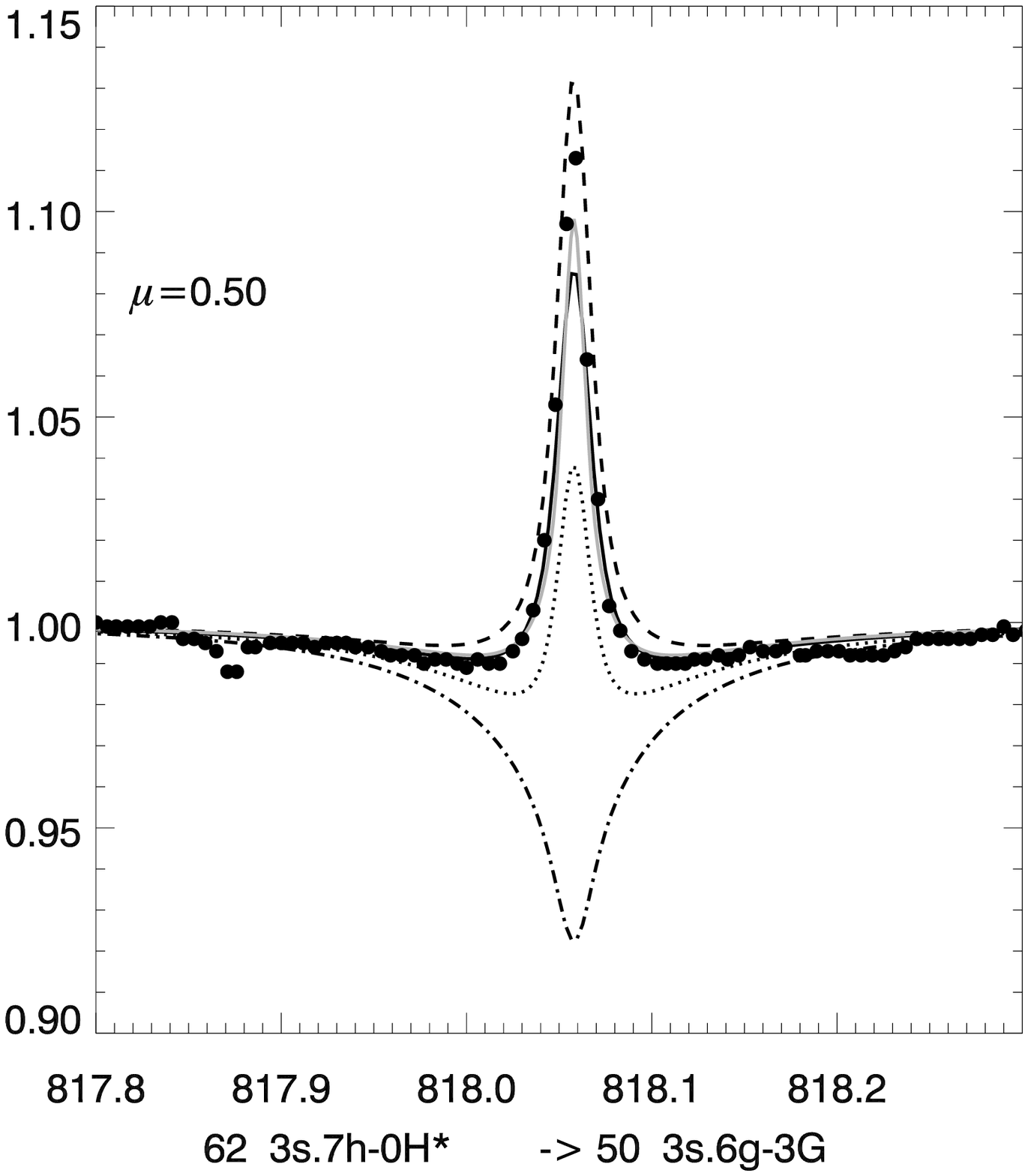}}
};
\draw [fill=white,white] (0.03\textwidth,0.00\textwidth) rectangle (0.25\textwidth,0.02\textwidth);
\end{tikzpicture}
&
\hspace{-0.02\textwidth}\begin{tikzpicture}
\node[anchor=south west, inner sep=0] (image) at (0,0) {
\subfloat{\includegraphics[width=0.32\textwidth]{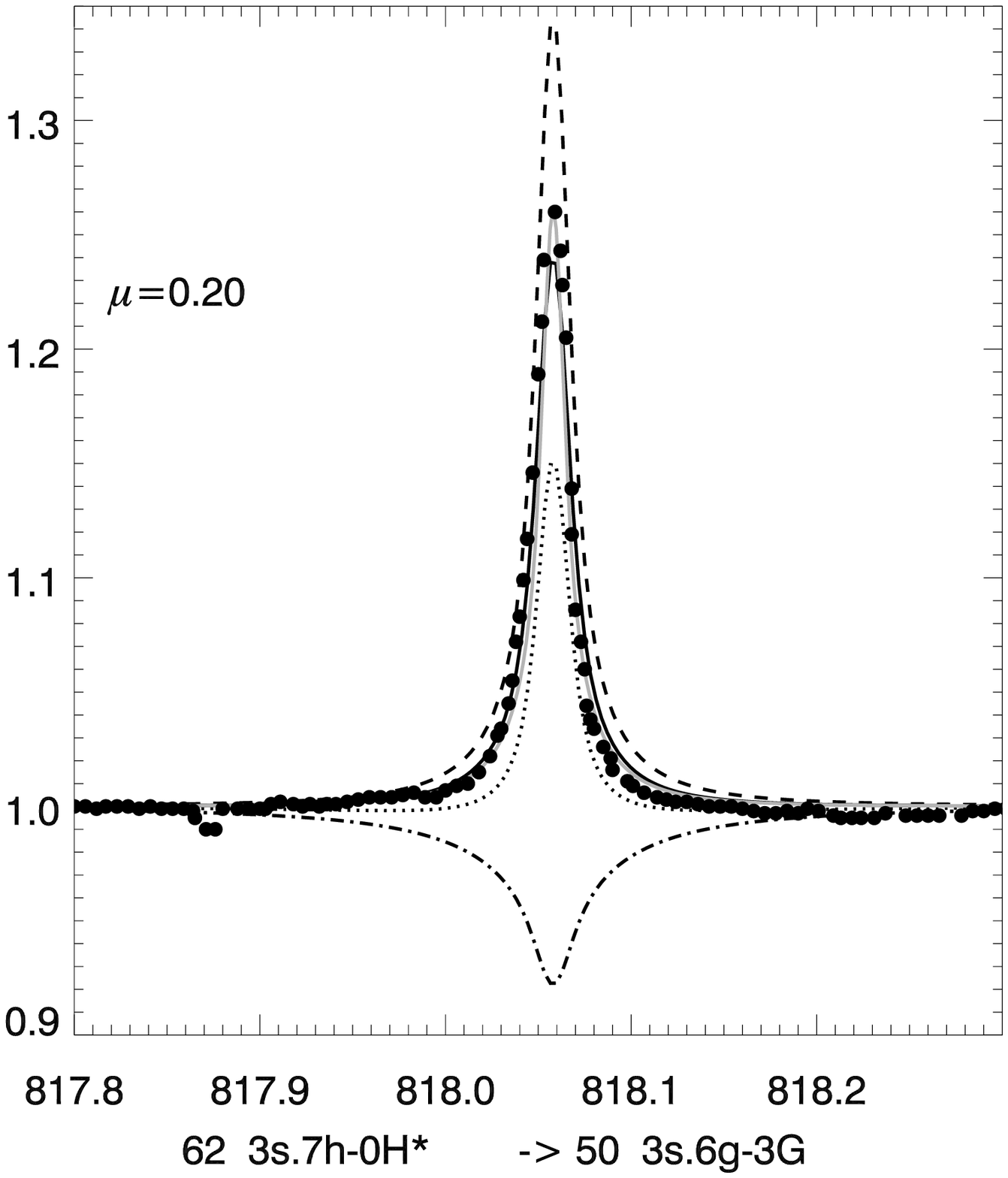}}
};
\draw [fill=white,white] (0.03\textwidth,0.00\textwidth) rectangle (0.25\textwidth,0.02\textwidth);
\end{tikzpicture} 
\\
\vspace{-0.02\textwidth}\hspace{-0.01\textwidth}\begin{tikzpicture}
\node[anchor=south west, inner sep=0] (image) at (0,0) {
\subfloat{\includegraphics[width=0.32\textwidth]{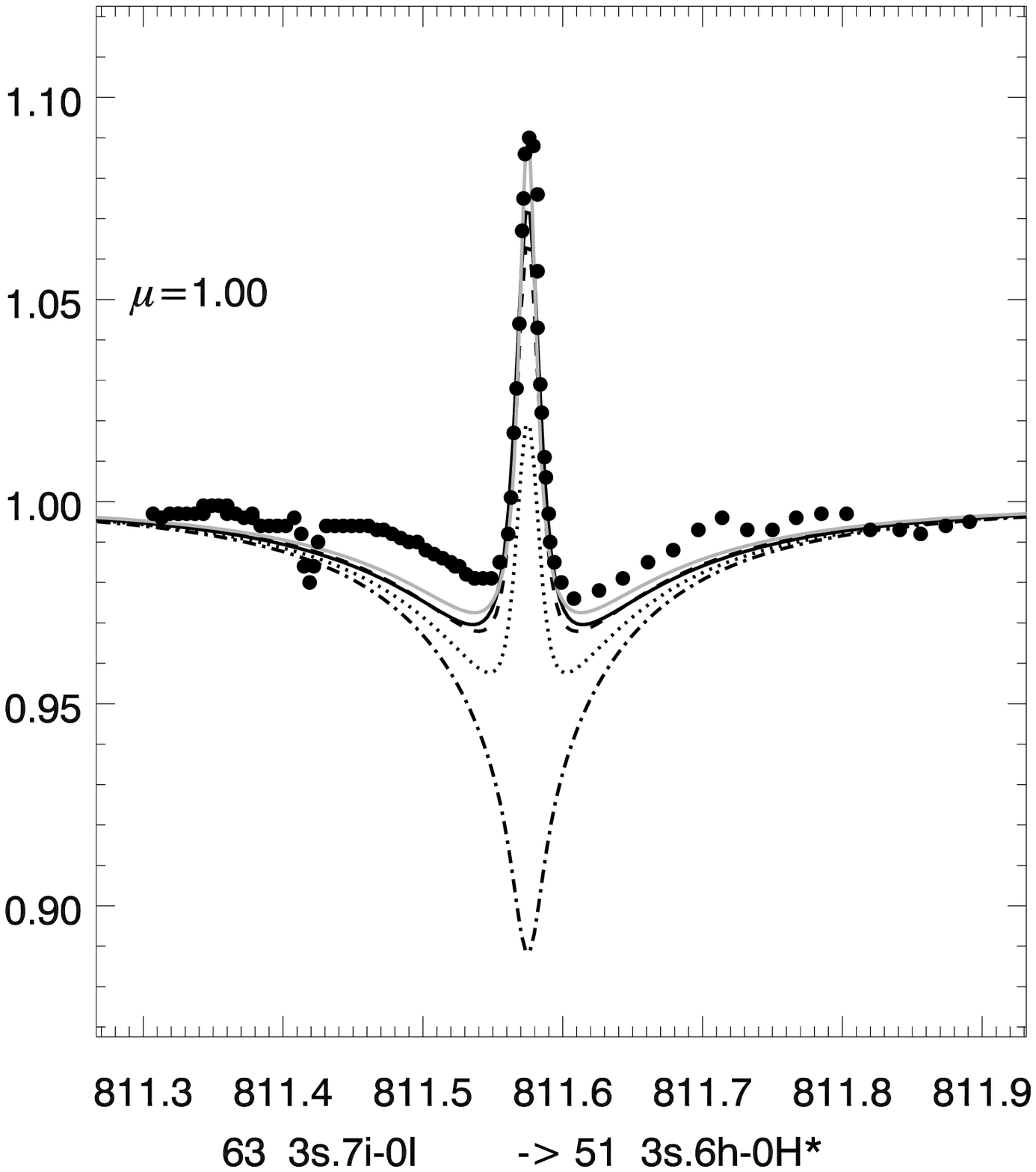}}
};
\draw [fill=white,white] (0.03\textwidth,0.00\textwidth) rectangle (0.25\textwidth,0.02\textwidth);
\node[left,rotate=90] at (-0.01\textwidth,0.25\textwidth) {\scalebox{0.7}{Normalized Intensity}};   
\node[left] at (0.23\textwidth,0.01\textwidth) {\scalebox{0.7}{Wavenumber (cm$^{-1}$)}};   
\end{tikzpicture}
&
\hspace{-0.02\textwidth}\begin{tikzpicture}
\node[anchor=south west, inner sep=0] (image) at (0,0) {
\subfloat{\includegraphics[width=0.32\textwidth]{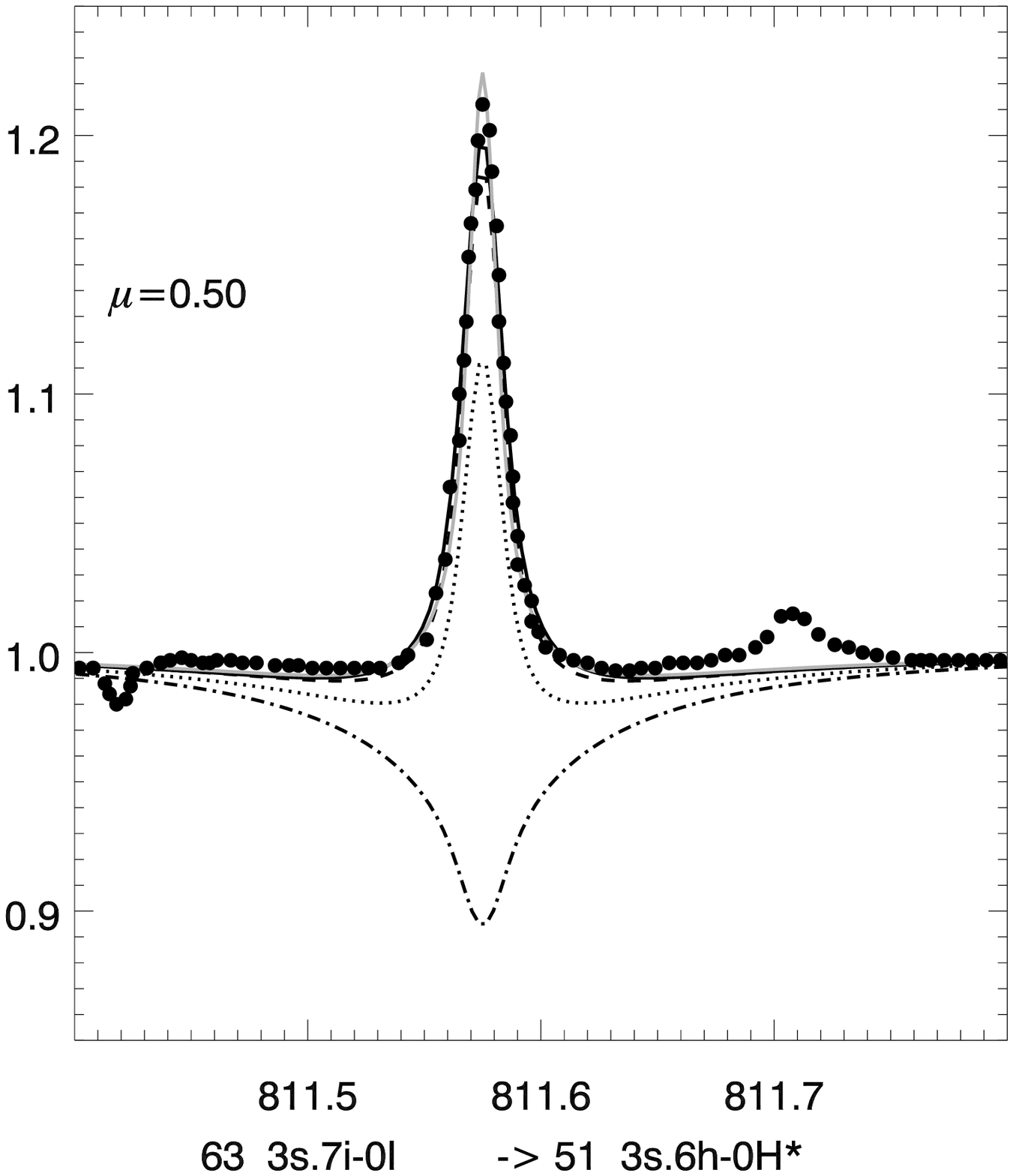}}
};
\draw [fill=white,white] (0.03\textwidth,0.00\textwidth) rectangle (0.25\textwidth,0.02\textwidth);
\node[left] at (0.23\textwidth,0.01\textwidth) {\scalebox{0.7}{Wavenumber (cm$^{-1}$)}}; 
\end{tikzpicture}
&
\hspace{-0.02\textwidth}\begin{tikzpicture}
\node[anchor=south west, inner sep=0] (image) at (0,0) {
\subfloat{\includegraphics[width=0.32\textwidth]{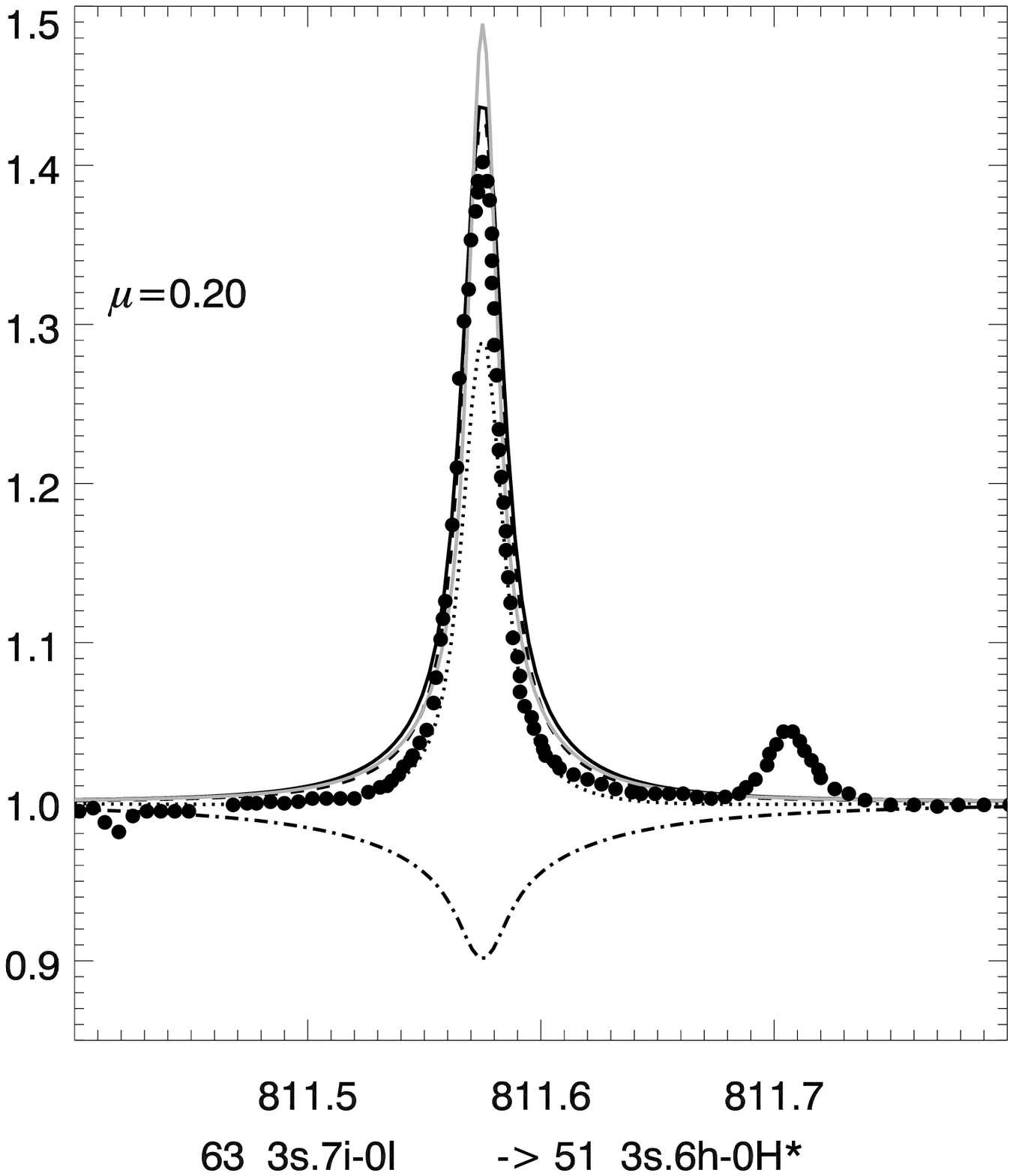}}
};
\draw [fill=white,white] (0.03\textwidth,0.00\textwidth) rectangle (0.25\textwidth,0.02\textwidth);
\node[left] at (0.23\textwidth,0.01\textwidth) {\scalebox{0.7}{Wavenumber (cm$^{-1}$)}}; 
\end{tikzpicture}\\
\end{tabular}
\caption{Solar IR line profiles at various positions on the disk.  The first row is the 7.3~$\mu$m line, the second row the 12.2~$\mu$m line, and the final row the 12.3~$\mu$m line.  The columns are $\mu = 1.0$, 0.5, and 0.2, in that order.  The circles are observations taken from \citet{1991ApJ...379L..79C} for the 7.3~$\mu$m line and from \cite{1983ApJ...269L..61B} for the two 12~$\mu$m lines. The synthetic spectral lines are calculated for various model atoms using A(\Mg)=7.6 and employing the $\langle$3D$\rangle _{1D}$ solar model atmosphere.  For the synthetic profiles, a macroturbulent velocity of 3.15 km/s and a rotational broadening of $v\sin i=1.6$~km/s are used.  For model F, the profile without macroturbulence is also shown (grey full line) to demonstrate its effect.}\label{IRlines}
\end{figure*}

Figure~\ref{IRlines} compares theoretical spectra for the 7.3, 12.2, and 12.3~$\mu$m emission lines with the observed solar spectrum across the solar disk.  The synthetic spectral lines are calculated for various model atoms, using the currently accepted solar photospheric Mg abundance \citep{Scott14}, A(\Mg)=7.59, and employing the $\langle$3D$\rangle _{1D}$ solar model atmosphere.  We note that our modelling does not include pressure-induced line shifts, and other atomic data for such shifts do not exist. Thus, we have no expectation to be able to model the asymmetries observed in the wings of these lines.

First, we note the well-known and practically obvious result that the lines cannot be reproduced in a photospheric model using LTE line formation.  Second, we see that our model F reproduces the observations both in the core and the wings rather well, but certainly not perfectly.  However, it certainly reproduces the observations better than other model atoms of interest.  As noted in Sect.~\ref{sect:electrons}, on physical grounds the vR formula is expected to overestimate the CE rates because it is based on the Born approximation.  The vR formula results in rates higher than those given by the IP method, producing stronger collisional coupling between levels of different $n$ and thus reducing the strength of the emission. The predicted line cores  for the model F(vR,$\Upsilon$) are in all cases far too weak.   In addition, we also see that a model atom with no hydrogen collision processes, model B$+$RM, produces line profiles that are generally too strong, both in the core and in the wings.  This supports the conclusion of earlier studies that additional collisional mechanisms are required.  Thus, we see that our model F employing the IP method for CE rates among Rydberg states, and the Kaulakys method for CH among Rydberg states, generally
produces the line profiles better than models using the vR method or ignoring CH.  The fact that the astrophysical comparison agrees with expectations based on the physical characteristics of the vR and IP methods, gives us confidence that our \emph{ab initio} modelling of collision rates among Rydberg states due to CE and CH is satisfactory. 

We note that the emission peaks are sensitive to macroturbulence (see \fig{IRlines}) and the wings can be better fitted with a change in abundance of order 0.1\,dex. As might be reasonably expected, the CH0 processes do not significantly affect these lines because CH0 processes only involve relatively low-lying states, and thus any effect must be indirect. 

\begin{figure}[t]
\hspace{-0.02\textwidth}\begin{tikzpicture}
\node[anchor=south west, inner sep=0] (image) at (0,0) {
\subfloat{\includegraphics[width=0.5\textwidth]{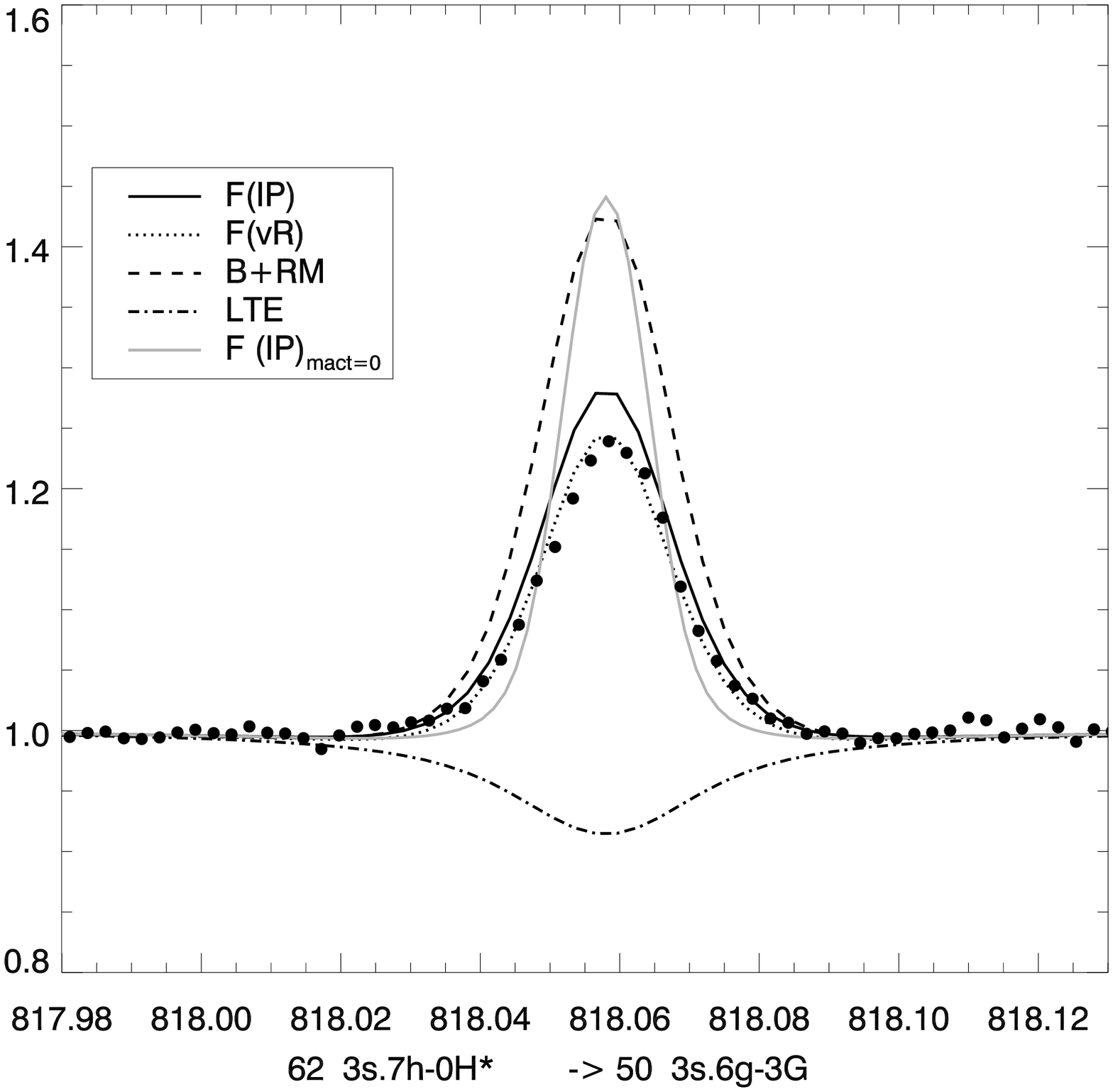}}
};
\draw [fill=white,white] (0.15\textwidth,0.01\textwidth) rectangle (0.40\textwidth,0.033\textwidth);
\node[left,rotate=90] at (0.045\textwidth,0.30\textwidth) {\scalebox{0.7}{Normalized Flux}};  
\node[left] at (0.35\textwidth,0.02\textwidth) {\scalebox{0.7}{Wavenumber (cm$^{-1}$)}}; 
\end{tikzpicture}
\caption{Profile of the 12.2$\mu$m line observed in Arcturus by Sundqvist et~al.\ (black circles) and compared with our calculations using the \avstar\ atmospheric model and the non-LTE abundance taken from Table \ref{benchmark}. The black lines show synthetic profiles using different model atoms (see Table \ref{tab:testatoms}), and are all convolved with a macroturbulent velocity of 5.2 km/s and a rotational broadening of $v\sin i=2.1$~km/s. The grey line is the profile using the F model atom but without macroturbulence. }\label{fig:aboo}
\end{figure}

\cite{2008AaA...486..985S} studied these lines in a small sample of K giants including Arcturus, in which they showed that the 12.2$\mu$m line was particularly sensitive to the modelling of hydrogen collisions.  This is due to the small influence of electron collisions in such atmospheres.  Figure \ref{fig:aboo} compares our results with their observations (kindly provided by Nils Ryde). The LTE calculation is again not able to form this line in emission. For our model atom F, the match in the wings is good, while the emission peak is stronger than in the observed spectra; however, we consider the agreement reasonable given, for example, the uncertainties in stellar parameters and that the macroturbulence and abundance have not been fine-tuned.  As is the case for the Sun, our model F reproduces the line better or as well as the other models of interest.  In particular, neglect of hydrogen collisions, model B$+$RM, again produces a peak considerably too strong.  We note that in this case, the F(vR) model performs marginally better than the F(IP) model; however, the differences are small and we judge that this is insignificant with respect to the other modelling uncertainties mentioned above.

\section{Spectral line behaviour}
\label{sect:lines}

\begin{table}[t!]
\caption{Stellar parameters of the atmospheric models used in Sec \ref{sect:lines}.}\label{teststars}
\begin{tabular}{crrrr}
{\bf Name } &  \teff & $\log g$ &   [Fe/H] & $v_{\rm{mic}}$ \\
            &  [K] & [cm/s$^2$] &   [dex] & [km/s] \\
Dwarf Poor & 6000 & 4.5 & $-$2.0 & 1.0 \\
Dwarf Rich & 6000 & 4.5 & 0.0 & 1.0 \\
Giant Poor  & 4500 & 1.0 & $-$2.0 & 2.0 \\
Giant Rich  & 4500 & 1.0 & 0.0 & 2.0 
\end{tabular} 
\end{table}

In this section we examine the behaviour of the modelled spectral lines, in particular the influence of different collisional processes on line strengths and the abundances that would be derived from the lines.  Tests were performed on a small grid of four 1D model atmospheres from the MARCS grid \citep{2008AaA...486..951G} with parameters described in Table \ref{teststars} where $v_{\rm{mic}}$ is the micro turbulent velocity used in calculating the model atmospheres. The same value is used in the line formation calculations.  Additionally, we performed a number of test calculations for the solar case, using the MARCS solar model.  We also performed calculations using the photospheric reference model with a chromosphere from \cite{1986ApJ...306..284M}, hereafter MACKKL, and the 1D-averaged 3D model for the Sun, \avsun, from \cite{2013AaA...560A...8M}.  Furthermore, to investigate possible effects of the chromosphere, we also made a hybrid model using MACKKL at $\tau_\mathrm{500 nm} > -3$, while replacing the upper layers with those from the \avsun\ model.  Where comparisons are made with solar spectra, we used a Gaussian macro-turbulence with $v_{\rm{mac}}=2.0$ km/s. 

As we are most interested in effects on the derived abundance of Mg, we studied the influence of the various collision processes on the line formation of \Mg\ lines through the non-LTE corrections, defined by 
\begin{equation}
\Delta A(\Mg)_{_{\mathrm{NLTE}-\mathrm{LTE}}}=A(\Mg)_{_\mathrm{NLTE}}-A(\Mg)_{_\mathrm{LTE}}.
\end{equation} 
The abundance corrections discussed in this section are based on matching equivalent widths.  We note that they therefore cannot be compared to the changes in abundances seen in the last section, where profile fitting was used.

For the lines to be discussed here, various non-LTE mechanisms interact to varying degrees to produce the final line profiles. Generally, overionization leads to an underpopulation of the low-lying levels of \Mgi, and a number of \Mg\ lines lead to photon losses, resulting in departures from LTE also among intermediate levels \citep[e.g. ][]{Carlsson12mu92}. These radiative mechanisms may be counteracted or even propagated by collisional processes. Charge exchange processes (CH0) drive the populations of specific intermediate levels, in particular those around 4s $^1$S, toward detailed balance with the ground state of \Mgii, which is generally close to LTE populations. CH couples low-lying and intermediate levels propagating departures from low-lying to higher levels. 

\begin{figure*}[t]
\begin{tabular}{cc}
\hspace{0.02\textwidth}\begin{tikzpicture}
\node[anchor=south west, inner sep=0] (image) at (0,0) {
\subfloat{\hspace{-0.04\textwidth}\includegraphics[width=0.5\textwidth]{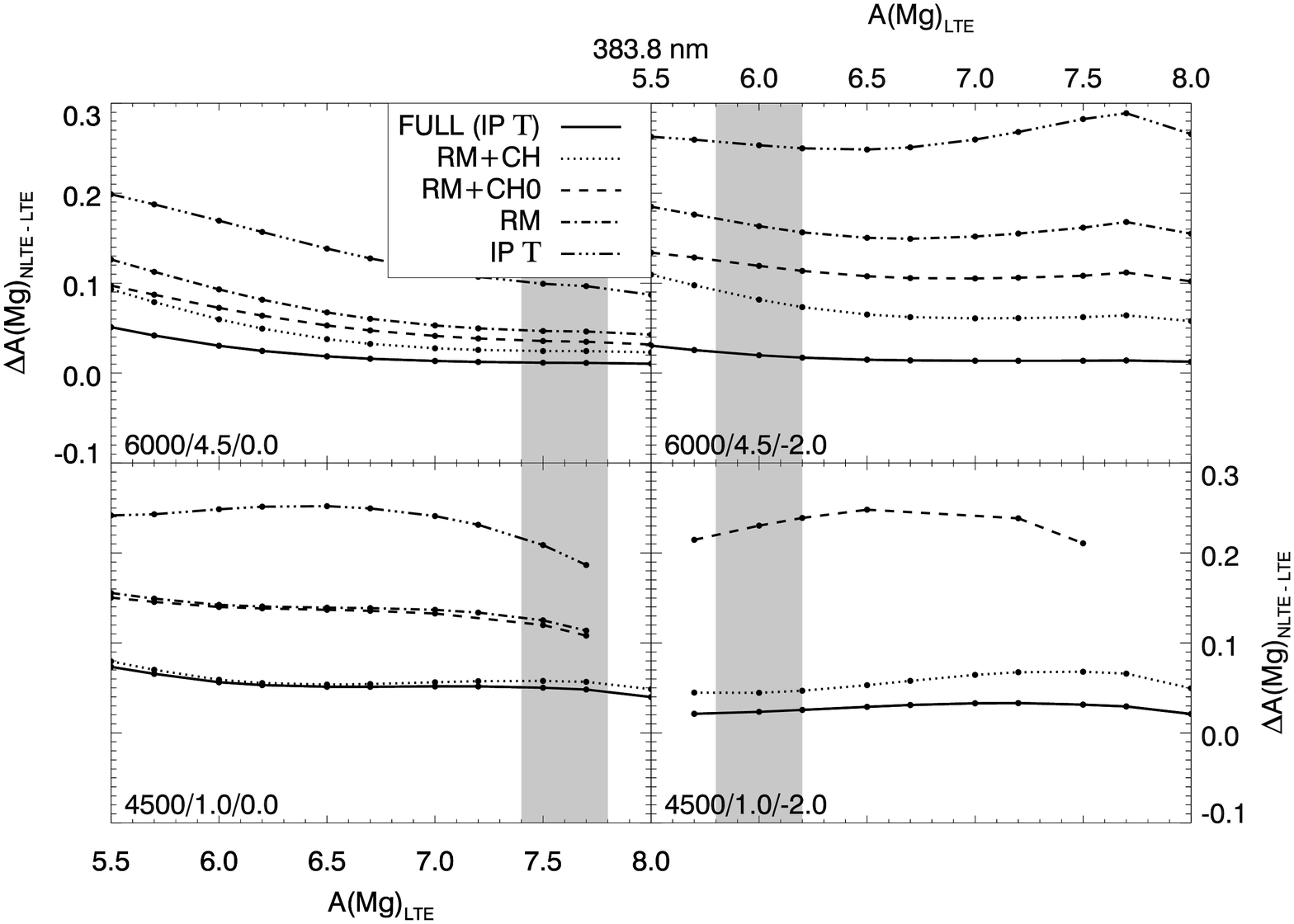}}
};
\draw [fill=white,white] (0.115\textwidth,0.240\textwidth) rectangle (0.175\textwidth,0.297\textwidth);
\node[left] at (0.180\textwidth,0.293\textwidth) {\scalebox{0.5}{F}};   
\node[left] at (0.180\textwidth,0.281\textwidth) {\scalebox{0.5}{B$+$RM$+$CH}};   
\node[left] at (0.180\textwidth,0.269\textwidth) {\scalebox{0.5}{B$+$RM$+$CH0}};   
\node[left] at (0.180\textwidth,0.257\textwidth) {\scalebox{0.5}{B$+$RM}};   
\node[left] at (0.180\textwidth,0.245\textwidth) {\scalebox{0.5}{B}};   
\end{tikzpicture}
&
\hspace{0.02\textwidth}\begin{tikzpicture}
\node[anchor=south west, inner sep=0] (image) at (0,0) {
\subfloat{\hspace{-0.04\textwidth}\includegraphics[width=0.5\textwidth]{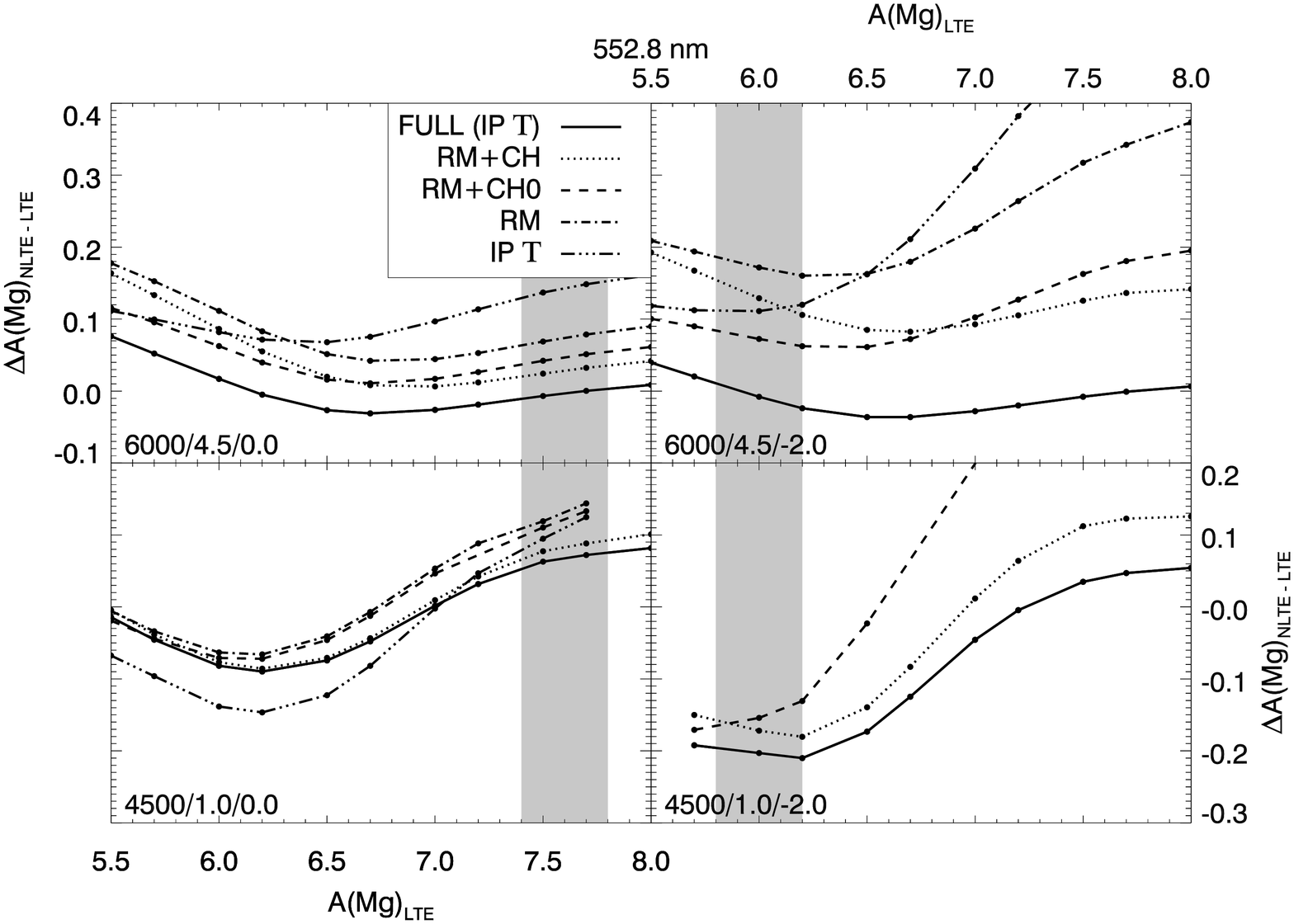}}
};
\draw [fill=white,white] (0.115\textwidth,0.240\textwidth) rectangle (0.175\textwidth,0.297\textwidth);
\node[left] at (0.180\textwidth,0.293\textwidth) {\scalebox{0.5}{F}};   
\node[left] at (0.180\textwidth,0.281\textwidth) {\scalebox{0.5}{B$+$RM$+$CH}};   
\node[left] at (0.180\textwidth,0.269\textwidth) {\scalebox{0.5}{B$+$RM$+$CH0}};   
\node[left] at (0.180\textwidth,0.257\textwidth) {\scalebox{0.5}{B$+$RM}};   
\node[left] at (0.180\textwidth,0.245\textwidth) {\scalebox{0.5}{B}};   
\end{tikzpicture}
\\
\hspace{0.02\textwidth}\begin{tikzpicture}
\node[anchor=south west, inner sep=0] (image) at (0,0) {
\subfloat{\hspace{-0.04\textwidth}\includegraphics[width=0.5\textwidth]{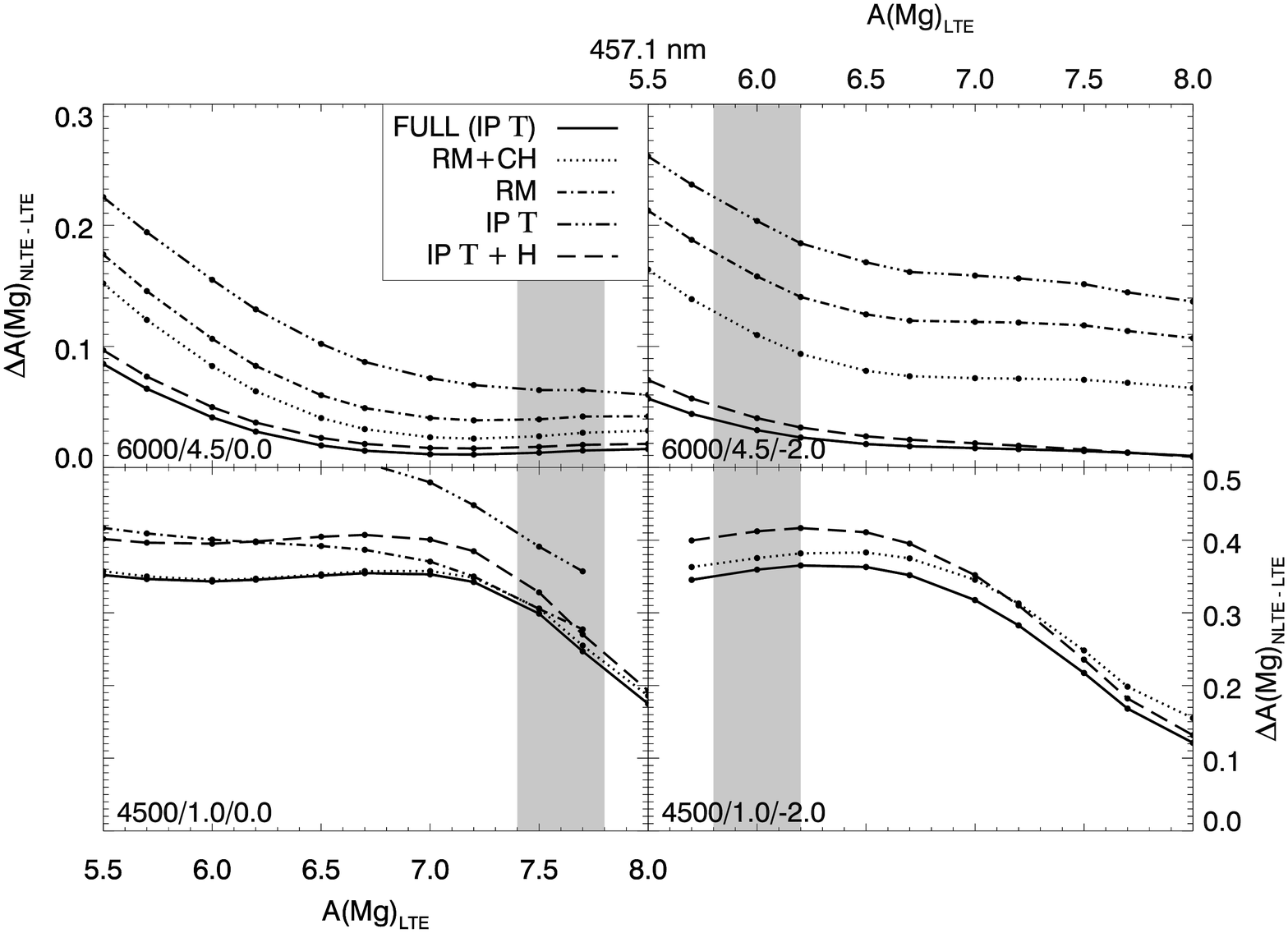}}
};
\draw [fill=white,white] (0.115\textwidth,0.240\textwidth) rectangle (0.175\textwidth,0.297\textwidth);
\node[left] at (0.180\textwidth,0.293\textwidth) {\scalebox{0.5}{F}};   
\node[left] at (0.180\textwidth,0.281\textwidth) {\scalebox{0.5}{B$+$RM$+$CH}};   
\node[left] at (0.180\textwidth,0.269\textwidth) {\scalebox{0.5}{B$+$RM}};   
\node[left] at (0.180\textwidth,0.257\textwidth) {\scalebox{0.5}{B}};   
\node[left] at (0.180\textwidth,0.245\textwidth) {\scalebox{0.5}{B$+$CH$+$CH0}};   
\end{tikzpicture}
&
\hspace{0.02\textwidth}\begin{tikzpicture}
\node[anchor=south west, inner sep=0] (image) at (0,0) {
\subfloat{\hspace{-0.04\textwidth}\includegraphics[width=0.5\textwidth]{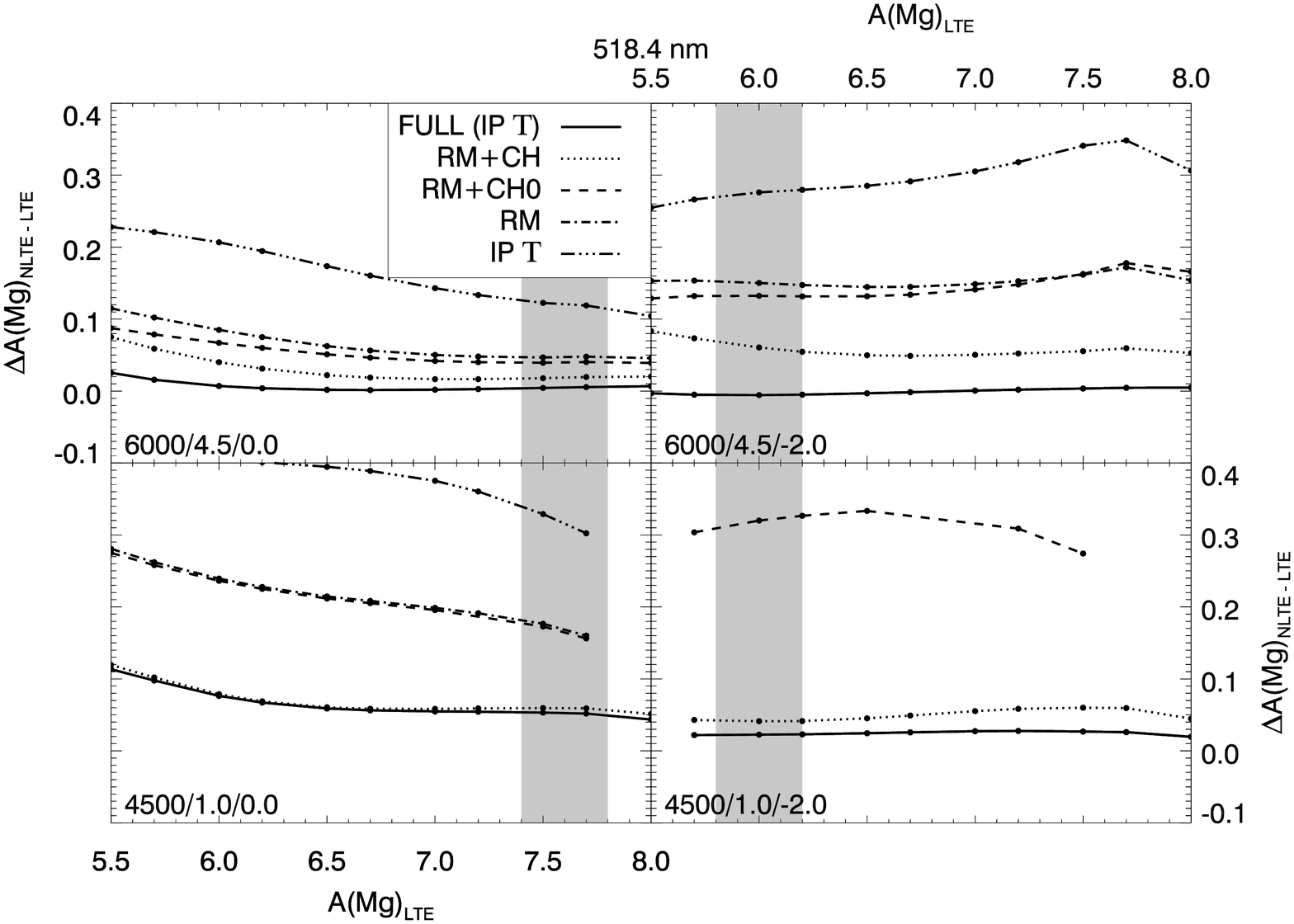}}
};
\draw [fill=white,white] (0.115\textwidth,0.240\textwidth) rectangle (0.175\textwidth,0.297\textwidth);
\node[left] at (0.180\textwidth,0.293\textwidth) {\scalebox{0.5}{F}};   
\node[left] at (0.180\textwidth,0.281\textwidth) {\scalebox{0.5}{B$+$RM$+$CH}};   
\node[left] at (0.180\textwidth,0.269\textwidth) {\scalebox{0.5}{B$+$RM$+$CH0}};   
\node[left] at (0.180\textwidth,0.257\textwidth) {\scalebox{0.5}{B$+$RM}};   
\node[left] at (0.180\textwidth,0.245\textwidth) {\scalebox{0.5}{B}};   
\end{tikzpicture}
\\
\hspace{0.02\textwidth}\begin{tikzpicture}
\node[anchor=south west, inner sep=0] (image) at (0,0) {
\subfloat{\hspace{-0.04\textwidth}\includegraphics[width=0.5\textwidth]{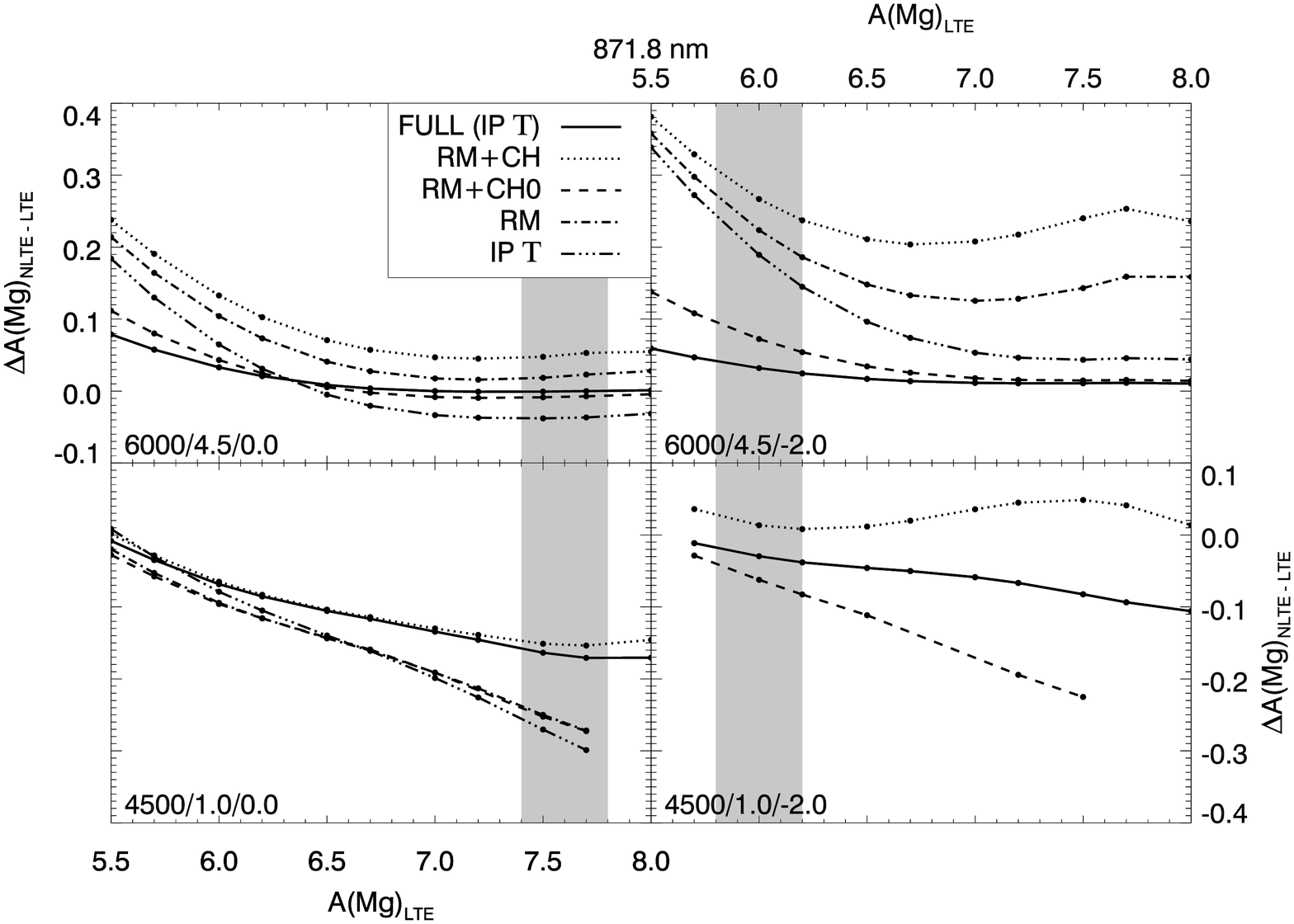}}
};
\draw [fill=white,white] (0.115\textwidth,0.240\textwidth) rectangle (0.175\textwidth,0.297\textwidth);
\node[left] at (0.180\textwidth,0.293\textwidth) {\scalebox{0.5}{F}};   
\node[left] at (0.180\textwidth,0.281\textwidth) {\scalebox{0.5}{B$+$RM$+$CH}};   
\node[left] at (0.180\textwidth,0.269\textwidth) {\scalebox{0.5}{B$+$RM$+$CH0}};   
\node[left] at (0.180\textwidth,0.257\textwidth) {\scalebox{0.5}{B$+$RM}};   
\node[left] at (0.180\textwidth,0.245\textwidth) {\scalebox{0.5}{B}};   
\end{tikzpicture}
&
\end{tabular} 
\caption{Abundance corrections for the 3838~\AA, 5528~\AA, 4571~\AA, 5184~\AA, and 8718~\AA\ lines in four different atmospheric models and using modified model atoms as described in the text.  The model parameters are given at the bottom left of each figure with \teff/\logg/[Fe/H].  The shaded region indicates the likely abundance of Mg for a normal star of given metallicity.  The black circles show the calculated abundances; some lines have fewer points because the calculations with those parameters did not converge.  Note that the model atoms shown in the 4571~\AA\ line case differ from the others.}\label{4panel_all}
\end{figure*}

\subsection{\Mgi\ lines}

\paragraph{3829, 3832, and  3838 \AA\ ($3d \, ^3\mathrm{D}$--$3p \, ^3\mathrm{P}^o$):}

The UV triplet is one of the strongest features of \Mgi\ and is often used for abundance determinations in metal-poor stars. In solar metallicity stars these lines are significantly blended, and also sensitive to chromospheric effects. The abundance corrections and their sensitivity to collisional processes are presented in \fig{4panel_all}.  Model B leads to very large corrections for these lines. However, in model F abundance corrections are general quite small.  In dwarfs the corrections are practically non-existent, but if collisions with \H\ were not included, we would find corrections of $\sim 0.2$ dex in the metal-poor case.  Thus, in dwarfs CH and CH0 play an important role in the abundance correction calculation.  In giants, corrections are also small, $\la$ 0.05 dex for abundance values of interest, and neglecting CH leads to much larger corrections, of about 0.2~dex in the metal-poor case. This means that in giants, CH stronglycontributes  to the abundance correction and CH0 has a negligible  effect. If hydrogen collisions are included, the use of more approximate models for CE, such as F(vR, $\Omega=1$) or F(IP, $\Omega=1$) does not change the results significantly.  
Spin change transitions have only a weak effect in the two dwarf stellar models, the correction changing by $\sim0.05$ dex in the metal-poor case for the F$-$S model.  In giants, the F$-$S model changes the corrections by $\sim0.07$ dex with the relative effect of SCE being strongest at solar metallicity and the effect of SCH strongest at low metallicity.

\paragraph{4167, 4703, 5528, and  8806 \AA\ ($nd \, ^1\mathrm{D}$--$3p \, ^1\mathrm{P}^o $): }

For a given stellar model, all four of these lines behave similarly, but the effects becomes stronger at longer wavelengths, where the lines form at shallower depths with larger population departures.  Figure~\ref{4panel_all} shows the abundance corrections for the 5528~\AA\ line.  Model B leads to large corrections in dwarfs, of about 0.2 dex at the abundances of most interest.  The corrections with model F are, however, small.  The results for 4167, 4703,
and 5528 were most sensitive to the introduction of hydrogen collisions generally (both CH and CH0), while for the 8806 line CH was particularly important.  This is because the populations of the levels involved are similarly affected by the introduction of CH0; the upper level of the 8806~\AA\ transition, 3d $^1$D, is low-lying and both levels are thermalized by CH0 to similar degrees in the line formation region, thus leaving S$_\nu$ unaltered.

\paragraph{4571 \AA\ ($3p\,^3\mathrm{P}^o_1$--$3s\,^1\mathrm{S}$):  }

The intercombination line, despite its weak oscillator strength, is observable across a wide range of stellar parameters because it involves the ground state of \Mgi.  The abundance corrections are shown in \fig{4panel_all}. In dwarfs, at abundances of typical interest, this line is formed close to LTE conditions as the collisional coupling of the two levels in different spin systems dominates the weak radiative coupling.  In giant models where collisional coupling is significantly decreased, this line shows significant departures from LTE, of the order of 0.2 dex at solar metallicity and almost 0.4 dex in the metal-poor giant model. In metal-rich giants, CH0 is much weaker than either CH and RM. This line is expected to be sensitive to spin change transitions. CE rates between the upper and lower levels are $10^8$ times higher than the corresponding CH rates, making this line most sensitive to spin change CE rates even in very metal-poor stars.

\paragraph{5167, 5172, and 5184 \AA\ ($4s\,^3\mathrm{S}$--$3p\,^3\mathrm{P}^o$):}        

The abundance corrections for one of the \Mgi\ b triplet lines are shown in \fig{4panel_all}. Model B gives large corrections for these lines, but model F gives corrections that are very small for dwarfs, and quite small for giants, lower than 0.05 dex.  As with the 8806~\AA\ line, the departure coefficients of the upper and lower levels are similarly affected by CH0, and thus CH0 has little effect on the line strength. The main change is due to the introduction of CH.  CH decreases the population of the upper level in the line formation region and increases the populations of the lower levels, bringing them closer to relative LTE.  

\begin{figure*}[ht!]
\begin{tabular}{cc}
\hspace{-0.03\textwidth}\begin{tikzpicture}
\centering
\node[anchor=south west, inner sep=0] (image) at (0,0) {
\subfloat{\hspace{-0.03\textwidth}\includegraphics[width=0.55\textwidth]{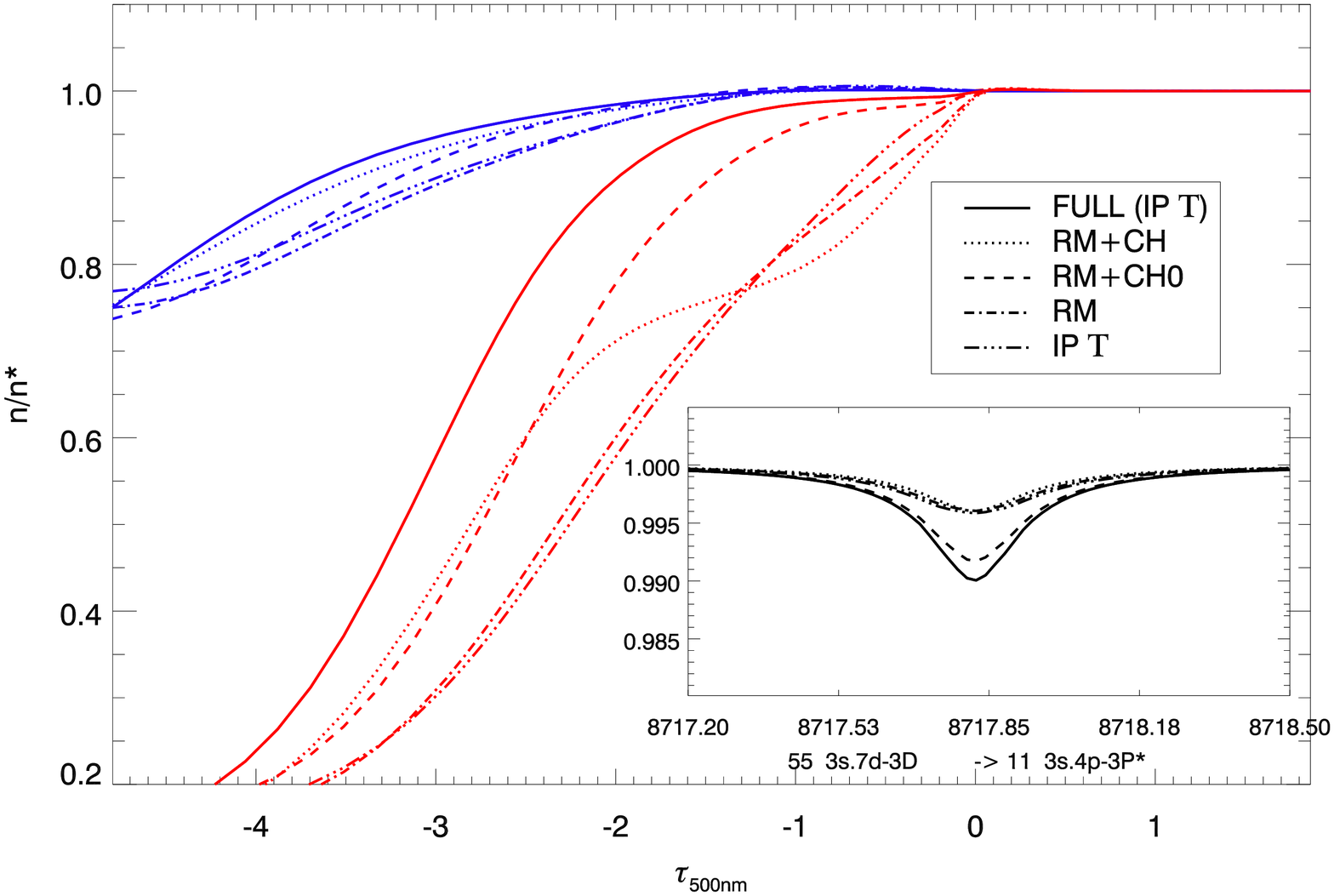}}
};
\draw [fill=white,white] (0.389\textwidth,0.210\textwidth) rectangle (0.46\textwidth,0.272\textwidth);
\node at (0.4\textwidth,0.320\textwidth) {\scalebox{0.7}{6000/4.5/-2.0 ; A[\ion{Mg}{}]=6.0}};
\node[right] at (0.385\textwidth,0.266\textwidth) {\scalebox{0.6}{F}};   
\node[right] at (0.385\textwidth,0.253\textwidth) {\scalebox{0.6}{B$+$RM$+$CH}};   
\node[right] at (0.385\textwidth,0.241\textwidth) {\scalebox{0.6}{B$+$RM$+$CH0}};   
\node[right] at (0.385\textwidth,0.228\textwidth) {\scalebox{0.6}{B$+$RM}};   
\node[right] at (0.385\textwidth,0.215\textwidth) {\scalebox{0.6}{B}};   
\end{tikzpicture}
& 
\hspace{-0.05\textwidth}\begin{tikzpicture}
\centering
\node[anchor=south west, inner sep=0] (image) at (0,0) {
\subfloat{\hspace{-0.03\textwidth}\includegraphics[width=0.55\textwidth]{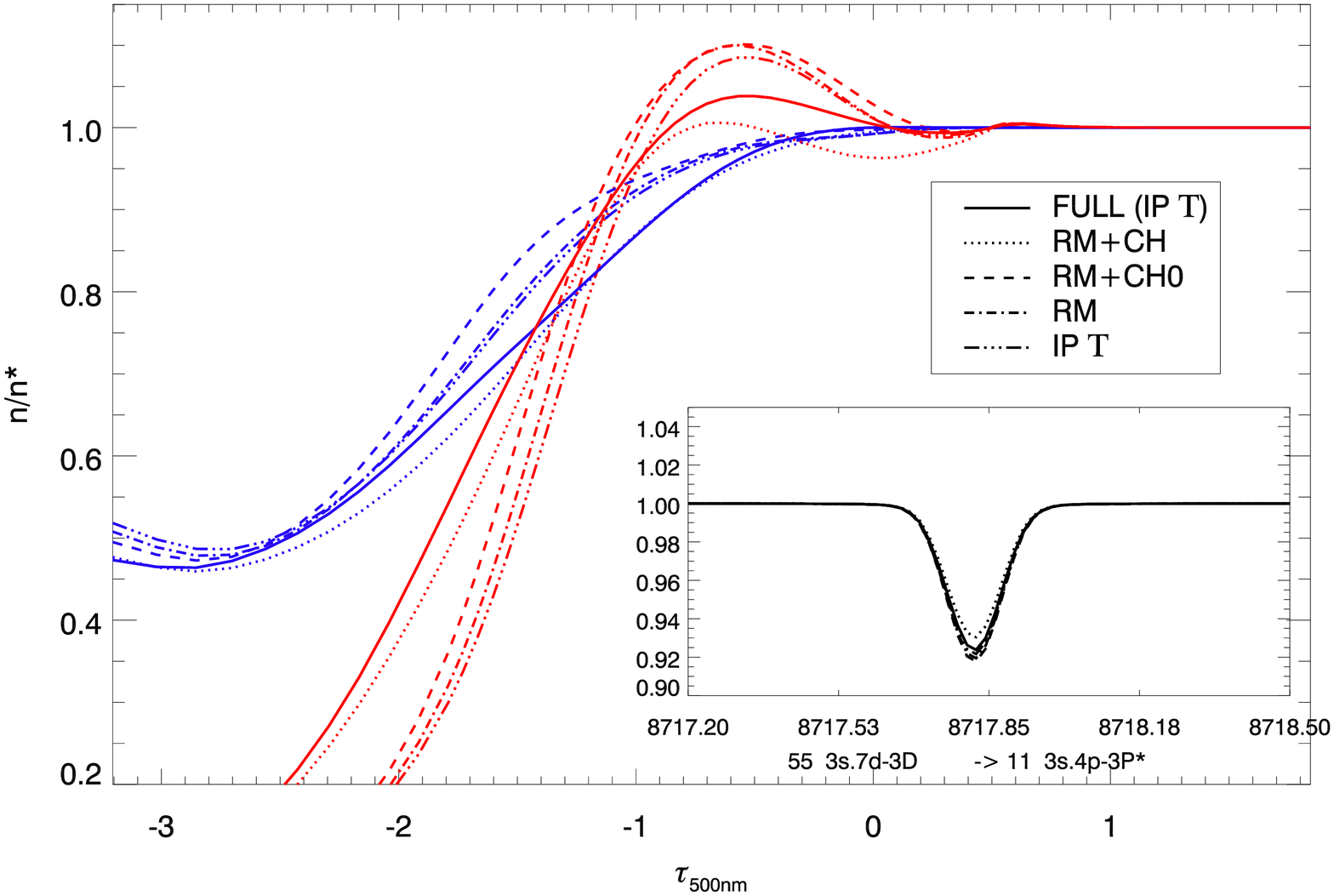}}
};
\draw [fill=white,white] (0.389\textwidth,0.210\textwidth) rectangle (0.46\textwidth,0.272\textwidth);
\node at (0.41\textwidth,0.320\textwidth) {\scalebox{0.7}{4500/1.0/-2.0 ; A[\ion{Mg}{}]=6.0}};
\node[right] at (0.385\textwidth,0.266\textwidth) {\scalebox{0.6}{F}};   
\node[right] at (0.385\textwidth,0.253\textwidth) {\scalebox{0.6}{B$+$RM$+$CH}};   
\node[right] at (0.385\textwidth,0.241\textwidth) {\scalebox{0.6}{B$+$RM$+$CH0}};   
\node[right] at (0.385\textwidth,0.228\textwidth) {\scalebox{0.6}{B$+$RM}};   
\node[right] at (0.385\textwidth,0.215\textwidth) {\scalebox{0.6}{B}};   
\end{tikzpicture}
\end{tabular}
\caption{Departure coefficients $b=n/n_{_\mathrm{LTE}}$ of the levels involved in the 8718~\AA\ line, 7d (blue) and 4p (red).  The left panel shows the metal-poor dwarf case, the right panel the metal-poor giant case.  The insets show the corresponding line profiles.  Note that in the giant case, abundance corrections corresponding to some models for which $b$ are shown here are not shown in \fig{4panel_all} because the models did not converge for at least three abundance values.}\label{bcoeff8718dp} 
\end{figure*} 

\paragraph{8736~\AA\  ($7f\,^3\mathrm{F}^o$--$3d\,^3\mathrm{D}$)  and 8710, 8713, and 8718~\AA\  (\mbox{$7d
\,^3\mathrm{D}$--$4p\,^3\mathrm{P}^o$}): }

These lines are $n=7 \leftrightarrow n'=3$ transitions involving triplet states, that is, between low-lying and high-lying levels. The abundance corrections for one of the lines are shown in \fig{4panel_all}.
These lines show examples of behaviour where the addition of collision processes can drive away from LTE, rather than towards it.  As these levels lie approximately 3--4 eV from the continuum, they are subject to overionisation from ultraviolet radiation.  In the solar case, \cite{Carlsson12mu92} have shown that the triplet $3d$ and $4p$ levels are among those involved in transitions that become optically thin in the photosphere, which in turn leads to photon losses and thus photon suction.  Now, together with these radiatively driven non-LTE processes, the CH and CH0 processes involving these states are among the strongest.  

The complex interplay of these processes leads to the situation seen in \fig{4panel_all}, where rather than the more usual convergence from model B towards the F model, the introduction of additional collisional processes has less predictable effects. For example, in the metal-poor dwarf case, with model B, rather large corrections are seen.  Introduction of the RM data via model B$+$RM leads to slightly increased corrections.  Then, contrary to the naive expectation that additional collisions drive towards LTE, the introduction of CH, model B$+$RM$+$CH actually increases the abundance corrections.  The left panel of \fig{bcoeff8718dp} shows departure coefficients of the levels involved in the formation of the 8718~\AA\ line in a metal-poor giant atmosphere, where we see that the introduction of CH leads to populations of the $4p$ level at around $\log(\tau_\mathrm{500 nm}) = -2$ to $-3$ closer to the LTE population, but the population at around $-1$, where the line forms, is in fact farther from LTE.  This occurs because deep in the atmosphere, CH processes lead to increased coupling with the lower-lying levels, especially $4s$ and $3p$ levels, which are subject to strong photoionisation and thus violent overionisation, leading to underpopulation.  However, we see in this model atmosphere that the dominant change is due to the introduction of CH0, which increases the coupling between the $4s$, $3d$ and $4p$ levels with the continuum and thus brings the population closer to relative equilibrium with Mg II and thus towards LTE populations, particularly in the line-forming region.  For the metal-poor giant model, with departure coefficients shown in the right panel of \fig{bcoeff8718dp}, model B shows an overpopulation of the $4p$ level in the line-forming region $\log(\tau_\mathrm{500 nm}) = -1$ caused by  photon losses.  The introduction of CH processes again pulls the populations towards those of the lower-lying levels, which are underpopulated between $\log(\tau_\mathrm{500 nm}) = -1$ and 0.5.  This causes an underpopulation at large depths, but brings the population to near LTE levels in the line-forming region.  Introduction of CH0 processes brings the population deep in the atmosphere back towards LTE, which in turn reintroduces a slight overpopulation in the line-forming region.

\subsection{\Mgii\ lines}

Since \Mgii\ is the dominant ion of \Mg\ in late-type stars, non-LTE effects are expected to be weaker in \Mgii\ than in \Mgi\ lines. In the four test model atmospheres studied here, the abundance corrections all had magnitudes lower than 0.1 dex for the lines
we analysed (see table \ref{linestable}).

\section{Conclusions}
\label{sect:conc}

We have presented a new model atom for Mg suitable for non-LTE line formation studies in late-type stars.  It includes data for electronic collision rates calculated in this paper by the $R$-matrix method, and we have further used these results to formulate a method to estimate transition rates for forbidden transitions not covered by the $R$-matrix calculations, and not possible to be estimated by the vR and IP methods.  Recent data for hydrogen collisions, including charge transfer processes, calculated by some of us, were also applied.  
Some new data for collisional broadening were calculated as well.  The new model atom, used together with 1D model atmospheres in the context of standard non-LTE modelling, compares well with observations.  The modelled spectra agree well with observed spectra from benchmark stars, showing much better agreement with line profile shapes than LTE modelling. The line-to-line scatter in the derived abundances shows some improvements compared to LTE  (where the cores of strong lines must often be ignored), particularly when coupled with averaged 3D models. The observed \Mg\ emission features at 7 and 12~$\mu$m in the spectra of the Sun and Arcturus, which are sensitive to the collision data, are reasonably well reproduced.

The spectral line behaviour and uncertainties were explored by extensive experiments in which sets of collisional data were changed or removed on a small grid of theoretical model atmospheres.  Charge transfer in collisions with \H\ is generally important as a thermalising mechanism in dwarfs, less so in giants.  Excitation due to collisions with \H\ was found to be quite important in both giants and dwarfs.  The $R$-matrix calculations for electron collisions also lead to significant differences compared to when approximate formulas are employed.  The modelling predicts non-LTE abundance corrections $\Delta A(\Mg)_{_{\mathrm{NLTE}-\mathrm{LTE}}}$ in dwarfs, both solar metallicity and metal-poor, to be very small (of about $0.01$dex), even smaller than found in previous studies.  In giants, corrections vary greatly between lines, but can be as large as 0.4 dex.  Our results emphasise the need for accurate data for electron and hydrogen collisions for precise non-LTE predictions of stellar spectra, but demonstrate that such data can be calculated and that \emph{ab initio} non-LTE modelling without resort to free parameters is possible.

Only two other elements, Li and Na, have collisional data for non-LTE modelling of comparable quality to that which has been used in this work.   RM calculations have been performed for excitation of low-lying levels of Li and Na by electron collisions by \cite{2011AaA...529A..31O} and \cite{Gao:2010gg}, respectively.  Before calculating the RM data, the data of \cite{park1971} were used in non-LTE modelling, and these data typically agree to within 80\% of the RM data, significantly better than the vR or IP methods.  Thus, the new RM data for electron collisions did not have a strong effect on non-LTE modelling, since the existing data were relatively good; see \cite{2011AaA...529A..31O} for Li, \cite{2011AaA...528A.103L} for Na.  Data for hydrogen collision processes, including charge transfer, have been calculated for \ion{Li}{} by \cite{2003AaA...409L...1B},  and for \ion{Na}{} by \cite{2010AaA...519A..20B}, and used in non-LTE studies by \citeauthor{2003AaA...409L...1B} and \cite{2009AaA...503..541L} in the case of Li and \cite{2011AaA...528A.103L} for Na.  These studies found for both elements that CH processes have practically no effect on the non-LTE modelling, while CH0, in particular that involving the first excited $S$-state, had a significant effect because it brought this level into closer detailed balance with the continuum.   These are, however, alkalis with only one spin system.  In contrast, \ion{Mg}{} has two spin systems, leading to the added feature of intersystem transitions. Furthermore, many more lines of \Mg\ are observable in stellar spectra from IR to UV.  Our modelling has shown that the situation for slightly more complex atoms such as \Mg\ is far more complicated than for alkalis, with the new data for CE, CH, and CH0 all affecting at least some lines.

We plan to use the current model atom to produce grids of departure coefficients and abundance corrections in 1D and  \avstar\ theoretical model atmospheres, specifically the MARCS \citep{2008AaA...486..951G} and STAGGER \citep{2013AaA...560A...8M} grids. Grids of abundance corrections can be directly applied to studies performed in LTE.  However, this depends on the use of equivalent widths, which is only strictly useful for weak lines.  Grids of departure coefficients allow non-LTE corrections to be made at the spectrum synthesis level, and thus is far more robust when using stronger lines. We have demonstrated that non-LTE calculations are necessary for Galactic archaeology studies with large spectroscopic surveys of the Milky-Way. In particular, Mg abundances of giant stars suffer from strong differential non-LTE effects with respect to dwarfs and subgiants, comparable to or exceeding the targeted precision. Using grids of departure coefficients computed with our model atom, on-going and future surveys can properly model these effects.

Finally, we note that a comparison with centre-to-limb variations of optical lines in the Sun would provide a strong complement to the tests of the modelling performed here. Such an observational campaign is underway with the Swedish Solar Telescope and will be the subject of future work. Ideally, this should be done in the context of full 3D non-LTE modelling.

\begin{acknowledgements} 
We thank Nils Ryde for providing the 12$\mu$m spectrum of Arcturus.  This work was supported by the G\"{o}ran Gustafssons Stiftelse, the Royal Swedish Academy of Sciences, the Wenner-Gren Foundation and the Swedish Research Council.  P.S.B. is a Royal Swedish Academy of Sciences Research Fellow supported by a grant from the Knut and Alice Wallenberg Foundation. P.S.B. was also supported by the project grant ``The New Milky'' from the Knut and Alice Wallenberg foundation. A.K.B. also gratefully acknowledges partial support from the Russian Ministry of Education and Science. A.S. and N.F. acknowledge support from the French CNRS-PNPS (Programme National de Physique Stellaire) and  the GAIA program of Paris Observatory.  


\end{acknowledgements}

\bibliographystyle{aa}                  
\bibliography{MgNLTE,papers2}           
\end{document}